\documentclass[referee,pdflatex,sn-basic]{sn-jnl}

\usepackage{graphicx}%
\usepackage{multirow}%
\usepackage{amsmath,amssymb,amsfonts}%
\usepackage{amsthm}%
\usepackage{mathrsfs}%
\usepackage[title]{appendix}%
\usepackage{xcolor}%
\usepackage{textcomp}%
\usepackage{manyfoot}%
\usepackage{booktabs}%
\usepackage{algorithm}%
\usepackage{algorithmicx}%
\usepackage{algpseudocode}%
\usepackage{listings}%
\usepackage{makecell}%

\theoremstyle{thmstyleone}%
\theoremstyle{thmstyletwo}%

\theoremstyle{thmstylethree}%

\begin{document}

\title[Venus Cloud Research: Progress and Perspectives]{Venus Cloud Research: Progress and Perspectives}


\author[1]{\fnm{Longkang} \sur{Dai}}

\author[2]{\fnm{Dmitrij V.} \sur{Titov}}

\author[3]{\fnm{Wencheng D.} \sur{Shao}}

\author[4]{\fnm{Xi} \sur{Zhang}}

\author[2]{\fnm{Jun} \sur{Cui}}

\author*[1]{\fnm{Siteng} \sur{Fan}}\email{fanst@sustech.edu.cn}

\affil*[1]{\orgdiv{Department of Earth and Space Sciences}, \orgname{Southern University of Science and Technology}, \orgaddress{\street{Xueyuan Avenue}, \city{Shenzhen}, \postcode{518055}, \state{Guangdong}, \country{China}}}

\affil[2]{\orgdiv{Planetary Environmental and Astrobiological Research Laboratory (PEARL), School of Atmospheric Sciences}, \orgname{Sun Yat-sen University}, \orgaddress{\street{Tangjiawan Town}, \city{Zhuhai}, \postcode{519028}, \state{Guangdong}, \country{China}}}

\affil[3]{\orgdiv{National Space Institute}, \orgname{Technical University of Denmark}, \orgaddress{\street{Anker Engelunds Vej}, \city{Lyngby}, \postcode{DK-2800}, \country{Denmark}}}

\affil[4]{\orgdiv{Department of Earth and Planetary Sciences}, \orgname{University of California Santa Cruz}, \orgaddress{\street{1156 High St}, \city{Santa Cruz}, \postcode{95064}, \state{CA}, \country{USA}}}


\abstract{Venus has regained attention on the international stage with the approval of three new missions by ESA and NASA. As the twin sister of Earth, Venus exhibits a distinct atmosphere, which casts a veil of mystery over the planetary evolution and is of great scientific significance. One of the most important components of Venus–the cloud–is believed to have significantly regulated its climate evolution and affect the environmental habitability. However, due to sparse in-situ measurements and the limitation of remote sensing, properties of these clouds remain largely unknown. Based on research conducted in past decades, this article reviews the observational structure of cloud properties, the progress of microphysical and simplified cloud model developments, and perspectives of future directions of this research field. Several possible solutions to the challenges associated with the coupling effect, ultraviolet absorption, and habitability are proposed and discussed in details, providing insights for future Venus’ explorations.}




\maketitle

\section{Introduction}\label{sec1}

As the twin-sister planet of the Earth, Venus has a similar size, mass, and age. Both planets received similar volatile inventories during their formations. However, Venus currently has an extremely harsh climate that is unsuitable for life, contrasting to that of the Earth. A variety of evidence indicates that Venus may have had a temperate surface environment similar to that on our planet at the early stages of its evolution, but their subsequent different evolution trajectories, regulated by the clouds to a great extent, arrived at distinct conditions \citep{Esposito1983, Gillmann2022}. 

The Venusian clouds gained increasing attention in planetary science research. From a scientific perspective, Venus is a natural laboratory for investigating terrestrial planets, and has great significance validating theories derived on the Earth. A deep understanding of the possible emergence of a habitable environment on early Venus, as well as its deterioration through evolution, provides valuable guidance in the exploration of the Earth and exoplanets, the criteria of planetary habitability, and the conditions for branching the route of planetary evolution. From a strategic perspective, the atmosphere of Venus is rich in natural resources, including energy and chemical substances \citep{Sanchez-Lavega2017, Marcq2018, Limaye2018, Titov2018}. In addition, understanding Venusian clouds is critical for monitoring, understanding, and forecasting Venus’ extreme environment. Research on the Venusian clouds provides a theoretical foundation and feasibility analysis for designing in-situ detectors and implementing remote sensing. 

The number of space missions sent to Venus is second only to the Moon and Mars. The earliest exploration of Venus can be traced back to Soviet “Venera-1” and American “Mariner-2” spacecraft in 1961-1962. Successful Soviet programs of Venera landers and balloons provided the unique and outstanding in-situ investigations. The early missions delivered first glimpses of the exotic planet, but the data was insufficient for understanding the Venus system. In 1978, the Pioneer Venus Orbiter (PVO) delivered four probes into the atmosphere of Venus and obtained the first near-complete in-situ observation data of the vertical profile of Venusian clouds \citep{Knollenberg+Hunten1980}. The orbiter also provided a large amount of data during its 11-year service. A wealth of observations about the clouds was collected by space missions, especially by Venus Express (VEX) (ESA) \citep{Svedhem2007, Titov2018} and Akatsuki (JAXA) \citep{Nakamura2016} as well as by space and ground-based telescopes. Nevertheless, due to the poor accessibility of the dense cloud to remote sensing, only radio-occultations are able to yield the vertical profiles of pressure and temperature in the cloud region \citep{Haus2013, Haus2014}, as well as a few trace species like sulfuric acid (H$_{2}$SO$_{4}$) \citep{Oschlisniok2012, Oschlisniok2021}, which provide poor constraints to the modeling efforts.

Venus has a thick atmosphere with a surface pressure of up to 92 bar and a surface temperature of $\approx$750 K \citep{Seiff1985}. Such an extreme environment has significantly challenged the few landers, let alone habitability. No Soviet “Venera” landers operated on the surface of Venus for more than two hours. Compared to the Earth, Venus's climate evolution seems to have developed toward an extreme. The hypothesis has been proposed that the dominant driving factor is the “runaway” greenhouse effect on Venus \citep{Ingersoll1969}. The surface liquid ocean that may have existed in the early days completely evaporated through this process, initiating a positive feedback related to the “runaway” greenhouse \citep{Ingersoll1969, Gillmann2022}. However, recent research suggested that the liquid ocean might have never formed on the surface of early Venus \citep{Turbet2021}. Both scenarios indicate the crucial role of the clouds in regulating the energy balance and the evolution of Venusian climate. On one hand, the subsolar clouds formed through water evaporation and upwelling transport may have significantly reflected the sunlight and maintained a low global average temperature, providing the necessary environment for the existence of surface liquid water \citep{Yang2013, Way2016}. On the other hand, the nightside clouds formed by day-night circulation can largely absorb and reemit the long-wave radiation from the surface and suppress planetary radiative cooling, which suggests that Venus may have never experienced a cooling process required by liquid water formation \citep{Turbet2021}.

This paper reviews physical and chemical distributions, spatial and temporal variabilities, and chemical and microphysical processes in Venus' sulfuric acid clouds. We collected the latest cloud models, analyzed their strong points, specific applications, and shortcomings, and presented perspectives of future developments. We also summarized and discussed three critical scientific issues, aiming to emphasize the role that the coupling effect of the clouds plays in understanding the structure and evolution of Venusian atmosphere and the potential value of the clouds in exploring the habitability of Venus in various periods and to provide insights for subsequent research on Venus and other planets. Finally, we outline clouds related results expected from the currently planned space missions (EnVision, VERITAS and DAVINCI) and provide an outlook for future studies.

\section{Observations of Venus Clouds}\label{sec2}

\subsection{Composition}\label{sec2.1}

The dense clouds on Venus are located in the altitude range of 47 to about 70 km with a thickness of over 20 km \citep{Knollenberg+Hunten1980, Titov2018}. There are also haze layers above and below these clouds \citep{Kawabata1980, Wilquet2009, Luginin2016, Luginin2024}. The planet is globally covered by clouds, which significantly reflect the solar radiation resulting with a high geometric albedo. Common understanding is that the main component of the cloud layer is concentrated sulfuric acid, which was first detected by \cite{Hansen+Hovenier1974}. The weight percentage of H$_{2}$SO$_{4}$ (acidity) can reach 75$\%$ to 98$\%$ \citep{Barstow2012, Krasnopolsky2015, McGouldrick2021, Dai2022a}. Water (H$_{2}$O) is the secondary component. It is worth noting that there are substantial uncertainties in the existing observations of Venusian cloud acidity due to the observation chanlleges in remote sensing of the dense cloud layer. This is mainly reflected in two aspects: 1) It is tough to determine the specific altitude range of observations within the clouds. Existing remote sensing observations of the cloud interior in the spectral transparency “windows” can only obtain decent signals in a narrow range of bands \citep{Palmer+Williams1975, Arney2014}. The signals in these “windows” are more sensitive to cloud opacity, while sensitivity on the cloud acidity is low. Consequently, the sensed altitudes of cloud acidity are highly uncertain. Therefore, in cloud observations, acidity is usually assumed to be constant throughout all altitudes \citep{Barstow2012, Arney2014}. 2) Significant uncertainties exist in the linear interpolation used to process derived spectral data. The remote sensing techniques provided the cloud acidity from the real part of the reflective index of cloud droplets, which was measured in a limited spectral range (visible and near infrared). Linear interpolation of laboratory spectral data of sulfuric acid solutions with three commonly used benchmark concentrations (75$\%$, 84.5$\%$, and 95.6$\%$) serves as the basis for the constraint \citep{Palmer+Williams1975}. However, the spectra of sulfuric acid solutions with different concentrations show significant nonlinear variations \citep{Barstow2012}. The simple linear interpolation may lead to significant deviations.

Both remote and in-situ observations suggest the presence of other than sulfuric acid species in the Venus cloud deck. Here we summarize the comprehensive reviews by \cite{Esposito1983}, \cite{Esposito1997}, and \cite{Titov2018}, and referr the reader to those papers and references therein for further details. The most enigmatic component is the UV-blue absorber responsible for  Venus’ low albedo at 0.32-0.5 $\mu$m and the remarkable cloud features moving routinely, which are used to study atmospheric circulation at the cloud tops. Observations constrain the location of this specie to the upper cloud layer ($\geq$57 km). Although the absence of sharp spectral features in this range indicates that the absorber is present in the cloud particles, gaseous candidates had also been proposed. Several of them provide good match to the spectral behavior of the Venus albedo which was the primary spectroscopic test for the candidates. From a chemical point of view, the candidates can be divided in two groups: those related to the sulfur chemical cycle and those not connected to the sulphur chemistry. Combination of various allotropes of sulfur S$_{x}$ results in good match of the planet spectrum. However, chemical models do not predict sufficient sulfur production within the cloud layer, but sulfur can form in large amount in the lower atmosphere with subsequent delivery to the clouds. Several gaseous candidates related to the sulfur cycle (S$_{2}$O, S$_{2}$O$_{2}$) show good agreement to the measured albedo (Figure \ref{fig11}), but their required abundance and life time are in contradiction with current chemical models. Non-sulfur-bearing candidates include iron chloride (FeCl$_{3}$) following detection of iron in cloud droplets by the Venera-14 X-ray fluorescent spectrometer. \cite{Krasnopolsky2017} developed detailed chemical, dynamical, and spectroscopic arguments in favor of this species confirming that FeCl$_{3}$ vertical distribution matches that measured by descent probes. Importantly, iron chloride is soluble in sulfuric acid and could act as a condensation nuclei. However spectral properties of the solutions do not match Venus albedo very well (Figure \ref{fig11}). Several other hybrid S-Cl bearing species like chlorosulfonic acid (ClSO$_{3}$H), sulfuryl chloride (SO$_{2}$Cl$_{2}$), and thionyl chloride (SOCl$_{2}$) have also been suggested \citep{Baines2013}, which have UV-blue absorptions at 0.21, 0.29, 0.39 and 0.47 $\mu$m. Similarly, this seems to still not quite match the observations.

In-situ investigations suggested presence of non-sulfur components in the main cloud deck. Analysis of the cloud material acquired by the aerosol collector/pyrolizeer on the VeGa probes by the mass-spectrometer suggested the presence of chlorine and sulfur. Venera-13 and -14 indicated the presence of sulfur, chlorine and iron, while similar experiments on Vega 1 and 2 found little iron but strong signatures of sulfur, chlorine, and phosphorus in the lower cloud, which indicating phosphoric acid (H$_{3}$PO$_{4}$) with P$_{3}$O$_{6}$ as a possible gaseous reservoir. Experimental evidences and chemical models both likely indicate presence of non-sulfur bearing species in the cloud deck. It seems that capabilities of remote sensing in composition analysis of Venus clouds are almost exhausted. Future progress is awaited from in-situ measurements on descent probes.

\begin{figure}[h]
\centering
\includegraphics[width=0.9\textwidth]{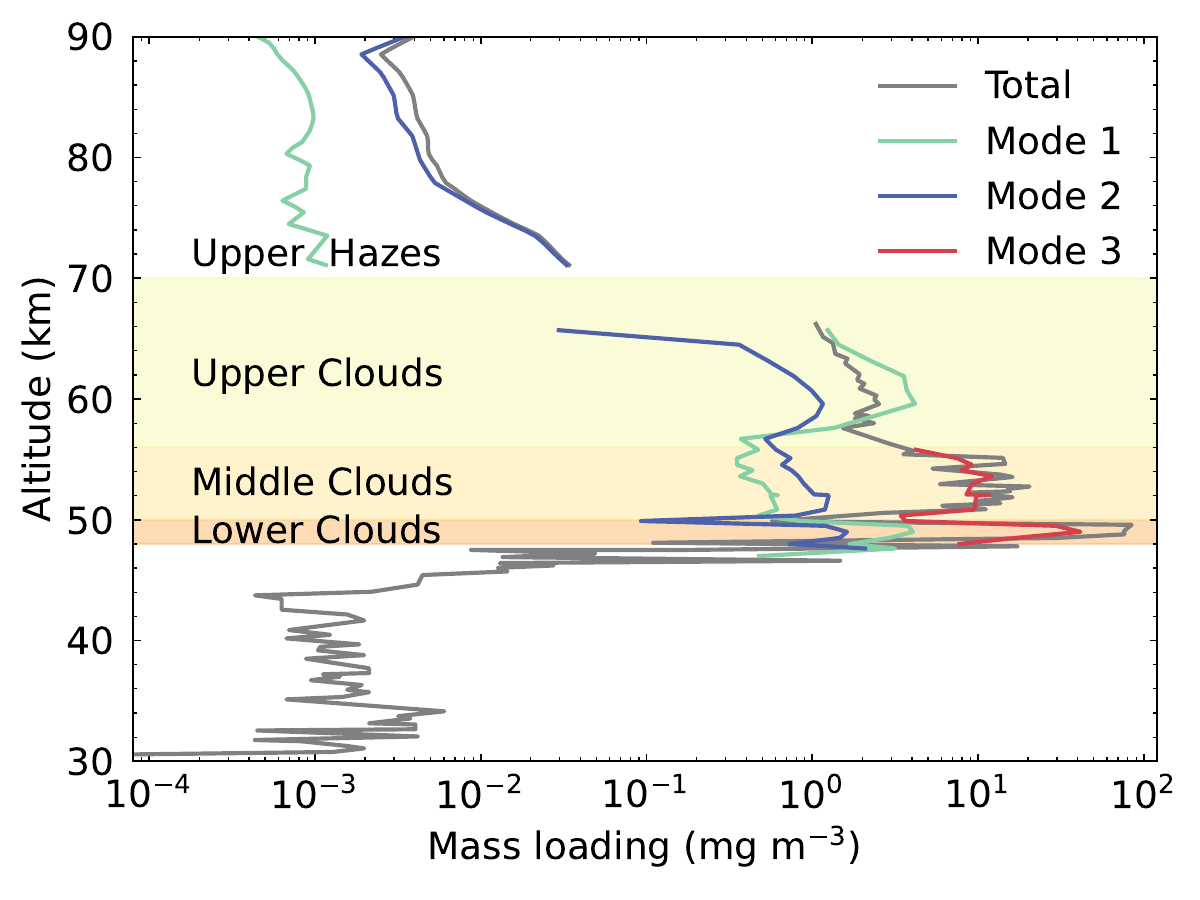}
\caption{The observed mass loading of the clouds and upper hazes from PVO and VEX. The grey solid line represents the total cloud mass loading. The green, blue, and red lines represent the mass loadings of Mode 1, 2, and 3 particles, respectively. Since their calculations are based on the number-density observations of these modes and the assumed constant mass density ($\rho=1.8 g/cm^{3}$) of particles, single-mode mass loading may be higher than the total cloud mass loading in some altitudes. Data in the cloud deck are adapted from \cite{Knollenberg+Hunten1980}. Data in the upper hazes are adapted from \cite{Wilquet2009}.}\label{fig1}
\end{figure}

\subsection{Microphysical Properties and Cloud Formation}\label{sec2.2}

The in-situ measurements of the cloud population on Venus were performed by the cloud particle size spectrometer aboard Large PVO probe \citep{Knollenberg+Hunten1980}. It revealed a significant layered structure of sulfuric acid clouds, as shown in Figure \ref{fig1}. Based on the distribution of the cloud mass loading in vertical, the cloud deck could be divided into three layers: lower clouds (about 48 to 50 km), middle clouds (about 50 to 56 km), and upper clouds (about 56 to 70 km). The cloud top is defined as the height at which the optical depth equals to 1. According to observations at various wavelengths and locations, the cloud top height ranges from 65 to 72 km \citep[e.g.,][]{Ignatiev2009, Cottini2015, Fedorova2015} (see Figure \ref{fig2}).

Cloud droplets are divided into three different modes according to their sizes: mode 1 with sub-micron particles (d$\textless$1 $\mu$m) have the largest number density in the entire cloud deck. Moderate-size mode 2 particles (d$\approx$2 $\mu$m) are also distributed throughout the clouds, but the number density is much less than mode 1. Besides, there is also mode 2’ with a bit larger particle size (d$\approx$2.8 $\mu$m) that is hard to separate from mode 2 particles. Mode 3 particles (d$\textgreater$7 $\mu$m) with the largest mass load are mainly distributed in the middle and lower clouds. However, the existence of large-size mode 3 particles in the middle and lower clouds has always been a controversial issue. \cite{Toon1984} pointed out that mode 3 particles may be just the “tail” of the mode 2 particle size distribution rather than a new mode with a larger size. This conclusion is consistent with the results of the particle size spectrometers onboard Soviet VeGa-1, -2 descent probes that measured bi-modal particle size distribution in the middle and lower clouds \citep{Gnedykh1987}. Interestingly, this experiment also indicated that about 20$\%$ of mode 1 particles have refractive index of 1.7-2.0, much larger than that of sulfuric acid. Based on these observations, the effective particle radius that dominates the cloud mass loading on Venus sites between 1 and 2 $\mu$m \citep{Petrova2015, Rossi2015, Lee2017, Markiewicz2018}, which is consistent with the size of mode 2 particles.

The formation process of the upper clouds differs from that of the middle and lower clouds. A variety of studies have pointed out that H$_{2}$SO$_{4}$ vapor is mainly formed in the photochemical reactions involving SO$_{2}$ and H$_{2}$O in the upper cloud region \citep{Krasnopolsky2012, Zhang2012a, Bierson+Zhang2020, Dai2024}. H$_{2}$SO$_{4}$ vapor rapidly condenses on the surface of the cloud condensation nuclei (CCN), which is mainly composed of dust, salts, and aerosols with sufficient size to overcome the potential surface tension barrier \citep{McGouldrick2017}. Therefore, the upper-cloud mass loading is regulated by photochemical processes leading to H$_{2}$SO$_{4}$ formation \citep{McGouldrick2017, Dai2022a}. The clouds in this region are usually called “photochemical clouds”. In the middle and lower clouds, the upward convection and eddy diffusion transport of species are balanced by precipitation and evaporation processes of large-size cloud droplets, maintaining the high mass loading \citep{McGouldrick2017, Dai2022a}. This area is also called “condensation clouds”.

\begin{figure}[h]
\centering
\includegraphics[width=0.9\textwidth]{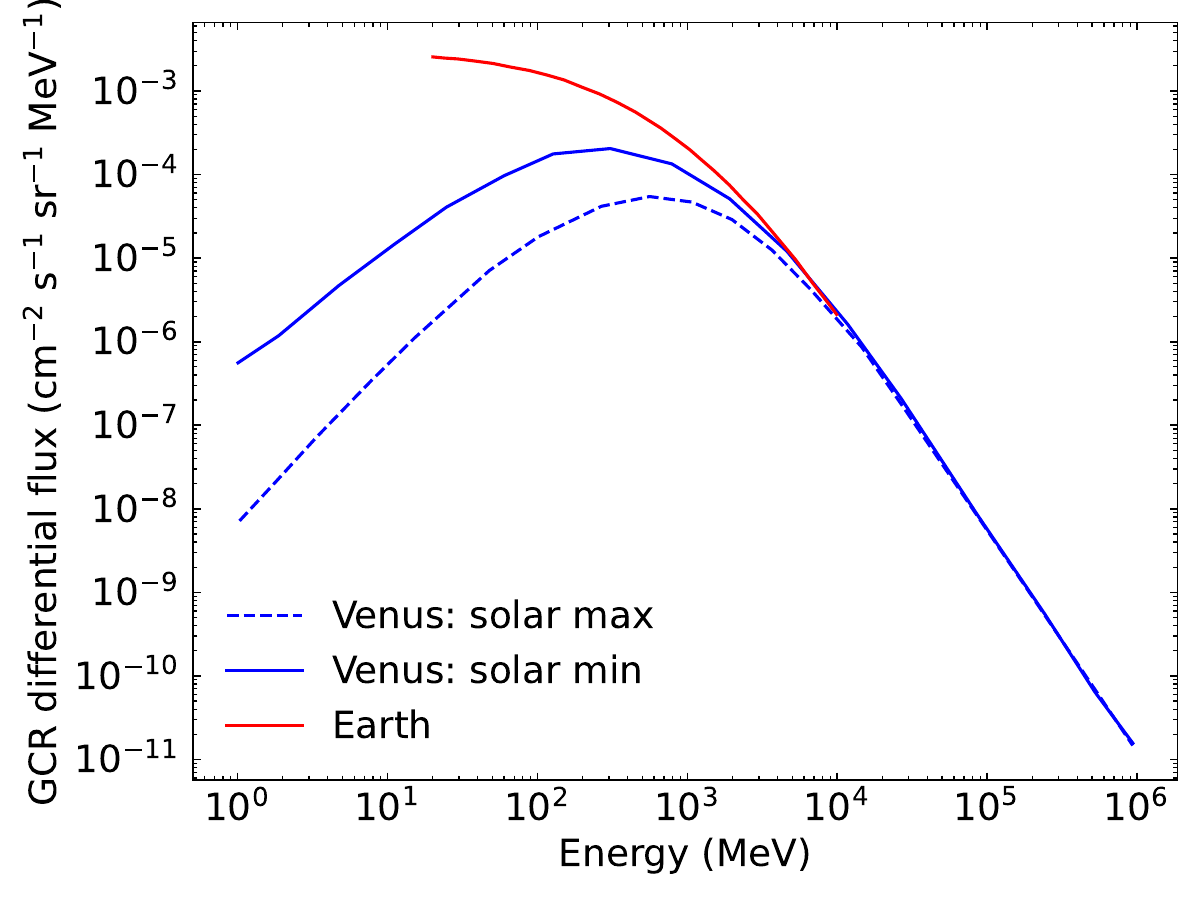}
\caption{The GCR differential fluxes on Earth and Venus. The GCR differential fluxes on Venus under both solar minimum (blue solid) and maximum (blue dashed) are adapted from \cite{Dartnell2015}. The GCR differential flux of the local surrounding interstellar medium on Earth (red solid) is adapted from \cite{Frisch+Mueller2013}.}\label{figadd1}
\end{figure}

\begin{figure}[h]
\centering
\includegraphics[width=0.9\textwidth]{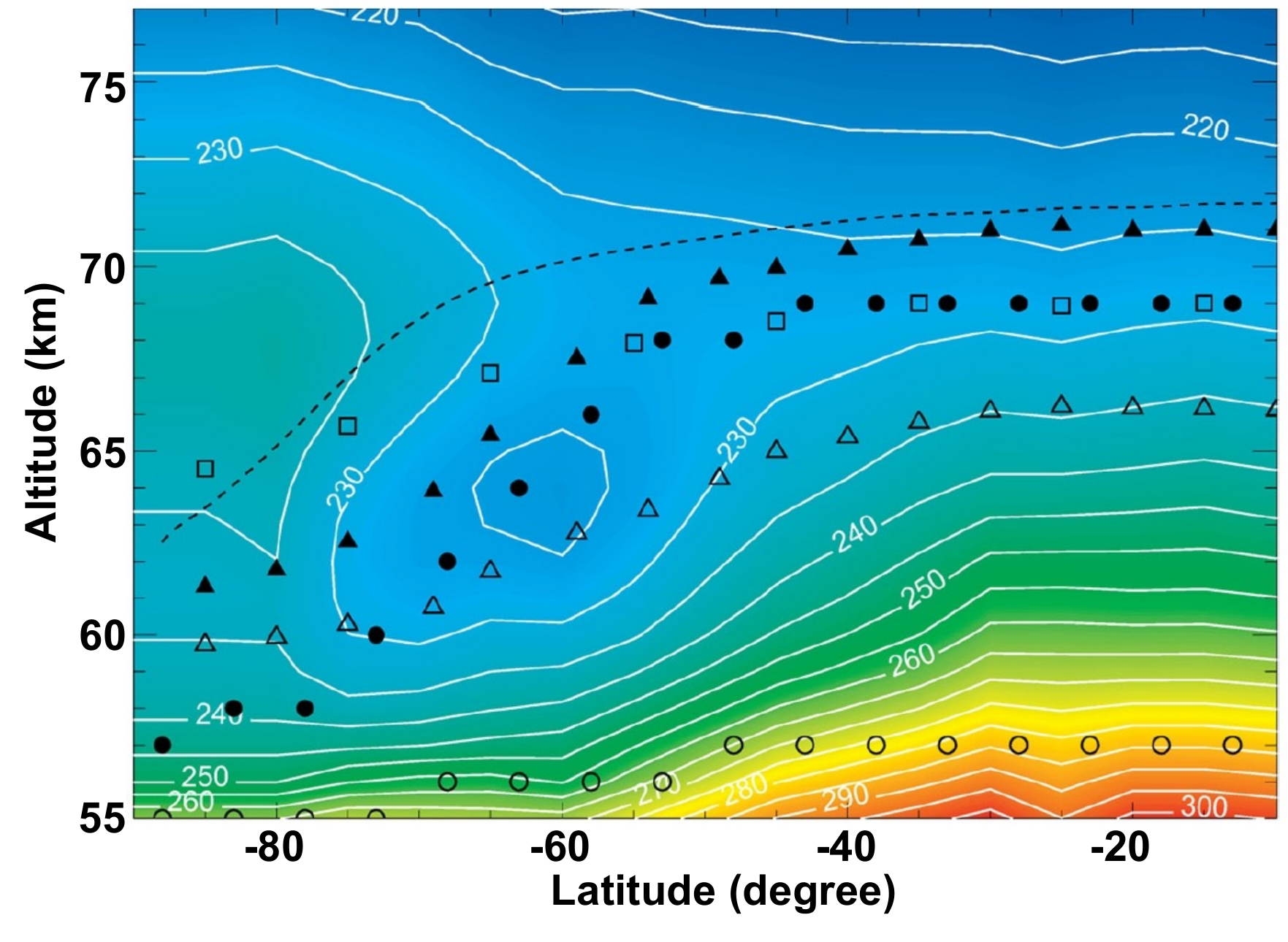}
\caption{The latitude dependence of the cloud top height. The black dashed line and dots in different shapes represent the cloud top heights observed by varying wavelengths, respectively: dashed line-1.5 $\mu$m, squares-2.5 $\mu$m, filled triangles-1 $\mu$m, empty triangles-5 $\mu$m, filled circles-8.2 $\mu$m, empty circles-27.4 $\mu$m. The filled contour map represents the temperature distribution. This figure is adapted from \cite{Titov2018}.}\label{fig2}
\end{figure}

In addition to photochemistry and dynamics, observations suggest that galactic cosmic rays (GCR) may influence cloud formation on Earth, as evidenced by the correlation between cosmic ray intensity and global average cloud cover over solar cycles \citep{Marsh+Svensmark2000, Carslaw2002, Svensmark2009}. Proposed hypotheses include ion-catalyzed nucleation of ultrafine condensation nuclei and the migration of highly charged aerosols at cloud boundaries, which possibly contributes to the formation of ice particles \citep{Carslaw2002}. However, this correlation still remains controversial as the results are highly methodology dependents \citep[e.g.][]{Laken2012}, and experiments reveal weak CCN response to the variations of cosmic rays \citep[e.g.][]{Gordon2017}. In contrast, the potential influence of cosmic rays on Venusian cloud formation remains poorly explored. Given Venus’s proximity to the Sun cosmic rays may play a diminished role in its cloud formation processes. Figure \ref{figadd1} shows the comparison between the GCR differential fluxes on Earth and Venus. The flux of GCR particles with energy lower than 4$\times$10$^{3}$ MeV on Venus is significantly lower than that on Earth. As the GCR-induced ionization in Venus’s atmosphere has been predicted \citep[e.g.][]{Nordheim2015} and recent analysis of VEX housekeeping data has revealed modulations in galactic cosmic rays \citep{Rimbot2024}, further comparisons between the variations of cosmic rays and Venusian clouds are essential to confirm or disapprove this scenario.

Venusian clouds show significant spatial and temporal variations. According to VEX observations (Figure \ref{fig2}), the altitude of the cloud top significantly decreases from the equator to high latitudes \citep{Cottini2012, Cottini2015, Haus2013, Haus2014}. \cite{Lee2012} proposed that this trend corresponds to the temperature distribution and believed that clouds are affected by changes in mesospheric temperature. \cite{Cottini2015} revealed a significant anti-correlation between the latitudinal variations of Venusian cloud top height and the H$_{2}$O vapor abundance at the same altitude. This indicates significant chemical and kinetic interactions between Venusian sulfuric acid clouds and the atmosphere.

In addition, ground-based telescope observations indicated that the large-size cloud particles are concentrated in the high-latitude regions of both the northern and southern hemispheres \citep{Carlson1993}. The band-like distribution structure of small cloud particles can be observed in both hemispheres. However, the mean particle size of cloud droplets in the northern hemisphere is larger than that in the southern hemisphere, which may be associated with large-scale dynamic processes \citep{Carlson1993}. Based on VEX Monitoring Camera (VMC) observations and Mie scattering calculations, \cite{Shalygina2015} proposed that cloud droplets in low latitudes of Venus have effective radius of about 1.2-1.4 $\mu$m, which is significantly larger than that of 0.9-1.05 $\mu$m near the southern pole. Moreover, the refractive index of cloud droplets at 40-60 degrees South is usually slightly lower than that in other locations. Sub-micron particles (with an effective size of about 0.23 $\mu$m) are always observed in the morning \citep{Shalygina2015}.

\begin{figure}[h]
\centering
\includegraphics[width=0.9\textwidth]{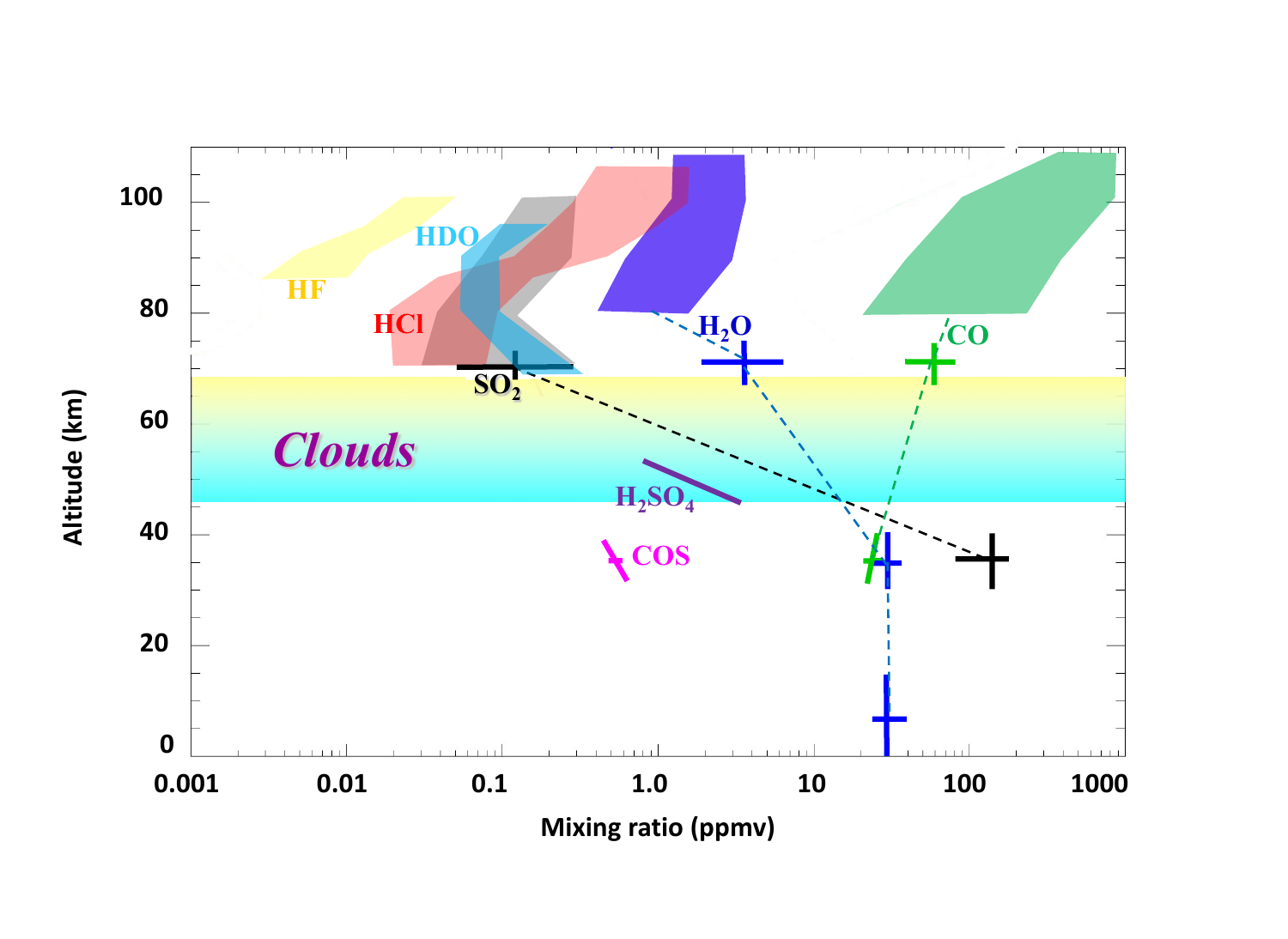}
\caption{The observations of crucial species near the cloud deck. The species are distinguished by different colors. The solid lines and fills represent the observations, and the dashed lines predict the trends inside the clouds.}\label{fig3}
\end{figure}

Based on VEX radio occultation observations, the latest altitude-latitude distribution of H$_{2}$SO$_{4}$ vapor in the cloud regions has been obtained \citep{Oschlisniok2012, Oschlisniok2021}. Its horizontal mean abundance decreases rapidly at the cloud base (46 km) (Figure \ref{fig3}), which is related to its partial pressure in presence of liquid sulfuric acid droplets and its dependence on temperature. The H$_{2}$SO$_{4}$ abundance below the cloud deck is significantly higher at low latitudes, which is consistent with the latitudinal variations of the clouds. The cloud-top sulfur dioxide (SO$_{2}$) abundance significantly fluctuated over a decadal period, as observed by PVO and VEX \citep{Marcq2013, Marcq2020}. It decreased from 400 ppb in 1980 to less than 50 ppb in 1995, then increased to over 300 ppb in 2007. A new decline started around this time, and the abundance reached about 50 ppb in 2012. This variation is speculated to be associated either with the convection and diffusion transport efficiency within the cloud or with volcanic activity \citep{Marcq2013, Marcq2020, Kopparla2020}. \cite{Encrenaz2020} also revealed a significant anti-correlation between this decadal fluctuation of SO$_{2}$ and cloud-top H$_{2}$O vapor abundance, based on ground-based remote sensing observations.

Based on a variety of observations from VEX, VEGA, and ground-based telescopes \citep{Vandaele2017a}, studies found that SO$_{2}$ abundance significantly decreases with increasing altitude within the cloud deck from 200 ppmv at 40 km to 0.1 ppmv at 70 km (Figure \ref{fig3}). This mechanism cannot be adequately explained by cloud formation since the decrease of H$_{2}$O abundance in the cloud deck reaches only 30 ppmv. \cite{Bierson+Zhang2020} tried to explain the observed SO$_{2}$ depletion by suppressing its upward transport from the lower atmosphere. However, the cloud layer shows relatively low static stabilities according to observations by VEX and Akatsuki \citep{Imamura2017, Ando2020a, Oschlisniok2021}, indicating the presence of strong convection in this region. This may contradict the scenario that the cloud eddy diffusion is weak \citep{Bierson+Zhang2020}. Besides, \cite{Rimmer2021} interpreted the SO$_{2}$ depletion by dissolving it into sulfuric acid droplets. A large amount of hydroxide salts are needed to facilitate this process through neutralization reactions, while their constant source in the cloud region is controversial.

\section{Modelling of Venus Sulfuric Acid Clouds}\label{sec3}

Venusian clouds play a crucial role in regulating the atmospheric environment structure and habitability evolution. Besides observations, numerical models are powerful tools supporting the studies of distributions and mechanisms of the clouds formation and explain their variability. 

On the one hand, the dense clouds exhibit high bolometric (A$\approx$0.77) albedo and large infrared opacity. While reflecting a large amount of solar radiation, they significantly absorb the long-wave thermal ($\lambda$$\textgreater$2.7 $\mu$m, mostly due to sulfuric acid) emission of the surface. By regulating the radiative transfer in Venusian atmosphere, the clouds greatly affect the global greenhouse effect and the temperature of the lower atmosphere. Based on model research, studies proposed that this might be one of the dominant factors branching the habitability evolution of early Venus, which is discussed in Section~\ref{sec4.3}.

On the other hand, the clouds are closely coupled with vertical transport. The vertical eddy mixing is proposed to be crucial in maintaining the evaporation-recondensation cycle of H$_{2}$SO$_{4}$ at the cloud base and accumulating the lower-cloud mass loading \citep{Imamura+Hashimoto2001, Dai2022a}. The clouds also affect the vertical transport via cloud-base water abundances \citep{Kopparla2020}. The vertical transport of angular momentum and the compensation effect of planetary waves play a crucial role in maintaining the superrotation \citep{Rossow+Williams1979, Gierasch1987, Lebonnois2016, Sanchez-Lavega2017, Horinouchi2020}. The influence of the surface topography on the zonal circulation was demonstrated by VMC and Akatsuki observations \citep{Bertaux2016, Patsaeva2019, Patsaeva2024}.

However, due to the lack of direct observations, there are still considerable differences in some issues resolved by models, e.g., the mixing efficiency in the clouds. One-dimensional planetary atmospheric model studies usually parameterize complicated three-dimensional atmospheric dynamic processes in terms of one-dimensional eddy diffusion efficiency ($K_{zz}$), which is always constrained by vertical motions and abundance gradients of trace species in the presence of sufficient observational constraints. In most Venus one-dimensional atmospheric chemistry-transport models $K_{zz}$ in the cloud layers is set at as low as $\approx$10$^{4}$ cm$^{2}$/s \citep[e.g.,][]{Krasnopolsky2012, Zhang2012a, Shao2020, Bierson+Zhang2020, Dai2024}. \cite{Bierson+Zhang2020} ever tried to further reduce the cloud $K_{zz}$ to explain SO$_{2}$ observations while the cloud static stability stays at a low level \citep{Imamura2017, Ando2020a, Oschlisniok2021}, indicating the presence of strong convective activity and hence higher $K_{zz}$ in this region. Although a recent modelling study has shown that cloud $K_{zz}$ has little influence on atmospheric chemical structure \citep{Dai2024}, it can still regulate the mass loading of the condensational clouds \citep{McGouldrick2017, Dai2022a}. Based on VEX observations of H$_{2}$SO$_{4}$ abundance and the assumption that condensation and diffusion regulate the abundance of H$_{2}$SO$_{4}$ in the cloud regions, \cite{Dai2023} obtained a much higher $K_{zz}$ value in the cloud ($\approx$10$^{8}$ cm$^{2}$/s) and suggestedsignificant latitudinal variations $K_{zz}$ in the cloud deck. By coupling the results of three-dimensional small-scale turbulence resolving model and chemical processes, \cite{Lefevre2022, Lefevre2024} inferred the cloud $K_{zz}$ and indicated a strong dependency of this parameter on specific species. They proposed that species with longer chemical lifetime tend to be well-mixed. The derived cloud $K_{zz}$ is 10$^{6}$-10$^{8}$ cm$^{2}$/s, significantly higher than most previous atmospheric chemistry-transport model studies.

Due to the limitations of remote-sensing observational capabilities, many dynamical and chemical issues within the clouds still need to be verified by future in-situmeasurements, such as the in-situ investigations aboard the coming DAVINCI+ mission \citep{Garvin2022}. At present, our understanding of the cloud formation process and its varying mechanism on Venus deeply relies on numerical simulations.

\subsection{Microphysical Models}\label{sec3.1}

Existing research on the Venusian cloud formation is mostly based on numerical simulations, among which microphysical modelling is the most common. Due to significant complexity of the microphysical processes and high computing requirements, this type of model mainly focuses on microphysical processes in the vertical direction, as well as several other cloud-related physical quantities. The governing equation of this model is listed below for further understanding \citep{Titov2018}:
\begin{align}
\frac{\partial N(m,z,t)}{\partial t} &= \frac{\partial}{\partial{z}}\left[K_{zz}(z) \rho (z) \frac{\partial}{\partial z}\frac{N(m,z,t)}{\rho (z)}\right] - \frac{\partial}{\partial z}\left\{N(m,z,t)\left[w(z) - v_{fall}(m,z,t)\right]\right\} \nonumber \\
&+ P(m,z,t) - L(m,z,t)N(m,z,t) \nonumber \\
&+ \int^{m}_{0}K_{coag}(m',m - m')N(m',z,t)N(m-m',z,t)dm' \nonumber \\
&- N(m,z,t)\int^{\infty}_{0}K_{coag}(m',m)N(m',z,t)dm' \nonumber \\
&- \frac{\partial}{\partial m}\left[N(m,z,t)G(m,z,t)\right] \label{eq1}
\end{align}
where $m$, $z$, and $t$ represent the particle mass, height, and time, respectively, $N$ is the particle number density, $\rho$ is the atmospheric density, $w$ and $v_{fall}$ represent the vertical velocities of wind and particle sedimentation, respectively, $P$ and $L$ represent chemical production and loss rates, $K_{coag}(m',m)$ is the coagulation kernel between cloud particles with mass $m'$ and $m$, and $G$ is the condensational growth rate. The lefthand side of the equation represents the rate of number density change for particles with specific sizes as function of altitude and time. The terms on the righthand side represent the transport by vertical eddy diffusion, sedimentation and winds, as well as nucleation, evaporation, growth by coagulation between particles (two terms including $K_{coag}$), and growth by condensation, respectively.

\begin{figure}[h]
\centering
\includegraphics[width=0.9\textwidth]{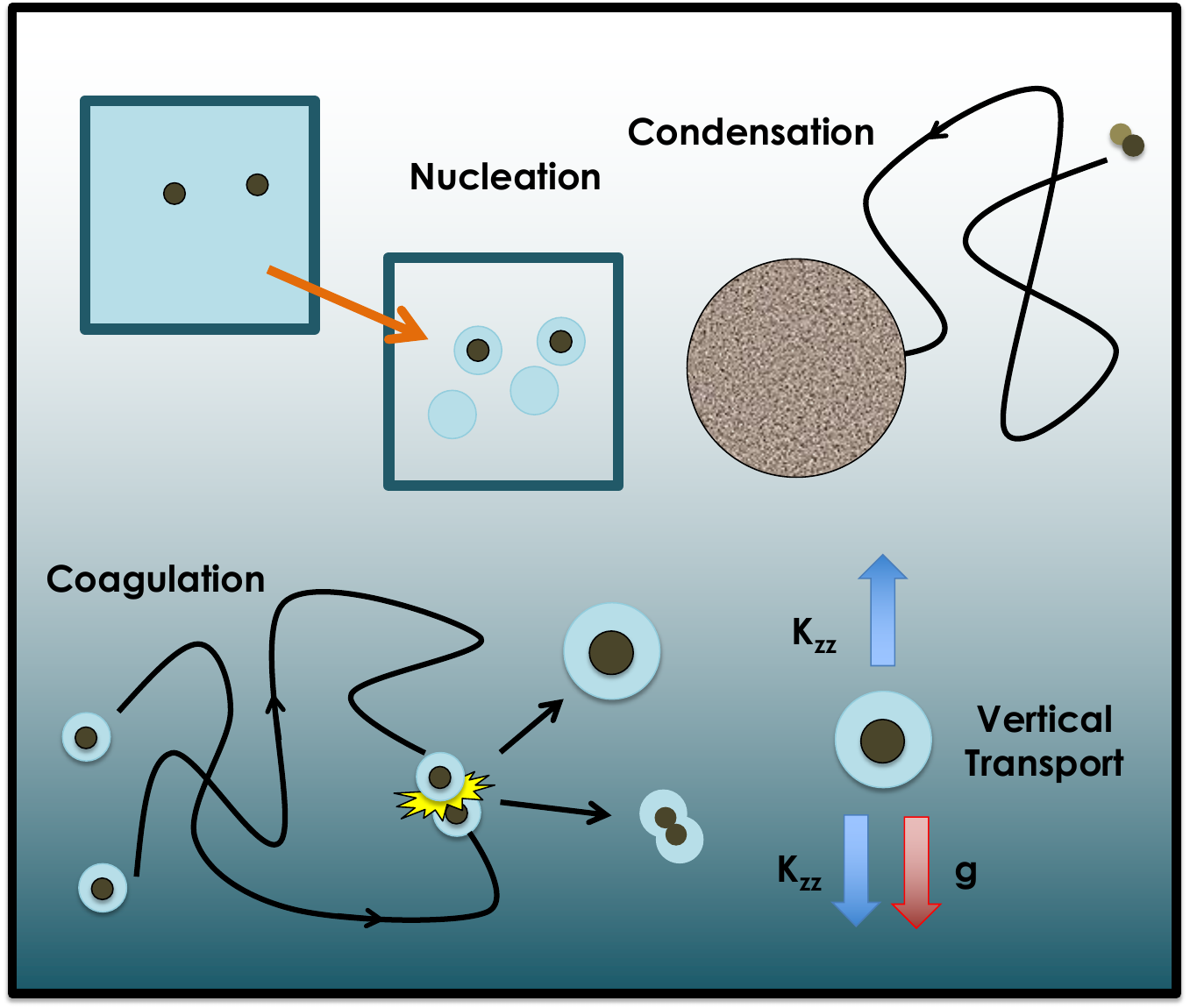}
\caption{A cartoon image of microphysics in the clouds. (Credit: Peter Gao)}\label{fig4}
\end{figure}

Figure \ref{fig4} intuitively illustrates several microphysical processes of the cloud formation. Among them, the vertical eddy diffusion transport is related to the mixing efficiency and vertical gradient of the abundance. Since the calculations of a single vertical direction cannot accurately retrieve the three-dimensional mixing motions such as turbulence, fluctuation, and convection, the model adopts the parameterized eddy diffusion to simulate such processes as mentioned at the beginning of Section~\ref{sec4}, although the physical significance and accuracy of this parameterization process are still tricky \citep{Zhang+Showman2018}. Transport caused by precipitation and winds is related to the vertical acceleration. As for the chemistry (not illustrated in the figure), most microphysical models use either fixed chemical rates or straightforward chemical networks limited by computational efficiency. Thus, there are still significant uncertainties in present coupling studies between Venusian clouds and atmospheric chemistry, as discussed in Section~\ref{sec4.1}. Coagulation (or coalescence) for solid (or liquid) particles process is always divided into two categories: Brownian coagulation for small particles and gravitational coagulation for large particles \citep{Seinfeld+Pandis2016}. This process is regulated by temperature, particle distribution, sedimentation velocity, and the collision between particles. Condensation and evaporation processes are associated with particle distribution, temperature, gas abundance, and cloud acidity \citep[in the case of Venus,][]{Zeleznik1991, Dai2022a}. Nucleation, both homogeneous and heterogeneous, largely relies on the CCN component, supersaturation, and collision.

\begin{figure}[h]
\centering
\includegraphics[width=0.9\textwidth]{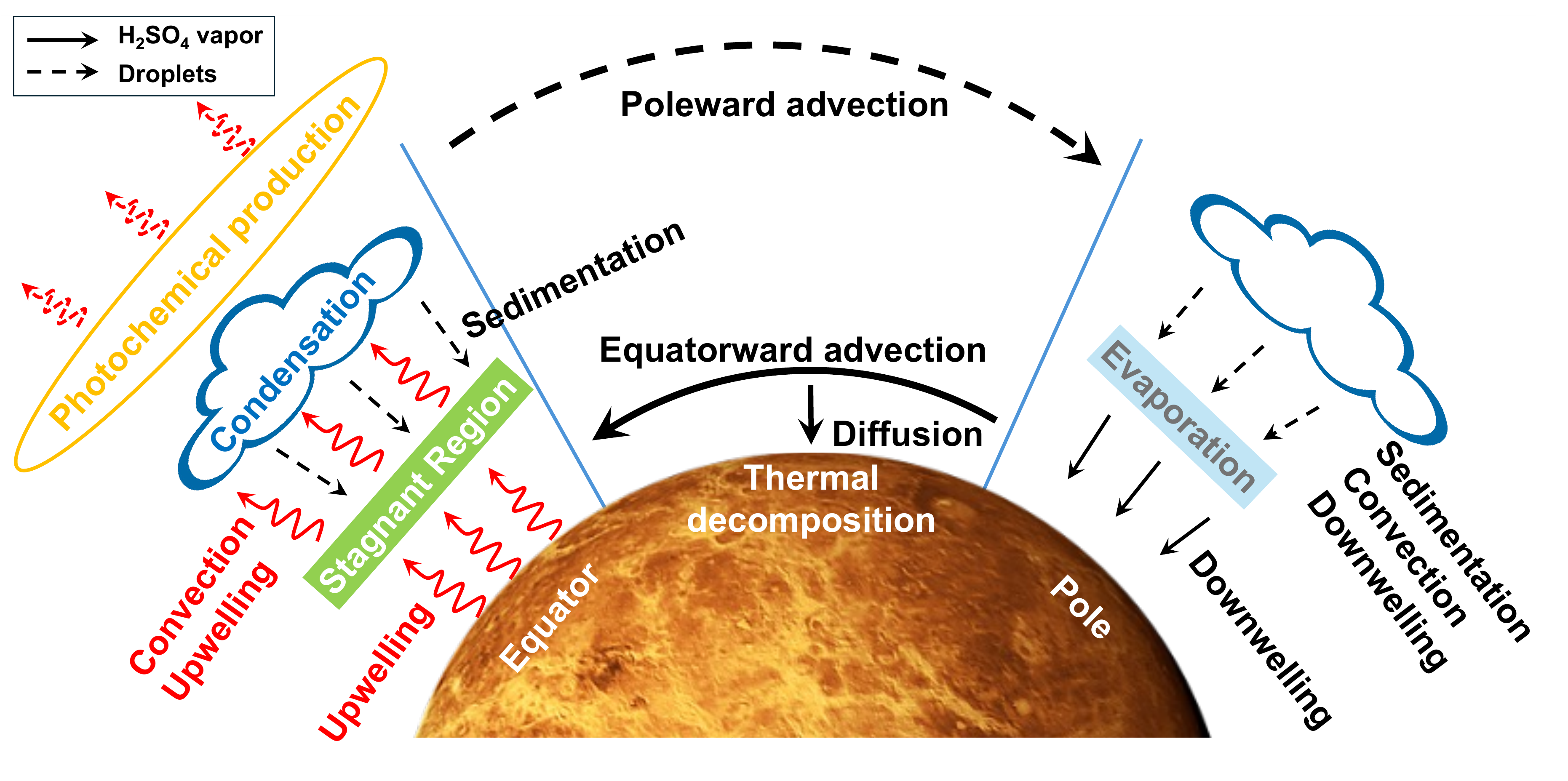}
\caption{A cartoon image of the cloud formation at the equator of Venus \citep[modified from][]{Imamura+Hashimoto2001}}\label{fig5}
\end{figure}

In recent decades, there have been several studies on Venus sulfuric acid clouds based on microphysical models, which pay more attention to convection, coagulation, and nucleation processes. \cite{Imamura+Hashimoto2001} calculated the coagulation of cloud droplets using a volume-conserving method. They assumed that the condensation nuclei were insoluble and ignored their dissolution effects. By conducting a 1D simulation study on the Venus equatorial clouds under the impact of Hadley-style circulation, they suggested different mechanisms regulating the upper and middle-lower clouds, as shown in Figure \ref{fig5}. The photochemically generated H$_{2}$SO$_{4}$ vapor in the upper cloud region rapidly condenses into droplets, which are subsequently removed by equatorial upwelling. The H$_{2}$SO$_{4}$ vapor in the middle and lower clouds comes from the large-scale dynamic transport below the cloud deck. This equatorward and upward flow facilitates the formation and growth of droplets in this region. The liquid H$_{2}$SO$_{4}$ is consumed by subsequent sedimentation, which conserves the sulfur system against the upwelling, maintaining the large mass loading in the middle and lower clouds.

In addition to large-scale circulations, the effects of radiative transfer and microphysical properties have also received attention. A similar and commonly used 1D microphysical model, the Community Aerosol and Radiation Model for Atmospheres (CARMA), was originally built by \cite{Toon1988} and applied to Venus clouds by \cite{James1997}. By performing CARMA simulations on Venusian clouds and coupling the results to a radiative-transfer model, \cite{McGouldrick+Toon2007} explored the radiative-dynamical feedback mechanism. Their model basically covers the same physical processes as the previous one of \cite{Imamura+Hashimoto2001}, while this model assumes soluble CCN, allowing them to include the solute effect of the nuclei in addition to the Kelvin effect. Besides, their H$_{2}$SO$_{4}$ supersaturation factor in the upper cloud region reaches only 2, indicating that the homogeneous nucleation in this model may be weaker than \cite{Imamura+Hashimoto2001}. This study proposed that the middle and lower clouds are heated by long-wave radiation absorption, maintaining the strong convective mixing motions within the cloud. This is consistent with the low static stability and large $K_{zz}$ in the clouds indicated by observation and model studies \citep{Imamura2017, Ando2020a, Oschlisniok2021, Dai2023, Lefevre2022, Lefevre2024}. The cloud opacity exhibits significant dependency on the efficiency of this vertical transport.

Subsequently, based on simulations of the same model, \cite{McGouldrick2017} proposed that the upper photochemical clouds have little short-timescale impact on the lower condensational clouds, which is consistent with the results of \cite{Imamura+Hashimoto2001}. By shutting down the coalescence efficiency of supercooled sulfuric acid, this study exhibited the growth of the particles in the upper clouds and found that the clouds occasionally exhibit an enhancement of small particles and rapid growth of the particles near the cloud base, indicating that the cloud near-infrared opacity variations could be partially explained by microphysics, in addition to dynamics.

The particle distributions in the clouds and hazes have significant issues. Based on CARMA simulations, \cite{Gao2014} retrieved the particle distributions in agreement with the observations in the cloud and haze layers. They found that the upper haze exhibited a unimodal distribution structure in a steady state, while a bimodal structure appeared when the upwelling was introduced. This suggests that the upper haze is a mixture of droplets formed from condensation of sulfuric acid vapor on smaller dust particles floating in this region and larger droplets upwelled from the underlying cloud decks. Such upward transport of the clouds and upper haze is hypothesized recently to have driven the distribution of the D/H (deuterium/hydrogen) ratio, which significantly increases with altitude above the clouds \citep{Mahieux2024}. The same model was applied by \cite{Parkinson2015a} to explore the dependency of cloud structures on latitude and local time. They further verified the unimodal distribution of upper hazes in the steady state. Besides, \cite{Gao2014} proposed that coagulation between tiny and mode-1 particles brought a low growth rate to the latter while they rapidly grew to mode 2 by condensation once they developed across the potential energy barrier of the Kelvin effect, which subsequently affected the formation of mode 3 particles by coagulation. This mechanism leads to a 6-month quasi-periodic oscillation in precipitation. 

Nucleation of CCN was also investigated. \cite{McGouldrick+Barth2023} revealed that the cloud opacity would significantly decrease and, under specific circumstances, perform a long-term (hundreds of Earth days) oscillation if the coagulation between CCN is suppressed. Both aforementioned studies indicate that the cloud layer of Venus is not stationary, which provides new insights for subsequent model research.

The latest microphysical model research was performed by \cite{Karyu2024}, who systematically tested the impact of various environmental parameters, including $K_{zz}$ and temperature, on Venusian clouds. They emphasized the significant regulating effect of vertical transport on H$_{2}$O vapor abundance, as well as the great impact of temperature on H$_{2}$SO$_{4}$ vapor abundance.

\subsection{Simplified Cloud Models}\label{sec3.2}

Several studies have simplified or developed specific aspects of the cloud model based on a variety of objectives and emphases. Thus, these studies on the Venusian cloud have shown a significant diversified trends of model development. \cite{Krasnopolsky+Pollack1994} devoted their attention to the feedback of cloud acidity on cloud formation. Based on the correlation between acidity and saturation state and the assumption of local thermodynamic equilibrium (LTE, here means that the timescale of condensation is negligible compared to other processes like diffusion and chemistry, so that the vapor abundances closely follow their saturation profiles), they constructed a set of analytical equations for gaseous species and acidity, greatly shortening the cloud simulation time. This model was subsequently updated by \cite{Krasnopolsky2015} using the latest observations and chemical kinetics. 

However, due to the LTE assumption, the above studies did not calculate the abundances of liquid species and cloud mass loading based on the condensational process. Several subsequent studies \citep{Imamura+Hashimoto2001, Dai2022a, Karyu2024} have proposed that H$_{2}$SO$_{4}$ in the upper clouds is significantly supersaturated and thus cannot meet the LTE requirement. For instance, \cite{Dai2022a}, the same as \cite{Krasnopolsky2015}, addressed their attention to the dependency of the gas saturation states on the cloud acidity in the Venusian sulfuric acid clouds. Nevertheless, they used different physical methods, including the condensation and evaporation processes relying on the differences between saturation and gas abundances, rather than analytical calculation. 

\begin{figure}[h]
\centering
\includegraphics[width=0.9\textwidth]{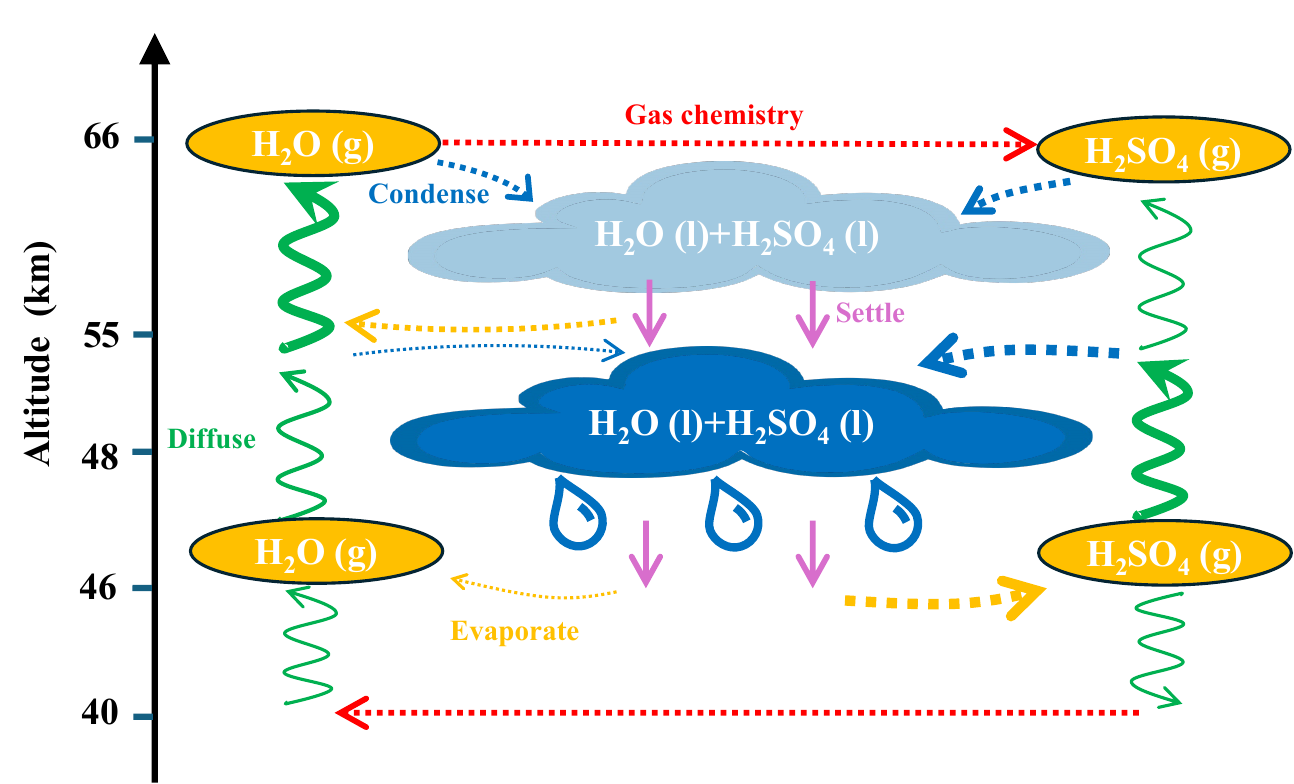}
\caption{The H$_{2}$SO$_{4}$ and H$_{2}$O cycles in the binary cloud system. The signs g and l in the brackets represent gas and liquid, respectively. The arrows in green, red, blue, gold, purple represent the diffusion, gas chemistry, condensation, evaporation, sedimentation processes, respectively. The thickness of the arrows shows the flux (thick arrow means high flux). This figure is modified from \cite{Dai2022a}.}\label{fig6}
\end{figure}

The model of \cite{Dai2022a} simultaneously covers condensation and evaporation, vertical eddy diffusion and sedimentation, and simple chemistry with respect to both gaseous and liquid species. The feedback of cloud acidity to the saturation pressure is also included. They assumed unimodal particles at each height to improve simulation efficiency, which keeps this model from obtaining the number density distribution of particles over varying sizes. As their derived unimodal size distribution is consistent with mode 2 observations \citep{Knollenberg+Hunten1980}, they suggested that the mode 2 particles regulate the cloud mass in the cloud deck except for the cloud base region, where their model failed to explain the rapid increase in cloud mass loading. They argued that mode 3 particles might take over near the cloud base. Since they used a relatively low $K_{zz}$ in the lower clouds (compared to othermicrophysical models), here we also propose a possibility that their underestimation of the lower-cloud mass loading may be attributed to a weak vertical transport. This study described in detail the H$_{2}$SO$_{4}$ and H$_{2}$O cycles within the cloud (Figure \ref{fig6}) and proposed that H$_{2}$SO$_{4}$ achieves significant supersaturation above 60 km, governed by the photochemical production of H$_{2}$SO$_{4}$. The supersaturation agrees with \cite{Imamura+Hashimoto2001} and \cite{Karyu2024} results and indicates that the LTE assumption is unsuitable for sulfuric acid vapor as the condensation rate of H$_{2}$SO$_{4}$ is suppressed by low CCN density above 60 km and H$_{2}$SO$_{4}$ abundance never closely follows its saturation profile.

Based on this conclusion, \cite{Dai2022b} established another cloud model on Venus. They expressed the supersaturation through the photochemical production rate of H$_{2}$SO$_{4}$ and used LTE assumption on H$_{2}$O, thereby simplifying the condensation and evaporation processes. By constructing the correlation between gas species and acidity, same as that in \cite{Krasnopolsky2015}, and iterating the cloud mass loading and acidity, this study presented a new semi-analytical cloud model. It greatly shortened the computing time of cloud simulation ($\approx$15 s) and simultaneously possessed the ability to obtain cloud mass loading and unimodal particle size, facilitating future coupling studies of the cloud formation and other chemical or dynamical processes.

A new zero-dimensional model called MAD-VenLA was recently applied to the Venusian cloud formation \citep{Maattanen2023}. Since the abundance of H$_{2}$O in the Venusian atmosphere is larger than H$_{2}$SO$_{4}$, this model maintained the H$_{2}$SO$_{4}$-H$_{2}$O binary condensational system in an adjusted equilibrium by tracking the transport and condensation of H$_{2}$O and using the constant mass of H$_{2}$SO$_{4}$. Affected by the hygroscopicity of concentrated sulfuric acid in the clouds, slight variations in H$_{2}$O could have a significant impact on the acidity and subsequent H$_{2}$O saturation pressure \citep{Zeleznik1991, Dai2022a}. Thus, the equilibrium of the binary system could be achieved by slight adjustments on H$_{2}$O. However, since H$_{2}$SO$_{4}$ significantly condenses and evaporates near the cloud base, the assumed constant H$_{2}$SO$_{4}$ mass may pose issues for this region. 

This model simplifies the condensation and evaporation, includes other microphysical processes like nucleation and coagulation through a variety of parameterization schemes, and excludes transport and radiative transfer by zero-dimensional simulation, allowing it to find a balance between model complexity and computational efficiency. This may be of great significance in the coupling of cloud formation in large-scale three-dimensional general circulation models (GCMs).

However, the existing coupling studies of GCMs usually adopt a significantly simplified cloud module to increase computational efficiency. \cite{Ando2020b} considered the effect of dynamic processes on clouds. They included the condensation and evaporation of H$_{2}$SO$_{4}$ and H$_{2}$O, along with the sedimentation and wind transport of particles, but ignored the acidity variations, particle size, surface tension, nucleation, and other processes associated with the cloud formation. This study obtained the altitude-latitude distributions of Venusian cloud mass loading and the global distribution of column density. They proposed that the formation of the thick clouds in high latitudes is attributed to the strong turbulent activity reflected by low static stability observations, while the thickness of clouds in low latitudes is mainly affected by the meridional-circulation transport of H$_{2}$SO$_{4}$ vapor.

Variations of the cloud optical depth in low latitudes are also in the focus of research. \cite{Karyu2023} also neglected the effect of acidity and complicated microphysical processes such as coagulation and nucleation in their model. They distinguished the clouds by comparing gas abundances and saturation conditions, ignoring the time scale of microphysical processes. This study pointed out that the variation of equatorial cloud optical depth at an altitude of 46-52 km is attributed to both gravity waves and an equatorial Kelvin wave, which induced rapid small-scale perturbances and a quasi-periodic zonal-wavenumber-1 variation. 

However, due to the unique cloud composition and thermal structure on Venus, the cloud acidity has a significant impact on the saturation state and subsequent condensation efficiency of H$_{2}$O \citep{Zeleznik1991, Dai2022a}. Ignoring the effect of cloud acidity may bring uncertainties to the cloud formation process. \cite{Stolzenbach2023} included the feedback of cloud acidity on the H$_{2}$SO$_{4}$-H$_{2}$O binary condensation in their three-dimensional planetary climate model (PCM) and involved a limited chemical network. Their solution on the coupled cloud formation was similar to the (semi-)analytical methods of \cite{Krasnopolsky2015} and \cite{Dai2022b}, except for that the latter used fixed total particle number densities to track the growth of unimodal particles through the variations of cloud mass loading, while this study used certain mass proportions between various modes based on observed log-normal size distributions and tracked the growth of particles by changes of total mass loading. The physical implications expressed by the studies above are basically the same. This study provided the altitude-latitude distributions of clouds and various species and emphasized that thermal tides and photochemistry above the clouds play a dominant role in large-scale meridional variations of the clouds.

\begin{figure}[h]
\centering
\includegraphics[width=0.9\textwidth]{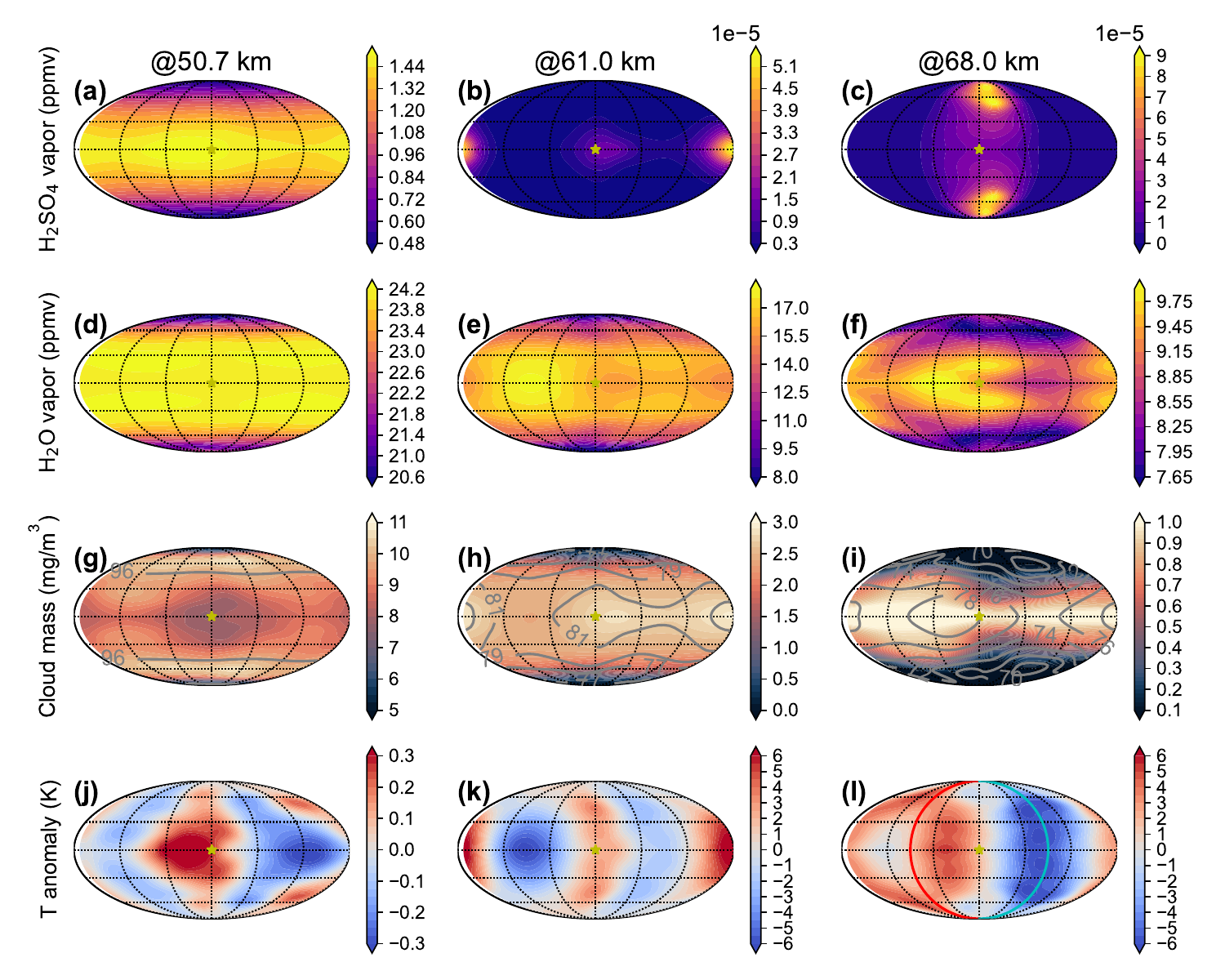}
\caption{The global distributions of (a-c) H$_{2}$SO$_{4}$ vapor abundance, (d-f) H$_{2}$O vapor abundance, (g-i) cloud mass loading, and (j-l) normalized temperature on Venus. The three columns represent the results at different altitudes: left—50.7 km, middle—61 km, and right—68 km. The star in every panel represents the subsolar point. The red and blue lines in panel (l) represent evening (18:00) and morning (06:00) terminators, respectively. This figure is modified from \cite{Shao2024}.}\label{fig7}
\end{figure}

To obtain a more detailed cloud structure, \cite{Shao2024} coupled the condensation model of \cite{Dai2022a} into their high-precision Venus GCM OASIS to achieve physical cloud formation by condensation, evaporation, and sedimentation processes and self-consistent acidity feedback. Since OASIS adopted high-precision icosahedral grids to simulate the polar-region dynamics, it had high computational efficiency requirements, and thus is hard to couple with a complete cloud formation model like CARMA while the model of \cite{Dai2022a} excluding coagulation and nucleation calculations seems to be feasible. This study provided detailed global distributions of sulfuric acid cloud characteristics (Figure \ref{fig7}). They found that the upper-cloud mass loading peaked in the equatorial region while the mid- and lower-cloud mass loading peaked near the poles and proposed that this phenomenon was regulated by meridional circulation and the chemical generation rate of H$_{2}$SO$_{4}$. The modelled increase of the cloud density and mass in the polar regions is consistent with enhanced total opacity in the polar regions derived from the VIRTIS near-IR imaging on the night side \citep{Cardesin-Moinelo2020}.

The relevant models for the sulfuric acid cloud formation on Venus have been summarized in Table \ref{tab1}. In general, the sulfuric acid clouds play a crucial role in the Venusian atmosphere, which is associated with many aspects such as atmospheric chemistry, transport, and radiative transfer. The developments of the one-dimensional model studies on the structures and states of individual Venusian clouds are increasingly specific and concentrated. The existing studies tend to simplify several aspects of the models based on specific objectives rather than using a comprehensive model covering a wide range of microphysical processes. On the other hand, the coupling studies of clouds with other physical and chemical processes, such as large-scale circulations, have come with more detailed in recent years. The impact of short-timescale processes like condensation, coagulation, and photochemistry on large-scale spatial and temporal variations has become increasingly clear, but it remains limited by the computational efficiency. With the rapid development of computing resources, this barrier has gradually shown a trend of being broken through.

\begin{sidewaystable}
\caption{Venus cloud models}\label{tab1}
\begin{tabular}{p{1.8cm}<{\centering} p{0.8cm}<{\centering} p{0.8cm}<{\centering} p{0.8cm}<{\centering} p{0.8cm}<{\centering} p{0.8cm}<{\centering} p{0.8cm}<{\centering} p{0.8cm}<{\centering} p{1.5cm}<{\centering} p{3cm}<{\centering} p{3cm}<{\centering} p{0.5cm}<{\centering}}
\toprule%
\multirow{3}*{\shortstack{Category \\ and Name}} & \multicolumn{7}{@{}c@{}}{Features\footnotemark[1]}& \multirow{3}*{\shortstack{Computation \\ efficiency}} &  \multirow{3}*{Focus of the study} & \multirow{3}*{Notes} &  \multirow{3}*{\shortstack{Refere- \\ nce\footnotemark[2]}} \\\cmidrule{2-8}%
~ & \multirow{2}*{\shortstack{Eddy \\ diffusion}} & \multirow{2}*{\shortstack{Sedime- \\ ntation}} & \multirow{2}*{Wind} & \multirow{2}*{\shortstack{Chem- \\ istry}} & \multirow{2}*{\shortstack{Conden- \\ sation}} & \multirow{2}*{\shortstack{Coagu- \\ lation}} & \multirow{2}*{\shortstack{Nucle- \\ ation}} & ~ & ~ & ~ & ~ \\
~ & ~ & ~ & ~ & ~ & ~ & ~ & ~ & ~ & ~ & ~ & ~ \\
\midrule
\multicolumn{12}{@{}l@{}}{\textbf{Microphysical model:}} \\
CARMA (1D) & $\checkmark$  & $\checkmark$  & $\bigcirc$  & $\bigcirc$  & $\checkmark$  & $\checkmark$  & $\checkmark$  & Slow & The impact of microphysical properties on particle distributions. & Several studies use soluble CCN and include the solute effect. & (1) (2) (3) (4) (5) \\
Imamura and Hashimoto's (1D) & $\checkmark$  & $\checkmark$  & $\bigcirc$  & $\bigcirc$  & $\checkmark$  & $\checkmark$  & $\checkmark$  & Slow & The effects of dynamical and thermal structures on cloud properties. & It uses insoluble CCN. & (6) (7) \\
\midrule
\multicolumn{12}{@{}l@{}}{\textbf{Simplified cloud model:}} \\
H$_{2}$SO$_{4}$-H$_{2}$O binary condensation model (1D/3D) & $\checkmark$ & $\checkmark$ & $\times$ & $\bigcirc$ & $\checkmark$ & $\times$ & $\times$ & Moderate & The impact of acidity on the binary condensation system and the effect of eddy diffusion and chemistry on cloud cycles and distributions. & It includes the feedback of acidity. & (8) (9) \\
Analytical/semi-analytical cloud model (1D) & $\checkmark$ & $\checkmark$ & $\times$ & $\bigcirc$ & $\bigcirc$ & $\times$ & $\times$ & Fast & The effect of temperature on the feedback of acidity and gas distributions. & Several studies obtain liquid abundances and cloud mass by either iterations on acidity and particle size or certain log-normal distributions. & (10) (11) (12) (13) \\
MAD-VenLA (0D) & $\times$ & $\checkmark$ & $\times$ & $\times$ & $\bigcirc$ & $\checkmark$ & $\checkmark$ & Moderate & A variety of microphysical parameters and preparation for future coupling in GCMs. & - & (14) \\
Parameterized cloud module in GCMs (3D) & $\bigcirc$ & $\checkmark$ & $\checkmark$ & $\bigcirc$ & $\bigcirc$ & $\times$ & $\times$ & Fast & The effect of large-scale circulations on the global distributions of cloud mass loading and gases. & Studies use highly parameterized cloud formation or simple cloud distinction. & (15) (16) \\
\midrule
\botrule
\end{tabular}
\footnotetext[1]{The $\checkmark$ refers to included feature and the $\times$ means excluded feature. The $\bigcirc$ means that the feature varies in different studies or it is highly simplified.}
\footnotetext[2]{References: (1) \cite{McGouldrick+Toon2007}; (2) \cite{Gao2014}; (3) \cite{Parkinson2015a}; (4) \cite{McGouldrick2017}; (5) \cite{McGouldrick+Barth2023}; (6) \cite{Imamura+Hashimoto2001}; (7) \cite{Karyu2024}; (8) \cite{Dai2022a}; (9) \cite{Shao2024}; (10) \cite{Krasnopolsky+Pollack1994}; (11) \cite{Krasnopolsky2015}; (12) \cite{Dai2022b}; (13) \cite{Stolzenbach2023}; (14) \cite{Maattanen2023}; (15) \cite{Ando2020b}; (16) \cite{Karyu2023}.}
\end{sidewaystable}

\section{Discussions of Key Scientific Issues}\label{sec4}

\subsection{The Coupling Effect of the Clouds and Atmospheric Chemistry}\label{sec4.1}

\begin{figure}[h]
\centering
\includegraphics[width=0.9\textwidth]{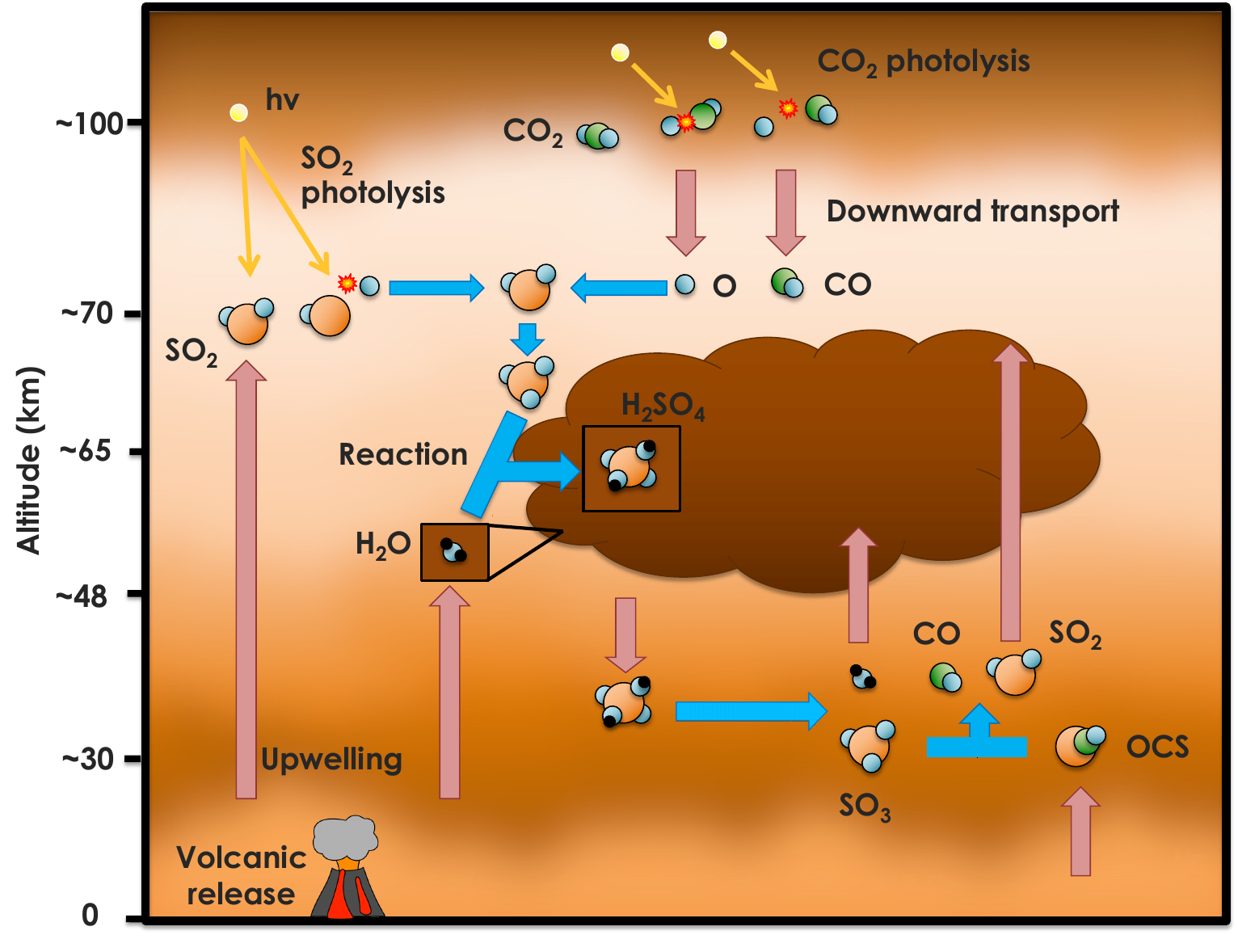}
\caption{A cartoon image of the chemical cycles associated with the cloud formation on Venus. The altitude scale is attached with a nonlinear distribution.}\label{fig8}
\end{figure}

The Venusian clouds are mainly composed of H$_{2}$SO$_{4}$, among which the sulfur element comes from photochemical processes associated with SO$_{2}$. Therefore, cloud formation is directly affected by the chemical structure of the atmosphere \citep{Yung+Demore1982, Imamura+Hashimoto2001, Krasnopolsky2012, Zhang2012a, Bierson+Zhang2020, Dai2022a, Dai2024}. The primary reservoir of sulfur element in the atmosphere of Venus is SO$_{2}$. It is lifted by atmospheric convection and diffusion, generating H$_{2}$SO$_{4}$ through a variety of (photo)chemical reactions in the cloud-top region (Figure \ref{fig8}):

\begin{equation}
SO_{2} + h\nu \to SO + O,\label{eq2}
\end{equation}
\begin{equation}
CO_{2} + h\nu \to CO + O,\label{eq3}
\end{equation}
\begin{equation}
SO_{2} + O + M \to SO_{3} + M,\label{eq4}
\end{equation}
\begin{equation}
SO_{3} + 2H_{2}O \to H_{2}SO_{4} + H_{2}O,\label{eq5}
\end{equation}
where $h\nu$ refers to solar photons, $M$ represents the total atmosphere acting as a catalyst facilitating the reaction. Most of the O atoms oxidizing SO$_{2}$ originate from the photolysis of carbon dioxide (CO$_{2}$) due to its dominant abundance in the atmosphere of Venus \citep{Rimmer2021}. The H$_{2}$SO$_{4}$ gas generated by the photochemical processes condenses at the upper-cloud altitudes, participating in cloud formation. The cloud particles sediment, reach the cloud base and rapidly evaporate, releasing a large amount of H$_{2}$SO$_{4}$ gas. About 86$\%$ of the H$_{2}$SO$_{4}$ gas is transported upward by eddy diffusion and re-condenses back into the cloud droplets \citep{Krasnopolsky+Pollack1994, Dai2022a}. The rest diffuses downward and undergoes thermal decomposition to produce sulfur trioxide (SO$_{3}$) at an altitude below 40 km \citep{Jenkins1994}, which is then reduced back to SO$_{2}$ by species such as carbonyl sulfide (OCS) \citep{Krasnopolsky2007, Bierson+Zhang2020}. Thus, the atmospheric (photo)chemistry plays a crucial role in the Venusian cloud formation.

Photochemical model studies \citep{Zhang2012a, Krasnopolsky2012, Shao2020} proposed that H$_{2}$SO$_{4}$ gas is mainly generated at an altitude range of 60-70 kilometers, while cloud model studies with sensitivity tests \citep{Imamura+Hashimoto2001, McGouldrick2017, Dai2022a, Karyu2024} pointed out that its photochemical generation efficiency significantly regulates the upper-cloud mass loading. Moreover, due to the relatively low condensation efficiency in this region, the H$_{2}$SO$_{4}$ chemical generation efficiency dominates its supersaturation in the upper cloud, which subsequently affects the homogeneous nucleation.

On the other hand, the presence of clouds also has significant feedback effect on the atmospheric chemistry, which can be reflected in several aspects:

1) The dense clouds indirectly affect atmospheric chemistry by regulating the radiative transfer. The thickness of the clouds on Venus exceeds 20 kilometers, resulting in its planetary albedo being as high as 0.8 in visible and near-IR \citep{Tomasko1979, Tomasko1980} and 0.42-0.52 in UV \citep{Bertaux2016} while the Earth’s average albedo is $\approx$0.3 \citep{Stephens2015}. The effects of clouds on radiative forcing have received wide attention on the Earth, let alone on Venus. The Venusian clouds significantly reflect the solar radiation and simultaneously absorb the long-wave ($\lambda$$\textgreater$2.7 $\mu$m) thermal emission from the surface via their high albedo and large infrared opacity, as mentioned above, affecting the radiative transfer on Venus \citep{Esposito1983, Titov2018}. This mechanism generates strong greenhouse effect, which is responsible for maintaining high temperature in the lower atmosphere and may have played a key role in Venusian climate evolution \citep{Yang2014, Turbet2021, Gillmann2022}. In other words, the sulfuric acid clouds act as a thick blanket covering the surface of Venus, blocking out both the sunlight and thermal IR radiation and dividing the middle- and lower-atmosphere into two distinct aspects: The atmosphere above the clouds is relatively thin and is exposed to solar radiation, and thus is regulated by photochemical processes. The temperature below the clouds rapidly increases as altitude decreases \citep{Seiff1985}. Since temperature significantly affects chemical efficiency and most solar radiation cannot penetrate the dense clouds, the chemistry in this region is regulated by thermochemical equilibrium in the gaseous phase and chemical buffering by the hot surface. 

\begin{figure}[h]
\centering
\includegraphics[width=0.9\textwidth]{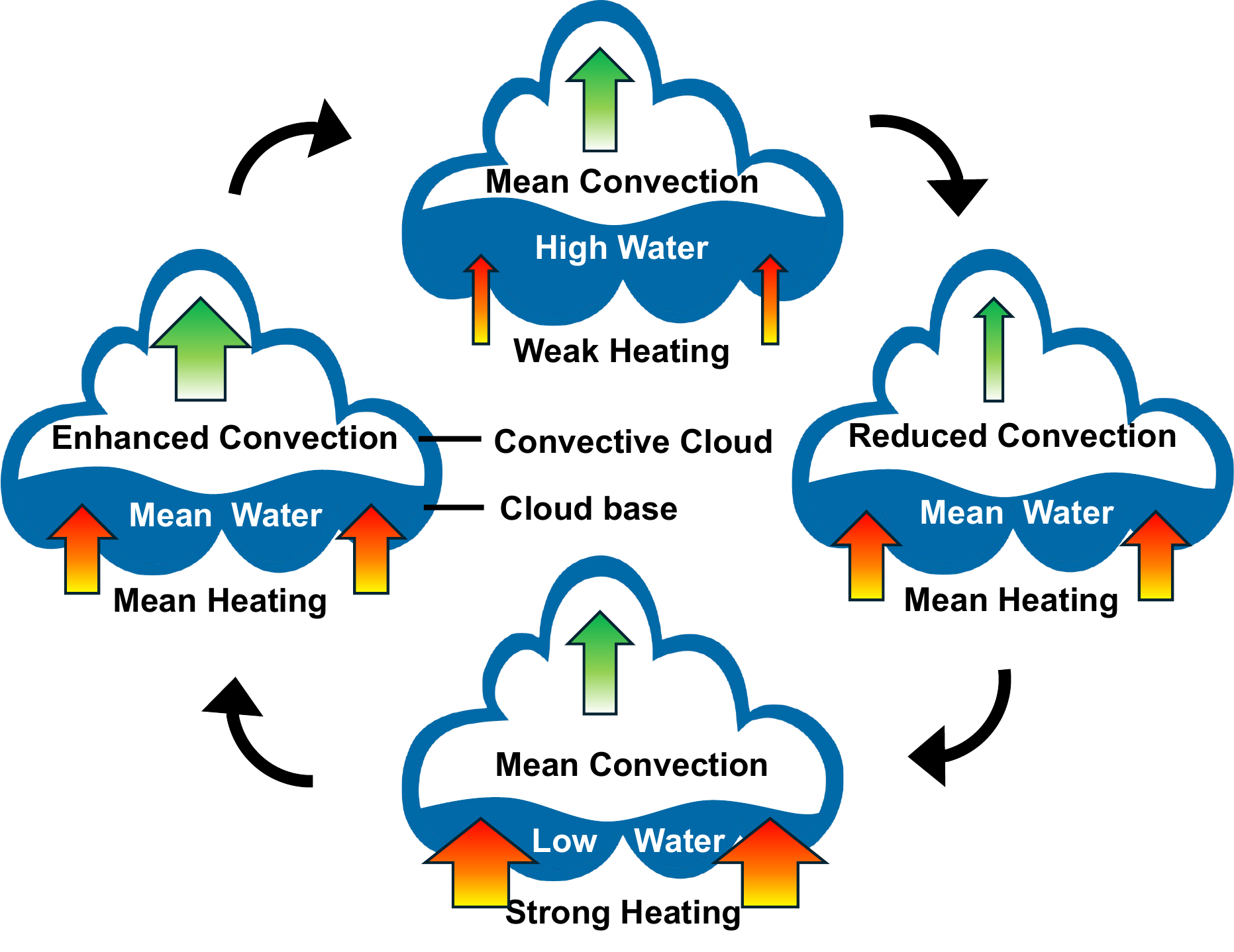}
\caption{The mechanism of the oscillation between cloud-base H$_{2}$O abundance and convection strength in the clouds. This figure is modified from \cite{Kopparla2020}.}\label{fig9}
\end{figure}

2) Clouds indirectly regulate atmospheric chemistry by affecting atmospheric dynamic processes. Significantly low static stability has been observed in Venusian cloud regions \citep{Imamura2017, Ando2020a, Oschlisniok2021}, indicating the presence of strong convective motions, which rapidly mix the atmospheres within the clouds. A recent model study has proposed a finite amplitude sustained oscillation between cloud convective strength and cloud-base water abundance for up to 3-9 years \citep{Kopparla2020} as shown in Figure \ref{fig9}. The high H$_{2}$O abundance at the cloud base weakens the cloud-base forcing from the lower atmosphere \citep{Lee2016}, suppressing the convective mixing in the clouds and decreasing the height of the convective layer. This further decreases the cloud-base H$_{2}$O abundance and starts the other half of the oscillation. Convection plays a significant role in the transport and mixing of gas species in the clouds and, therefore, regulates atmospheric chemistry. In addition, the strong zonal wind at the cloud level with velocity of up to 100 m/s affects the atmospheric chemistry by transporting the clouds and gas species horizontally. 

3) The clouds harbor internal aqueous chemistry and directly influence atmospheric chemistry through the gas-liquid exchange. The Venusian clouds are mainly composed of liquid H$_{2}$SO$_{4}$ and H$_{2}$O, along with some minor species \citep{Krasnopolsky1985}. The H$_{2}$SO$_{4}$-H$_{2}$O cycles within the clouds show significant spatial variations \citep{Dai2022a}. Regulated by low temperature, the H$_{2}$SO$_{4}$ and H$_{2}$O vapors condense into the droplets in the upper-cloud region, forming significant sinks of the two species. The acidity in this region is relatively low ($\sim$77 wt$\%$ of H$_{2}$SO$_{4}$). As the droplets settle, the temperature rises rapidly, and most of the liquid H$_{2}$O in the clouds continues to evaporate and release water vapor, leading to an acidity of 98$\%$ at the cloud base. Moreover, liquid sulfuric acid evaporates when the droplets fall below the cloud base, releasing a large amount of H$_{2}$SO$_{4}$ gas. These cycles directly affect the H$_{2}$SO$_{4}$ and H$_{2}$O chemistry at different altitudes. Besides, as one of the most important life indicators, H$_{2}$O plays a key role in constructing habitability environments on planets. Venusian clouds are suspected to act as a medium, transporting liquid H$_{2}$O and its isotopic molecule HDO up to 110 km, where evaporation and fractionation by photolysis are triggered, significantly increasing the D/H ratio with altitude \citep{Mahieux2024}. As large and non-gaseous particles, the droplets of the Venusian clouds may provide a platform facilitating the heterogeneous chemistry on the surface of the particles. Although few model studies have explored this aspect of Venus in depth, the significant impact of the Martian ice cloud heterogeneous chemistry on the stability and composition of the Martian atmosphere has been demonstrated by studies \citep[e.g.,][]{Lefevre2008}. Inclusion of heterogeneous chemistry may be one of the future directions of cloud modelling on Venus. As a liquid solvent, Venusian sulfuric acid clouds can provide reaction zones for some aqueous chemical processes \citep{Rimmer2021}. The stored SO$_{2}$ in the liquid droplets is proposed to buffer the gases and further contribute to the depletion of O$_{2}$ in the clouds \citep{Dai2024}. 

The atmospheric chemical cycle provides the material basis for forming sulfuric acid clouds. It is also affected by cloud interactions, radiative transfer, and dynamics. There are significant coupling effects between these two processes. Nevertheless, the current understanding of their temporal and spatial variations is limited, and there are even uncertainties in grasping their basic characteristics. This may justify the need for coupling model studies between the clouds and middle- and lower atmospheric chemistry. 

\begin{figure}[h]
\centering
\includegraphics[width=0.9\textwidth]{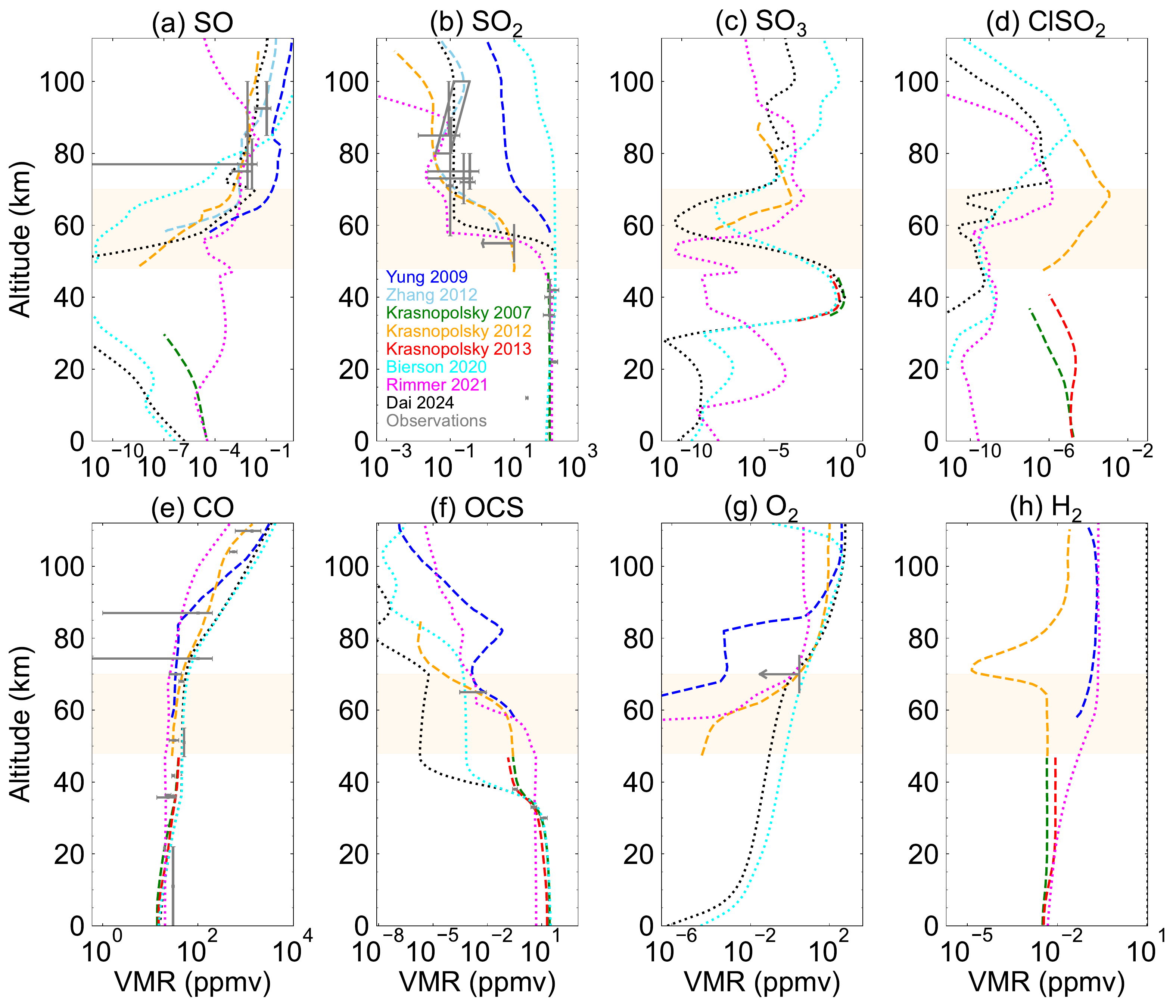}
\caption{The abundances of several crucial species in distinct model studies. The data are adapted from \cite[][darkblue]{Yung2009}, \cite[][skyblue]{Zhang2012a}, \cite[][green]{Krasnopolsky2007}, \cite[][orange]{Krasnopolsky2012}, \cite[][red]{Krasnopolsky2013a}, \cite[][cyan]{Bierson+Zhang2020}, \cite[][pink]{Rimmer2021}, and \cite[][black]{Dai2024}. The observations are marked in grey. The region filled by light yellow shows the cloud layer.}\label{fig10}
\end{figure}

Most of the present model studies on Venusian atmospheric (photo)chemistry use clouds as their model’s fixed boundaries to avoid the strong convective motions and complicated microphysics within the clouds. They separately analyzed the chemical structures of the atmospheres above or below the cloud layer \citep{Yung2009, Krasnopolsky2007, Krasnopolsky2012, Krasnopolsky2013a, Zhang2012a, Shao2020}. In several studies \citep[e.g.,][]{Bierson+Zhang2020, Dai2024} the species associated with the clouds like H$_{2}$SO$_{4}$ and H$_{2}$O were artificially controlled to maintain the equilibrium of chemistry-transport system throughout their domain, and thus the feedback effects from the clouds are inevitably neglected. By integrating the results of these studies, it can be found that the predicted atmospheric chemical structures in either regions above or below the clouds are discontinuous. Significant discrepancies exist at the junction, that is, the cloud region. The abundance profiles of several crucial species cannot effectively connect between distinct studies \citep[e.g.,][]{Yung2009, Krasnopolsky2007, Krasnopolsky2012, Krasnopolsky2013a, Zhang2012a}, especially sulfur-bearing species (Figure \ref{fig10}). The observed large depletion of SO$_{2}$ in the cloud region \citep{Vandaele2017a} cannot be adequately interpreted through independent studies of atmospheric chemistry due to the lack of H$_{2}$O \citep{Rimmer2021}. This indicates that unknown physical and chemical processes have yet to be included in these studies. Although \cite{Bierson+Zhang2020} and \cite{Dai2024} conducted chemical studies throughout the middle and lower atmosphere, they artificially fixed H$_{2}$SO$_{4}$ and H$_{2}$O distributions and excluded the cloud formation and the subsequent feedback effects in their models. \cite{Rimmer2021} calculated the cloud formation in their chemistry-transport model study by a highly parameterized scheme based on Henry’s law and aqueous ion chemistry, while their assumption that large amounts of hydroxide salts are present in the cloud region is controversial. 

Although the timescale of the cloud formation could be very short in the lower cloud region, the condensation rate significantly decreases with increasing altitude due to the rapid decrease of particle number density. The chemical generation rate of H$_{2}$SO$_{4}$ keeps in the same magnitude as the loss rate of it by condensation in a wide region near 66 km \citep{Dai2022a}, where several species like SO, SO$_{3}$, NO$_{2}$, and so on have similar chemical timescales \citep{Dai2024}. This indicates that the chemistry of these species has the potential to interact with sulfuric acid clouds. Note that the sulfur cycle is widely implicated in the atmospheric chemical structure on Venus. In addition to the cloud formation, it has essential effects on the carbon, oxygen, and chlorine cycles \citep{Yung+Demore1982, Zhang2012a, Bierson+Zhang2020}. In addition to the gas-phase chemistry, heterogeneous chemistry may further alter the lifetime of the species, which leads to multiple possibilities of the coupling state.

Generally, a close coupling between Venusian sulfuric acid clouds and atmospheric chemistry has yet to be comprehensively explored and understood. Due to the significant spatial and temporal variations and uncertainties of the atmospheric chemical structure on Venus, especially the sulfur and water cycles, research on the coupling effects between the clouds and the chemistry in the middle and lower atmospheres may become the key to deeply understanding the Venusian atmosphere.

\subsection{The “Unknown” UV Absorber}\label{sec4.2}

About 80$\%$ of the incident solar radiation energy is scattered back to space by Venus dense clouds \citep{Tomasko1980, Moroz1981}. This high albedo in the visible spectral range decreases at the short-wavelength end and in UV (0.5-0.32 nm) \citep{Barker1975}. Simultaneously, the Venusian clouds also exhibit a strong absorption in the ultraviolet range, among which the absorption spectrum with wavelengths below 320 nm highly coincides with the pattern displayed by SO$_{2}$. Thus, SO$_{2}$ is thought to be primarily responsible for the absorption in this wavelength region \citep{Esposito1980, Marcq2020}. However, there is still another absorption enhancement in the 320-400 nm range \citep{Perez-Hoyos2018, Titov2018}. Since no suitable species has been found to match it, this second absorber is usually called the “unknown” UV absorber \citep{Titov2018}.

The UV absorber significantly contributes to the heating and dynamical processes of the clouds and is one of the most mysterious unsolved issues on Venus. Observational constraints on vertical distribution of this substance are relatively scant and are based on the measurements of solar flux on descent probes. \cite{Tomasko1980} and \cite{Ekonomov1984} found that UV light absorption mainly occurs in the upper-cloud region ($\textgreater$57 km), while it is negligible in the middle and lower clouds. Vertical distribution of the UV absorbing specie at the cloud top was assessed using phase angle dependence of UV contrasts. Several studies indicated that the UV absorber is located below the cloud top \citep[e.g.,][]{Pollack1980}. VEX spectral imaging in the near-IR CO$_{2}$ bands suggested that in low and middle latitudes the cloud top is located at 72 $\pm$ 1 km decreasing to 61-67 km poleward of $\pm$50$^{\circ}$ \citep{Titov2018} with no considerable local time variations observed. Importantly, \cite{Cottini2015} did not find any systematic correlations between the cloud top altitude, water vapor abundance and brightness at 0.375–0.385 $\mu$m. However, dark UV features several hundred kilometers in size tend to coincide with enhanced cloud density (or, equivalently, with higher cloud tops) and bright features with less dense (or deeper) clouds. This result seems to contradict the general understanding derived from global Pioneer-Venus and Venus Express imaging that the UV-dark material is located deeper in the upper cloud and is being brought to the cloud top by dynamical mixing \citep{Esposito1983, Titov2008}. This conclusion, though, was related to the global UV pattern and may not take into account local processes.

Based on the VMC/VEX observations and model results, \cite{Molaverdikhani2012} predicted a highly mixed layer of UV absorber either co-located with H$_{2}$SO$_{4}$ producing region (above $\approx$63 km) or near the upper clouds ($\approx$71 km). UV images show that the clouds at the equator are significantly darker than those at middle and high latitudes, indicating a continuous supply of the UV absorber from below in the low latitudes on Venus \citep{Titov2012}. Based on the observations of VEX and model predictions, the UV absorber is responsible for $\approx$50$\%$ of the solar heating rate at the cloud top \citep{Crisp1986, Lee2015a}. The vertical distribution of this substance exhibited significant variations. It seems to keep moving from the cloud top level to the lower layers based on 5-year observations of VEX in 2006-2011 \citep{Lee2015b}. In recent years, research has believed that the UV absorber might be well mixed with mode 1 particles \citep{Perez-Hoyos2018, Marcq2020}. Based on the dependence of the full-disk brightness on the scattering direction at 365 nm, \cite{Lee2021} predicted the presence of this UV absorber in either the entire clouds or a thin layer near but below the cloud top. \cite{Lee2019} revealed strong decadal variation of Venus albedo at 365 nm in 2006-2017, indicating that the abundance of the UV absorber has varied by a factor of two. They also proposed that this variation greatly affected the solar heating, altering the strength of the Hadley circulation on Venus. This mechanism may also explain long-term variations of the circulation at the cloud top \citep[e.g.][]{Khatuntsev2013}.

\begin{figure}[h]
\centering
\includegraphics[width=0.9\textwidth]{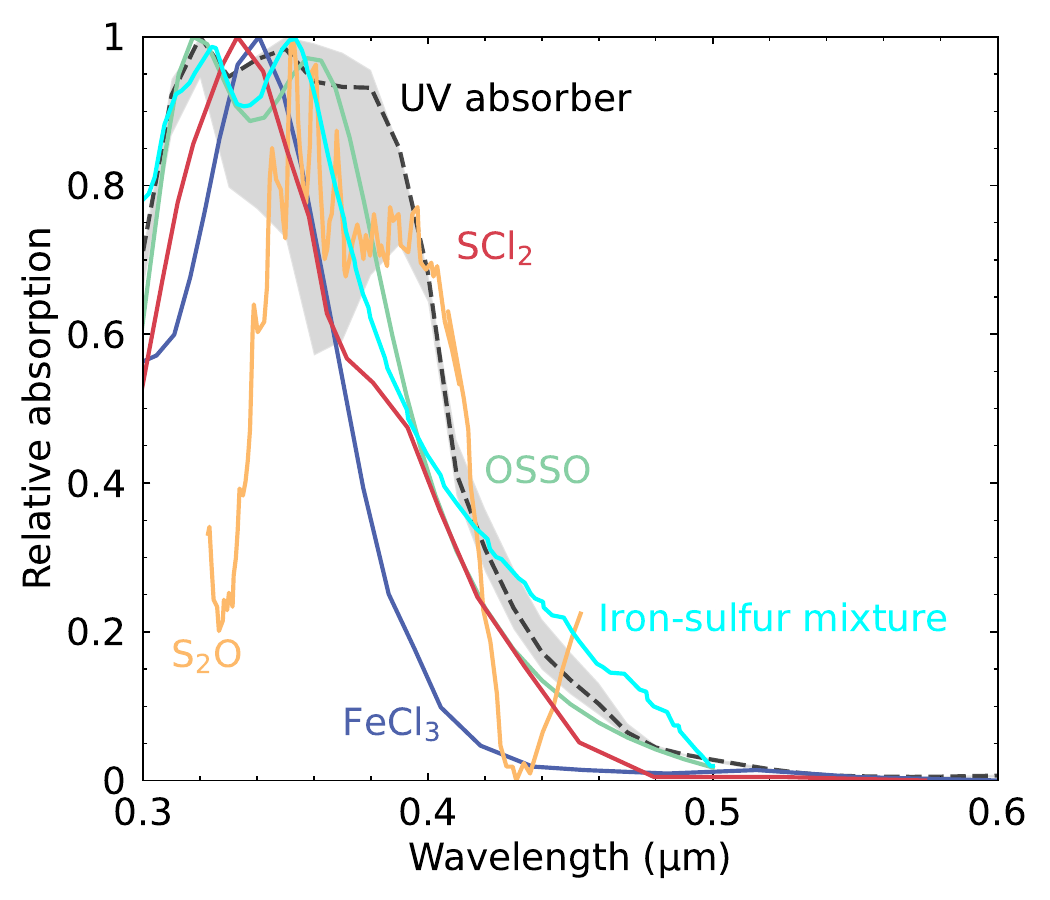}
\caption{The comparison between the absorption spectrum of distinct candidates and the UV absorber. The cyan solid line represents the mixture of 1 wt$\%$ rhomboclase in 74 wt$\%$ H$_{2}$SO$_{4}$ and 1.25 wt$\%$ AFS in 84 wt$\%$ H$_{2}$SO$_{4}$. The other lines represent the observation of UV absorber \citep[the black dashed line represents the best fitting values in the model of][and the gray area indicates the maximum and minimum values.]{Perez-Hoyos2018} and several tested candidates, including SCl$_{2}$ (red), OSSO (green), S$_{2}$O (orange), and FeCl$_{3}$ (blue). The data are adapted from \cite{Perez-Hoyos2018} and \cite{Jiang2024}}\label{fig11}
\end{figure}

Various studies have been devoted to exploring the nature of the UV absorber, but they have yet to obtain a reasonable explanation. The candidates suitable from the viewpoint of spectral properties are presented in Figure \ref{fig11}. Polysulfur (S$_{x}$) was originally considered as one of the candidates for the absorber \citep{Hapke+Nelson1975}, but all the allotropes failed to match the observed absorption spectra. Although several aerosol particles formed by coagulation between S$_{x}$ molecules were proposed to fit the observations well \citep{Carlson2010, Carlson2016}, it is still of great uncertainty whether the distribution of S$_{x}$ abundance is consistent with the observation. Based on the results of a photochemistry-transport model, \cite{Krasnopolsky2013b} proposed that a large amount of S$_{x}$ was produced in the lower atmosphere below the cloud base. \cite{Krasnopolsky2016} indicated an aerosol layer below the cloud base with a mass loading of 10$\%$ of the lower clouds by calculations on S$_{x}$ abundance. However, presence of large amounts of sulfur in the lower cloud would contradict spectral measurements on descent probes \citep{Tomasko1980, Ekonomov1984}. In addition, S$_{x}$ is insoluble in sulfuric acid. Thus, it would migrate at the surface of the spherical droplet and suppress the glory feature see in both polarimetric and intensity scattering phase functions \citep{Rossi2015, Petrova2018}. Thus, \cite{Krasnopolsky2016} believed that S$_{x}$ cannot interpret the UV absorption in the upper clouds. A recent study proposed a novel mechanism of S$_{x}$ generation by the coupling between the S and Cl atmospheric chemistries \citep{Frances-Monerris2022}, which could have an impact on the current understanding of the vertical distribution structure of sulfur in Venusian atmosphere. 

\cite{Zasova1981} proposed that 1$\%$ FeCl$_{3}$ dissolved in sulfuric acid mode 2 particles has spectral behavior close to that of the unknown UV absorber (Figure \ref{fig11}). This result was confirmed by \cite{Krasnopolsky2017}. As a soluble substance in sulfuric acid solution, FeCl$_{3}$ could act as a CCN in cloud formation. The subsequently formed mixture would have higher refractive index than sulfuric acid. Since the VMC often detected a relatively high real part of the refractive index for the 1-$\mu$m mode of cloud particles on Venus, this mixture of FeCl$_{3}$ may be a potential component with a high value of the refractive index required to be present in the cloud droplets, which significantly absorbs the radiation at the range of 320-400 nm \citep{Petrova2015, Markiewicz2018}. However, recent observations have suggested that the absorption band of the UV absorbers is shifted toward the blue wavelengths compared to previous observations \citep{Perez-Hoyos2018}. The absorption band of FeCl$_{3}$ is too narrow compared to that required for the UV absorber to match the spectral albedo of Venus. It presents a weak absorption in the 400-500 nm range, showing disagreement with observations (Figure \ref{fig11}). 

In contrast to FeCl$_{3}$, disulfur monoxide (S$_{2}$O) provides a good match to the planet spectral albedo at 400-500 nm \citep{Lo2003}. However, its absorption rapidly weakens with decreasing wavelength in the 320-340 nm range, where the observed UV absorber still has strong absorption \citep{Perez-Hoyos2018}.

Studies have recently concentrated on sulfur monoxide dimers (OSSO). \cite{Frandsen2016} proposed that two isomers of OSSO (S$_{2}$O$_{2}$) – $cis$-OSSO and $trans$-OSSO – could be generated in the atmosphere of Venus and primarily lost through near-ultraviolet photolysis. By calculating the photolysis rate and exploring the dependence of its opacity on wavelength and altitude, this study indicated that these two isomers have optical properties similar to those required for the “unknown” UV absorber and are potential candidates. Subsequently, \cite{Frandsen2016} conducted further analysis on the photochemical processes associated with SO$_{2}$, figuring out that radiation in 340-400 nm can transform SO$_{2}$ into triplet-ground-state $^{3}$SO$_{2}$ ($^{3}$B$_{1}$) and subsequently produce OSSO based on a series of chemical reactions \citep{Frandsen2020}:

\begin{equation}
SO_{2} + hv \to ^{3}SO_{2},\label{eq6}
\end{equation}
\begin{equation}
^{3}SO_{2} + CO \to ^{3}SO + CO,\label{eq7}
\end{equation}
\begin{equation}
^{3}SO_{2} + SO_{2} \to ^{3}SO + SO_{3},\label{eq8}
\end{equation}
\begin{equation}
^{3}SO + ^{3}SO \to OSSO.\label{eq9}
\end{equation}

Based on the comparison between the spectra of OSSO allotropes and the UV absorber, \cite{Frandsen2020} found that a great spectral consistency of them in the 330-420 nm range when $cis$-OSSO and $trans$-OSSO have comparable abundances in the mixture (1:1). However, even with similar absorption spectra, whether the abundance of OSSO in the Venusian atmosphere is sufficient to support it in reaching the observed absorption intensity remains a key issue. The latest photochemical-transport model study of the middle and lower atmosphere of Venus pointed out that OSSO is the main chemical product of SO in the clouds, but its chemical lifetime is extremely low. It is quickly decomposed back to SO once generated. Therefore, the volume mixing ratio of OSSO at the cloud tops does not exceed 10$^{-14}$, which is insufficient to explain the observed UV spectral albedo of Venus \citep{Dai2024}.

Based on laboratory studies, \cite{Jiang2024} focused on two iron-bearing compounds – rhomboclase [(H$_{5}$O$_{2}$)Fe(SO$_{4}$)$_{2}$·3H$_{2}$O] and acid ferric sulfate [(H$_{3}$O)Fe(SO$_{4}$)$_{2}$, AFS]. The experiments showed that under specific laboratory conditions, ferric iron (Fe$^{3+}$) in concentrated sulfuric acid solution will gradually generate various proportions of anhydrous ferric sulfate [Fe$_{2}$(SO$_{4}$)$_{3}$], rhomboclase, and AFS. Among them, rhomboclase presents high stability in 50-75 wt$\%$ H$_{2}$SO$_{4}$, while AFS is more stable in a larger weight percent of H$_{2}$SO$_{4}$. Copiapite [Fe$^{2+}$Fe$^{3+}$$_{4}$(SO$_{4}$)$_{6}$(OH)$_{2}$·20H$_{2}$O] and other hydrated iron-sulfur minerals would be formed if the environmental H$_{2}$SO$_{4}$ is lower than 50 wt$\%$. This study proposed that a linear combination of roughly equal parts 1 wt$\%$ rhomboclase in 74 wt$\%$ H$_{2}$SO$_{4}$ and 1.25 wt$\%$ AFS in 84 wt$\%$ H$_{2}$SO$_{4}$ exhibits a significant absorption in the wavelength range of 300-500 nm, which is consistent with the observations. Nevertheless, both iron-bearing compounds are formed under specific laboratory conditions rather than in a nature-like environment at the height of Venusian clouds. Thus, whether these species exist in Venusian clouds and what their abundance is remain uncertain.

Understanding the composition and distribution of the UV absorber in the Venusian clouds is of greatsignificance. To date, the composition of the UV absorber observed in the Venusian clouds is still unknown. Several other hypotheses have also been speculated like the large acid-soluble organic molecule "red oil" \citep{Spacek+Benner2021, Spacek2021, Spacek2024} or merely the fact that the UV absorber may consist not in a sole compound, but also in a mixture of several UV-absorbing molecules, which may explain the large width and lack of spectral structure in the absorption spectrum. It may be necessary to conduct a comprehensive analysis combining various factors, including in-situ measurements of the cloud species, laboratory investigations, and numerical simulation of cloud formation and atmospheric chemistry. The key points are its ability to explain the spectral albedo of Venus, the distribution of abundance in the cloud layer, and the reasonable and stable source to keep its abundance from being removed by chemistry, diffusion, and other mechanisms. 

\subsection{The Role of Clouds in Venus’ Climate and Habitability}\label{sec4.3}

\begin{figure}
\centering
\includegraphics[width=0.8\textwidth]{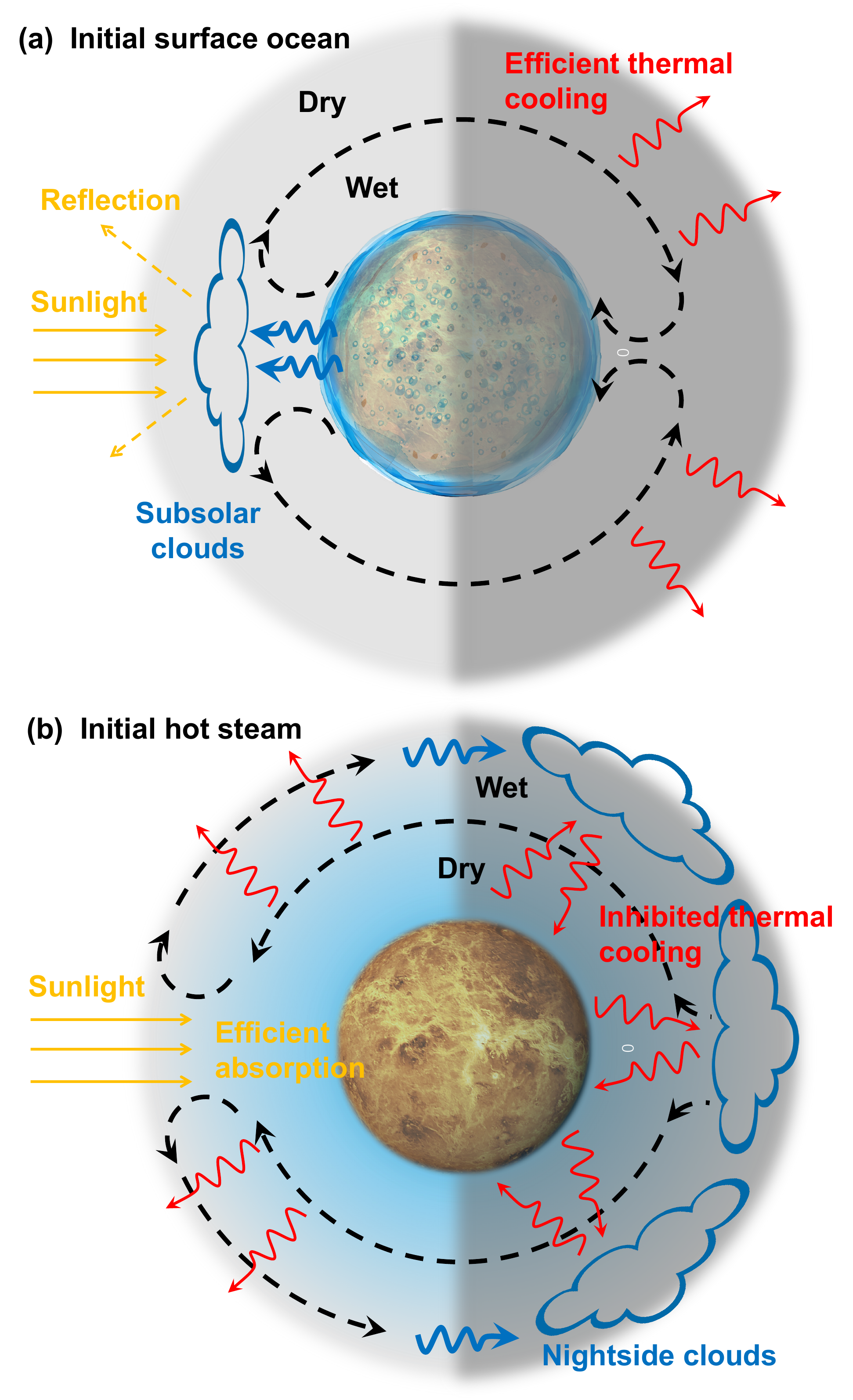}
\caption{The cartoon images of the cloud feedback mechanisms in distinct hypothesis about the climate evolution on Venus. (a) The simulations starting with a cool surface and liquid ocean; (b) The simulations starting with hot steams. This figure is modified from \cite{Turbet2021}.}\label{fig12}
\end{figure}

Habitability searching has always been one of the ultimate goals in planetary science. The ratio of deuterium (D) to hydrogen (H) in the Venusian atmosphere is about 150 times that of the Earth, as observed by early studies \citep[e.g.,][]{deBergh1991, Bertaux2007}. Recent research has even found this ratio increasing to 1,500 times larger than that of the Earth \citep{Mahieux2024}, which indicates that a large amount of hydrogen escape has occurred during the billion-year evolution on Venus. This provides potential evidence for the presence of a large amount of water vapor in the atmosphere of early Venus, and even oceans \citep{Donahue1982, McElroy1982}. Therefore, it was commonly believed that early Venus may have had a habitable environment similar to that of the Earth.

In the early days of Venus and Earth, the impact collisions with a large number of small bodies continuously heated the planets, forming the hot magmatic surfaces \citep{Hamano2013}. The subsequent rapid cooling in this state allowed abundant water vapor to condense and gather into liquid oceans on the surface \citep{Way+DelGenio2020}. However, the distance between Venus and the sun is 0.72 times that of the Earth, which means it receives nearly twice as much solar radiation as the Earth. Thus, keeping the surface ocean from being evaporated by efficient solar heating becomes a crucial issue. Several studies \citep{Yang2013, Way2016, Salvador2017, Way+DelGenio2020} have hypothesized that the subsolar clouds could be maintained due to the slow rotation of Venus. The H$_{2}$O vapor from the surface ocean may transport upward by convection and condensed to form the clouds as the temperature decreases. These subsolar clouds significantly reflect solar radiation and inhibit the heating, maintaining the surface ocean (Figure \ref{fig12}a). In other words, this hypothesis believes that early Venus followed a similar evolutionary route as the Earth during this period \citep{Gillmann2022}. 

Due to a variety of possible causes, like the increase in solar radiation and the release of CO$_{2}$ by geological activity during billion-year evolution, the evaporation of the ocean was facilitated, and the atmosphere was filled with water vapor. The strong greenhouse effect completely took over the planetary thermal emission. When the solar insolation had reached a critical level, the short-wave solar radiation penetrating the clouds was stronger than this thermal emission, the “runaway” greenhouse effect was triggered resulting in vaporing all the surface water to the atmosphere \citep{Ingersoll1969, Pierrehumbert2010}. This mechanism is thought to be responsible for disappearance of the ocean. The heating could continue until the near-infrared cooling is enhanced towards the shorter wavelengths following Planck's law, balancing the deposited solar energy. According to this hypothesis, complete evaporation of the oceans on the Earth in future may just be a matter of time as the solar luminosity grows \citep{Kasting1988, Goldblatt2013}. 

Other studies have put forward different views \citep{Turbet2021, Turbet2023, Salvador2023} that the liquid ocean has never been formed on the surface of Venus. They argued that the surface of early Venus was too hot for a liquid ocean to condense and all water inventory remained in the hot atmosphere, and thus, the clouds could not be formed. The H$_{2}$O vapor was transported upward by convection and subsequently flowed to the nightside following the atmospheric subsolar-to-antisolar circulation (SSAS), condensing to form clouds. The nightside clouds significantly absorbed the long-wave thermal emission from the surface and consequently inhibited the planetary cooling, in addition to the efficient absorption of the solar radiation by the dayside atmosphere. This kept the mean surface temperature at a high level, suppressing the formation of the liquid ocean on Venus (Figure \ref{fig12}b). The atmosphere remained hot and dry. The strong dynamical transport significantly facilitated water escape \citep{Gillmann2009} and the subsequent isotope fractionation of hydrogen. In this hypothesis, Venus never went through the ocean loss processes and the associated warming driven by the "runaway" greenhouse effect.

\cite{Bullock+Grinspoon2001} modelled the recent evolution of Venus climate related to the global resurfacing event 600–1100 Myr ago. Their results show sensitivity of the clouds to the abundance of water and sulfur dioxide in the atmosphere. Intense volcanic outgassing during this event could have led to formation of thick sulfuric acid clouds that gradually almost disappeared following loss of sulfur dioxide due to reactions with surface minerals. This resulted in significant raise of surface temperatures to $\approx$900 K. Subsequent evolution to current conditions had occurred due to loss of atmospheric water at the top of the atmosphere, ongoing low-level volcanism, and the reappearance of sulfuric acid clouds. 

The Venusian clouds played a crucial role in the evolution of the planet habitability. Settling this open issue relies on large-scale circulation simulation involving highly coupled surface-ocean-atmosphere processes, as well as assumptions and boundary conditions adopted in the simulation. The composition and proportion of the clouds on early Venus may differ from today. Therefore, more refined and physical simulations of the cloud coupling effects are of great significance for understanding the evolution of Venus' habitability.

Although the surface environment of Venus is harsh and uninhabitable, the temperature and pressure conditions in the clouds are moderate (300-350 K, 0.1-1 bar) \citep{Seager2021}. Therefore, over the past decades, there have been many attempts to search for potential life indicators in this region \citep{Limaye2021, Bains2024, Seager2023, Petkowski2024}.

The “unknown” UV absorber in the clouds was hypothesized by recent studies \citep{Limaye2018, Limaye2021} to be possibly associated with the biochemical components of microorganisms helping to absorb solar radiation. If all the extinct solar radiation in this band is used for photosynthesis, it could provide a large amount of metabolic energy \citep{Grinspoon1997}. This seems reasonable from the perspective of life evolution. The radiation filters life, resulting in the ability of the survivors to absorb and utilize the light. However, this hypothesis is secondary to confirming the existence of life on Venus, which, however, is still ongoing. Based on the absorption spectrum measurement of the products of different organic molecules in concentrated sulfuric acid solution, the latest studies believed that the organic molecules could be responsible for the UV absorption in the clouds, and the precursor simple organic matter may come from meteorite, photochemistry, and even life activities \citep{Spacek2021, Spacek2024}. It is worth noting that although inferring products from life is easy, it is necessary to be extremely cautious with inferring life from the products; after all, they may not be exclusively associated to biotic processes. There are few remote sensing observations inside the Venusian clouds, and these studies may be limited by the complicated structure of organic molecules and the diversity of potential candidates.

Water is an essential element of life. Although the Venusian clouds are in the liquid phase, they are mainly composed of H$_{2}$SO$_{4}$ while the weight percent of H$_{2}$O is only 2-15$\%$. The H$_{2}$O vapor in the cloud altitudes on Venus is about 1 ppmv, which is even 1-2$\%$ of that in the Atacama Desert, one of the driest regions on Earth, and is far below the minimum water activity that known extremophiles can withstand \citep{Hallsworth2021}.

\begin{figure}
\centering
\includegraphics[width=0.9\textwidth]{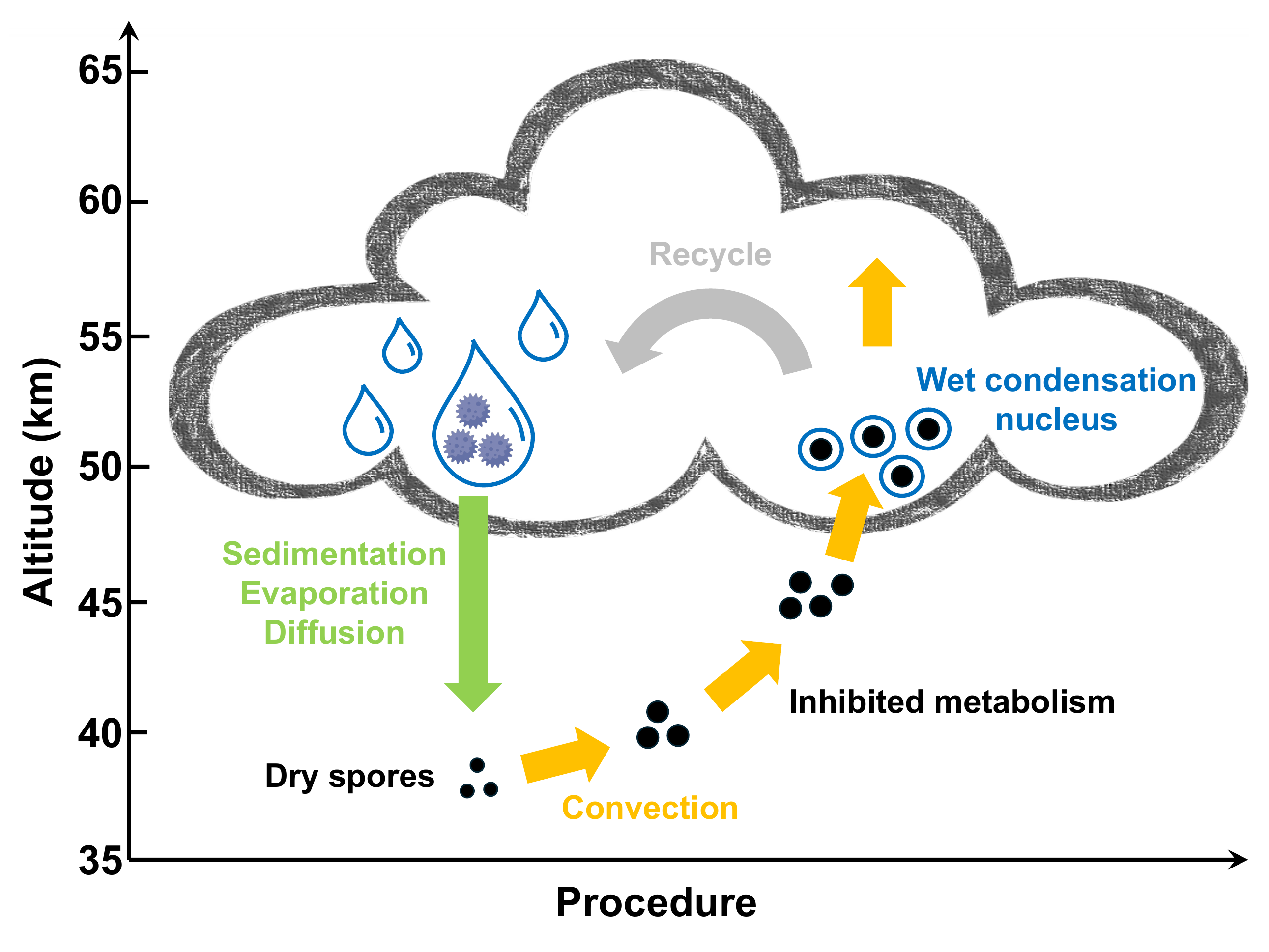}
\caption{A cartoon image of hypothesized life states in the cloud cycle. This figure is modified from \cite{Seager2021}.}\label{fig13}
\end{figure}

Oxygen (O$_{2}$) is one of the most essential substance for human survival. Although in-situ observations obtained by PVO and Venera probes exhibited the presence of tens of ppmv of O$_{2}$ in Venusian clouds \citep{Oyama1980, Mukhin1982}, remote sensing in recent decades has never directly detected O$_{2}$ in the atmosphere of Venus \citep{Trauger+Lunine1983, Marcq2018}. Such abundant O$_{2}$ cannot be maintained under the temperature and chemical conditions of the Venusian atmosphere \citep{Krasnopolsky2006a}. Therefore, these in-situ measurements are commonly considered to be questionable. Based on the instrument accuracy, \cite{Trauger+Lunine1983} derived an upper limit of 0.3 ppmv in the cloud top region, which was later revised to a column density of 8×10$^{17}$ cm$^{-2}$ \citep{Krasnopolsky2006a}. Detections of O$_{2}$ airglows \citep{Connes1979}, cloud-top ozone (O$_{3}$) at high latitudes \citep{Marcq2019}, and abundant O atoms in 90-120 km \citep{Hubers2023}, and atmospheric chemical model results \citep[e.g.,][]{Bierson+Zhang2020, Rimmer2021, Dai2024} all indicated the presence of O$_{2}$ above the Venusian clouds. Thus, the O$_{2}$ abundance and its potential sources and sinks in this region may be issues worth further study.

Phosphine (PH$_{3}$) was proposed to have been detected recently with an abundance of tens ppbv in Venusian clouds \citep{Greaves2021} bringing the topic of environmental habitability on Venus to the forefront. Due to the strong reducing property of PH$_{3}$, its chemical lifetime in the highly oxidized Venusian atmosphere is short. Thus, the direct detection indicates the presence of continuous and strong sources of this species in the clouds, which was suspected to be life activities by a few studies \citep[e.g.,][]{Greaves2021}. However, most studies have questioned phosphine detection, explaining it by either incorrect data processing \citep{Akins2021, Thompson2021, Villanueva2021}, signals mistaken with SO$_{2}$ \citep{Lincowski2021, Villanueva2021}, or wrong sensitive heights \citep{Lincowski2021}. Based on infrared observations, the upper limit of cloud PH$_{3}$ was subsequently revised to sub-ppbv range \citep{Encrenaz2020, Trompet2021, Cordiner2022}.

Besides, several other species, like SO$_{2}$, have also been marked as life indicators. Based on an atmospheric chemical model, \cite{Jordan2022} constructed the correlation between various biological metabolic activities and SO$_{2}$ consumption and proposed that the SO$_{2}$ depletion in Venusian clouds were more likely the results of abiotic processes. They suggested the upper limit of sulfur-metabolizing life in the clouds to be 10$^{-5}$-10$^{-3}$ mg/m$^{3}$.

If life exists in the droplets, it must go through the tough environment with high temperature and without liquid venues and light once the droplets fall below the cloud base and evaporate. \cite{Seager2021} proposed a novel and interesting hypothesis that microorganisms can exist in the lower hazes in the form of dry "spores" with inhibited metabolism and act as condensation nuclei after being transported to the cloud layer by convection, regaining the environmental conditions and restoring activity (Figure \ref{fig13}). We note that this hypothesis has not been verified yet, which may need the support of in-situ sampling research on Venusian clouds.

This section discussed Venusian clouds and their role in climate evolution and habitability. Clouds significantly regulate the radiative transfer and may play a crucial role in Venusian climate evolution. Understanding this impact mechanism relies on the large-scale dynamical models, which are significantly affected by assumptions on initial and boundary conditions, along with involved coupling processes. The studies searching for life indicators in the clouds tend to use lateral evidence like products of potential life activities, essential elements for life, and stability of organic molecules to support large-scope views with significant uncertainties. There is still a long way to go in exploring the clouds and environmental habitability on Venus.

\section{Summary and Outlooks}\label{sec5}

The clouds on Venus exhibit unique features and behaviors, which significantly regulate its atmospheric chemistry, energy balance and dynamics, as well as the evolution of the environmental habitability. This article reviewed the progress and perspectives of observations and model studies of Venusian sulfuric acid clouds, focusing on the spatiotemporal distribution and variation mechanisms of cloud mass loading, cloud-top height, particle size, and gas abundances. It analyzed the structure, emphasis, and development trends of existing Venusian cloud models. The modelling efforts increasingly focus on a particular feature like microphysical properties, the relationship between cloud acidity and gases, the impact from chemistry and dynamics, and the simulation efficiency as well as coupling between the processes. 

Moreover, this article provided specific considerations and analyses of the crucial issues in Venusian clouds, including the coupling effect of the clouds and atmospheric chemistry, the “unknown” UV absorber, and the environmental habitability in distinct periods. Based on observational, experimental, and modelling studies, we concluded that the coupling effects between the clouds and the chemistry in the middle and lower atmospheres may become a key for deep understanding the Venusian atmosphere. The key points of investigating the “unknown” UV absorber include analyzing the consistency of the absorption spectrum and intensity of potential candidates with observational data, examining the distribution of their abundance within the cloud layer, and exploring the possibility of their natural formation or alternative source mechanisms. The impact of the clouds on climate evolution largely relies on assumptions on initial and boundary conditions, along with involved coupling processes.

Despite of the achievements in the Venus clouds research and understanding the clouds chemistry and dynamics several fundamental questions are still open. The expected progress is associated with the future missions to Venus, EnVision, VERITAS and DAVINCI+, and supporting modelling efforts. Here, we summarize the insights and directions for future research on Venusian clouds.

\begin{enumerate}[1.]
\item The coupling between the clouds and atmospheric chemistry. The feedback of the clouds through radiative transfer, dynamics, and gas-liquid exchange. These studies should take into account strong spatial and temporal variability of the clouds and gaseous species suggested by the recent Venus Express and Akatsuki observations, as well as the new data expected from the future.

\item Identification of non-sulfuric acid components and especially the “unknown” UV absorber(s) in the clouds and study of their influence on the radiative energy budget, atmospheric chemistry, cloud formation, and habitability. It appears that capabilities of remote sensing methods in this domain are almost exhausted. The hope to unveil the enigma of the cloud composition rests with future state-of-the-art in-situ measurements on descent probes or, even better, balloons with changeable floating altitude.

\item The mechanism of climate evolution on Venus regulated by clouds. Other factors that may have potential effects on Venusian evolution. Correlations between life indicators and cloud habitability. The past modelling of the recent Venus climate \citep{Bullock+Grinspoon2001, Hashimoto+Abe2005} indicated strong coupling between the cloud properties and surface composition and volcanism. A breakthrough in our understanding of the surface mineralogy, geology and evolution is associated with imaging spectroscopy in the spectral transparency “windows” around 1 $\mu$m and synthetic aperture radar mapping by EnVision and VERITAS missions \citep[e.g.][]{Helbert2023}. Their results would help constrain geochemical and compositional models that significantly affect the clouds evolution. However, capabilities of the remote spectroscopy are strongly limited by accessible spectral domain. Thus, in-situ measurements of the surface composition in several locations, including tesserae, would be highly desirable for understanding of the surface composition and its interaction with the atmosphere. 

\item Distribution and variability of trace gases in the clouds, including SO$_{2}$, O$_{2}$, and H$_{2}$O. This includes impact of cloud aqueous chemistry on gas transport and gas-liquid exchange and implications of these processes for isotope distribution. This topic was deeply addressed by the composition investigations of the mesosphere and lower atmosphere by solar occultation spectroscopy and spectral imaging in the near-IR transparency “windows” onboard Venus Express \citep{Marcq2018, Marcq2023}. One of the goals of the EnVision mission is to establish connection between volcanic activity and atmospheric composition.

\item The uncertainty of atmospheric mixing efficiency in clouds. The effects of gas transport on cloud formation and atmospheric chemistry. Modelling of minor species transport should take into account detailed characterization of the atmospheric circulation at the clouds level \citep{Sanchez-Lavega2017} and temperature structure of the neutral atmosphere \citep{Limaye2018}.

\item The effects of short-term cloud microphysics on long-term properties of the cloud, atmospheric chemistry, and dynamics. Current understanding of the cloud microphysics is based on few early in-situ measurements by the Pioneer-Venus and Venera descent probes. Better insight in the microphysical properties can be provided by detailed in-situ investigations preferably on balloons with changeable floating altitude. However, these investigations are not yet in the plans of the world space agencies.

\item The regulation of cloud properties by particle size distribution and the composition of CCN. The large-size particles could control most of the cloud mass in the lower clouds while their presence is still controversial. The particle size distribution affects the clouds' microphysical properties, as well as the short-term and long-term variations in cloud morphology. Understanding the composition of CCN can provide constrains on its sources like salts, dusts, meteoroids and heterogeneous nucleation of H$_{2}$SO$_{4}$ and its physico-chemical properties like catalytic capacity and solubility. These are of great significance for improving the current simulation of the Venusian clouds.

\end{enumerate}

We note that exploration of the Venus clouds should be verified by observations. In view of strong limitations of remote sensing methods, in-situ detection and analysis are increasingly essential to understanding the sulfuric acid clouds on Venus. Venus has regained attention on the international stage with the approval of three new missions: EnVision by ESA and DAVINCI+ and VERITAS by NASA, which would be able to significantly advance our knowledge of Venus clouds. Whereafter, the Indian mission Shukrayaan-1 is just approved, and a variety of prospective mission concepts like Venera-D (Russia), Venus Life Finder (USA), and Rocket Lab Mission (USA) are being developed. China also develops a Venus mission concept, which may include a sample return, for the decade of 2030s. This sampling mission might be conducive to providing the most direct evidence for the habitability within the cloud, the distribution of cloud droplet size, the composition of the clouds and CCN, and so on. We hope and believe that in the near future, the mysteries of the Venusian clouds could be uncovered solving the long-standing enigma.

\bmhead{Acknowledgements}

L.D. acknowledges support from the National Natural Science Foundation of China through grant No. 42305135 and Natural Science Foundation of Hunan Province through grant No. 2023JJ40664. S.F. acknowledges funding from the Stable Support Plan Program for the Higher Education Institutions of the Shenzhen Science and Technology Innovation Commission through grant No. 20231115103030002 and High Level Special Funds through grant No. G03050K001. We are pleased to thank Hiroki Karyu for helpful comments on the Venusian cloud modelling. We appreciate receiving the materials from Peter Gao for Figure \ref{fig4} and \ref{fig8}. We thank the anonymous reviewer for the thorough and constructive comments to greatly improve this review.

\section*{Declarations}
\bmhead{Conflict of Interest Statement}
The authors declare that they have no conflict of interest.

\bibliography{Reference}

\begin{thebibliography}{180}
\providecommand{\natexlab}[1]{#1}
\providecommand{\url}[1]{{#1}}
\providecommand{\urlprefix}{URL }
\providecommand{\doi}[1]{\url{https://doi.org/#1}}
\providecommand{\eprint}[2][]{\url{#2}}
 \bibcommenthead

\bibitem[{{Akins} et~al.(2021){Akins}, {Lincowski}, {Meadows}, and
  {Steffes}}]{Akins2021}
{Akins} AB, {Lincowski} AP, {Meadows} VS, et~al (2021) {Complications in the
  ALMA Detection of Phosphine at Venus}. The Astrophysical Journal Letters
  907(2):L27. \doi{10.3847/2041-8213/abd56a},
  {\href{https://arxiv.org/abs/2101.09831}{{arXiv:2101.09831}}} {[astro-ph.EP]}

\bibitem[{{Ando} et~al.(2020{\natexlab{a}}){Ando}, {Imamura}, {Tellmann},
  {P{\"a}tzold}, {H{\"a}usler}, {Sugimoto}, {Takagi}, {Sagawa}, {Limaye},
  {Matsuda}, {Choudhary}, and {Antonita}}]{Ando2020a}
{Ando} H, {Imamura} T, {Tellmann} S, et~al (2020{\natexlab{a}}) {Thermal
  structure of the Venusian atmosphere from the sub-cloud region to the
  mesosphere as observed by radio occultation}. Scientific Reports 10:3448.
  \doi{10.1038/s41598-020-59278-8}

\bibitem[{{Ando} et~al.(2020{\natexlab{b}}){Ando}, {Takagi}, {Sugimoto},
  {Sagawa}, and {Matsuda}}]{Ando2020b}
{Ando} H, {Takagi} M, {Sugimoto} N, et~al (2020{\natexlab{b}}) {Venusian Cloud
  Distribution Simulated by a General Circulation Model}. Journal of
  Geophysical Research (Planets) 125(7):e06208. \doi{10.1029/2019JE006208}

\bibitem[{{Arney} et~al.(2014){Arney}, {Meadows}, {Crisp}, {Schmidt}, {Bailey},
  and {Robinson}}]{Arney2014}
{Arney} G, {Meadows} V, {Crisp} D, et~al (2014) {Spatially resolved
  measurements of H$_{2}$O, HCl, CO, OCS, SO$_{2}$, cloud opacity, and acid
  concentration in the Venus near-infrared spectral windows}. Journal of
  Geophysical Research (Planets) 119(8):1860--1891. \doi{10.1002/2014JE004662}

\bibitem[{{Baines} and {Delitsky}(2013)}]{Baines2013}
{Baines} KH, {Delitsky} ML (2013) {Aqueous Chemistry in the Clouds of Venus: A
  Possible Source for the UV Absorber}. In: AAS/Division for Planetary Sciences
  Meeting Abstracts \#45, p 118.08

\bibitem[{{Bains} et~al.(2024){Bains}, {Petkowski}, and {Seager}}]{Bains2024}
{Bains} W, {Petkowski} JJ, {Seager} S (2024) {Venus' Atmospheric Chemistry and
  Cloud Characteristics Are Compatible with Venusian Life}. Astrobiology
  24(4):371--385. \doi{10.1089/ast.2022.0113},
  {\href{https://arxiv.org/abs/2306.07358}{{arXiv:2306.07358}}} {[astro-ph.EP]}

\bibitem[{{Barker} et~al.(1975){Barker}, {Woodman}, {Perry}, {Hapke}, and
  {Nelson}}]{Barker1975}
{Barker} ES, {Woodman} JH, {Perry} MA, et~al (1975) {Relative spectrophotometry
  of Venus from 3067 - 5960 {\r{A}}.} Journal of the Atmospheric Sciences
  32:1205--1211. \doi{10.1175/1520-0469(1975)032<1205:RSOVFT>2.0.CO;2}

\bibitem[{{Barstow} et~al.(2012){Barstow}, {Tsang}, {Wilson}, {Irwin},
  {Taylor}, {McGouldrick}, {Drossart}, {Piccioni}, and
  {Tellmann}}]{Barstow2012}
{Barstow} JK, {Tsang} CCC, {Wilson} CF, et~al (2012) {Models of the global
  cloud structure on Venus derived from Venus Express observations}. Icarus
  217(2):542--560. \doi{10.1016/j.icarus.2011.05.018}

\bibitem[{{Bertaux} et~al.(2007){Bertaux}, {Vandaele}, {Korablev}, {Villard},
  {Fedorova}, {Fussen}, {Qu{\'e}merais}, {Belyaev}, {Mahieux}, {Montmessin},
  {Muller}, {Neefs}, {Nevejans}, {Wilquet}, {Dubois}, {Hauchecorne},
  {Stepanov}, {Vinogradov}, {Rodin}, {Bertaux}, {Nevejans}, {Korablev},
  {Montmessin}, {Vandaele}, {Fedorova}, {Cabane}, {Chassefi{\`e}re},
  {Chaufray}, {Dimarellis}, {Dubois}, {Hauchecorne}, {Leblanc}, {Lef{\`e}vre},
  {Rannou}, {Qu{\'e}merais}, {Villard}, {Fussen}, {Muller}, {Neefs}, {van
  Ransbeeck}, {Wilquet}, {Rodin}, {Stepanov}, {Vinogradov}, {Zasova}, {Forget},
  {Lebonnois}, {Titov}, {Rafkin}, {Durry}, {G{\'e}rard}, and
  {Sandel}}]{Bertaux2007}
{Bertaux} JL, {Vandaele} AC, {Korablev} O, et~al (2007) {A warm layer in Venus'
  cryosphere and high-altitude measurements of HF, HCl, H$_{2}$O and HDO}.
  Nature 450(7170):646--649. \doi{10.1038/nature05974}

\bibitem[{{Bertaux} et~al.(2016){Bertaux}, {Khatuntsev}, {Hauchecorne},
  {Markiewicz}, {Marcq}, {Lebonnois}, {Patsaeva}, {Turin}, and
  {Fedorova}}]{Bertaux2016}
{Bertaux} JL, {Khatuntsev} IV, {Hauchecorne} A, et~al (2016) {Influence of
  Venus topography on the zonal wind and UV albedo at cloud top level: The role
  of stationary gravity waves}. Journal of Geophysical Research (Planets)
  121(6):1087--1101. \doi{10.1002/2015JE004958}

\bibitem[{{Bierson} and {Zhang}(2020)}]{Bierson+Zhang2020}
{Bierson} CJ, {Zhang} X (2020) {Chemical Cycling in the Venusian Atmosphere: A
  Full Photochemical Model From the Surface to 110 km}. Journal of Geophysical
  Research (Planets) 125(7):e06159. \doi{10.1029/2019JE006159}

\bibitem[{{Bullock} and {Grinspoon}(2001)}]{Bullock+Grinspoon2001}
{Bullock} MA, {Grinspoon} DH (2001) {The Recent Evolution of Climate on Venus}.
  Icarus 150(1):19--37. \doi{10.1006/icar.2000.6570}

\bibitem[{{Cardes{\'\i}n-Moinelo} et~al.(2020){Cardes{\'\i}n-Moinelo},
  {Piccioni}, {Migliorini}, {Grassi}, {Cottini}, {Titov}, {Politi}, {Nuccilli},
  and {Drossart}}]{Cardesin-Moinelo2020}
{Cardes{\'\i}n-Moinelo} A, {Piccioni} G, {Migliorini} A, et~al (2020) {Global
  maps of Venus nightside mean infrared thermal emissions obtained by VIRTIS on
  Venus Express}. Icarus 343:113683. \doi{10.1016/j.icarus.2020.113683},
  {\href{https://arxiv.org/abs/2003.00228}{{arXiv:2003.00228}}} {[astro-ph.EP]}

\bibitem[{Carlson(2010)}]{Carlson2010}
Carlson R (2010) Venus’ ultraviolet absorber and sulfuric acid droplets. In:
  International Venus Conference, Aussois, France, pp 4--4

\bibitem[{Carlson(2016)}]{Carlson2016}
Carlson R (2016) In: International Venus Conference, Oxford, UK

\bibitem[{{Carlson} et~al.(1993){Carlson}, {Kamp}, {Baines}, {Pollack},
  {Grinspoon}, {Encrenaz}, {Drossart}, and {Taylor}}]{Carlson1993}
{Carlson} RW, {Kamp} LW, {Baines} KH, et~al (1993) {Variations in Venus cloud
  particle properties: a new view of Venus's cloud morphology as observed by
  the Galileo near-infrared mapping spectrometer}. Planetary and Space Science
  41(7):477--485. \doi{10.1016/0032-0633(93)90030-6}

\bibitem[{{Carslaw} et~al.(2002){Carslaw}, {Harrison}, and
  {Kirkby}}]{Carslaw2002}
{Carslaw} KS, {Harrison} RG, {Kirkby} J (2002) {Cosmic Rays, Clouds, and
  Climate}. Science 298(5599):1732--1737. \doi{10.1126/science.1076964}

\bibitem[{{Connes} et~al.(1979){Connes}, {Noxon}, {Traub}, and
  {Carleton}}]{Connes1979}
{Connes} P, {Noxon} JF, {Traub} WA, et~al (1979) {O$_{2}$($^{1}$DELTA )
  emission in the day and night airglow of Venus.} The Astrophysical Journal
  Letters 233:L29--L32. \doi{10.1086/183070}

\bibitem[{{Cordiner} et~al.(2022){Cordiner}, {Villanueva}, {Wiesemeyer},
  {Milam}, {de Pater}, {Moullet}, {Aladro}, {Nixon}, {Thelen}, {Charnley},
  {Stutzki}, {Kofman}, {Faggi}, {Liuzzi}, {Cosentino}, and
  {McGuire}}]{Cordiner2022}
{Cordiner} MA, {Villanueva} GL, {Wiesemeyer} H, et~al (2022) {Phosphine in the
  Venusian Atmosphere: A Strict Upper Limit From SOFIA GREAT Observations}.
  Geophysical Research Letters 49(22):e2022GL101055.
  \doi{10.1029/2022GL101055},
  {\href{https://arxiv.org/abs/2210.13519}{{arXiv:2210.13519}}} {[astro-ph.EP]}

\bibitem[{{Cottini} et~al.(2012){Cottini}, {Ignatiev}, {Piccioni}, {Drossart},
  {Grassi}, and {Markiewicz}}]{Cottini2012}
{Cottini} V, {Ignatiev} NI, {Piccioni} G, et~al (2012) {Water vapor near the
  cloud tops of Venus from Venus Express/VIRTIS dayside data}. Icarus
  217(2):561--569. \doi{10.1016/j.icarus.2011.06.018}

\bibitem[{{Cottini} et~al.(2015){Cottini}, {Ignatiev}, {Piccioni}, and
  {Drossart}}]{Cottini2015}
{Cottini} V, {Ignatiev} NI, {Piccioni} G, et~al (2015) {Water vapor near Venus
  cloud tops from VIRTIS-H/Venus express observations 2006-2011}. Planetary and
  Space Science 113:219--225. \doi{10.1016/j.pss.2015.03.012}

\bibitem[{{Crisp}(1986)}]{Crisp1986}
{Crisp} D (1986) {Radiative forcing of the Venus mesosphere I. Solar fluxes and
  heating rates}. Icarus 67(3):484--514. \doi{10.1016/0019-1035(86)90126-0}

\bibitem[{{Dai} et~al.(2022{\natexlab{a}}){Dai}, {Zhang}, and {Cui}}]{Dai2022b}
{Dai} L, {Zhang} X, {Cui} J (2022{\natexlab{a}}) {A fast, semi-analytical model
  for the Venusian binary cloud system}. Monthly Notices of the Royal
  Astronomical Society 515(1):817--827. \doi{10.1093/mnras/stac1803},
  {\href{https://arxiv.org/abs/2207.10243}{{arXiv:2207.10243}}} {[astro-ph.EP]}

\bibitem[{{Dai} et~al.(2022{\natexlab{b}}){Dai}, {Zhang}, {Shao}, {Bierson},
  and {Cui}}]{Dai2022a}
{Dai} L, {Zhang} X, {Shao} WD, et~al (2022{\natexlab{b}}) {A Simple
  Condensation Model for the H$_{2}$SO$_{4}$-H$_{2}$O Gas-Cloud System on
  Venus}. Journal of Geophysical Research (Planets) 127(3):e07060.
  \doi{10.1029/2021JE007060},
  {\href{https://arxiv.org/abs/2207.10238}{{arXiv:2207.10238}}} {[astro-ph.EP]}

\bibitem[{{Dai} et~al.(2023){Dai}, {Shao}, {Gu}, and {Sheng}}]{Dai2023}
{Dai} L, {Shao} W, {Gu} H, et~al (2023) {Determination of the eddy diffusion in
  the Venusian clouds from VeRa sulfuric acid observations}. Astronomy \&
  Astrophysics 679:A155. \doi{10.1051/0004-6361/202347714}

\bibitem[{{Dai} et~al.(2024){Dai}, {Shao}, and {Sheng}}]{Dai2024}
{Dai} L, {Shao} W, {Sheng} Z (2024) {An investigation into Venusian atmospheric
  chemistry based on an open-access photochemistry-transport model at
  0{\textendash}112 km}. Astronomy \& Astrophysics 689:A55.
  \doi{10.1051/0004-6361/202450552}

\bibitem[{{Dartnell} et~al.(2015){Dartnell}, {Nordheim}, {Patel}, {Mason},
  {Coates}, and {Jones}}]{Dartnell2015}
{Dartnell} LR, {Nordheim} TA, {Patel} MR, et~al (2015) {Constraints on a
  potential aerial biosphere on Venus: I. Cosmic rays}. Icarus 257:396--405.
  \doi{10.1016/j.icarus.2015.05.006}

\bibitem[{{de Bergh} et~al.(1991){de Bergh}, {Bezard}, {Owen}, {Crisp},
  {Maillard}, and {Lutz}}]{deBergh1991}
{de Bergh} C, {Bezard} B, {Owen} T, et~al (1991) {Deuterium on Venus:
  Observations from Earth}. Science 251(4993):547--549.
  \doi{10.1126/science.251.4993.547}

\bibitem[{{Donahue} et~al.(1982){Donahue}, {Hoffman}, {Hodges}, and
  {Watson}}]{Donahue1982}
{Donahue} TM, {Hoffman} JH, {Hodges} RR, et~al (1982) {Venus Was Wet: A
  Measurement of the Ratio of Deuterium to Hydrogen}. Science
  216(4546):630--633. \doi{10.1126/science.216.4546.630}

\bibitem[{{Ekonomov} et~al.(1984){Ekonomov}, {Moroz}, {Moshkin}, {Gnedykh},
  {Golovin}, and {Crigoryev}}]{Ekonomov1984}
{Ekonomov} AP, {Moroz} VI, {Moshkin} BE, et~al (1984) {Scattered UV solar
  radiation within the clouds of Venus}. Nature 307(5949):345--347.
  \doi{10.1038/307345a0}

\bibitem[{{Encrenaz} et~al.(2020){Encrenaz}, {Greathouse}, {Marcq}, {Sagawa},
  {Widemann}, {B{\'e}zard}, {Fouchet}, {Lef{\`e}vre}, {Lebonnois}, {Atreya},
  {Lee}, {Giles}, {Watanabe}, {Shao}, {Zhang}, and {Bierson}}]{Encrenaz2020}
{Encrenaz} T, {Greathouse} TK, {Marcq} E, et~al (2020) {HDO and SO$_{2}$
  thermal mapping on Venus. V. Evidence for a long-term anti-correlation}.
  Astronomy \& Astrophysics 639:A69. \doi{10.1051/0004-6361/202037741}

\bibitem[{{Esposito}(1980)}]{Esposito1980}
{Esposito} LW (1980) {Ultraviolet contrasts and the absorbers near the Venus
  cloud tops}. Journal of Geophysical Research 85:8151--8157.
  \doi{10.1029/JA085iA13p08151}

\bibitem[{{Esposito}(1997)}]{Esposito1997}
{Esposito} LW (1997) {Chemistry of lower atmosphere and clouds.} In: Venus II:
  Geology, geophysics, atmosphere, and solar wind environment. University of
  Arizona Press, p 415–458

\bibitem[{{Esposito} et~al.(1983){Esposito}, {Knollenberg}, {Marov}, {Toon},
  and {Turco}}]{Esposito1983}
{Esposito} LW, {Knollenberg} RG, {Marov} MI, et~al (1983) {The clouds and hazes
  of Venus.} In: Venus. University of Arizona Press, p 484--564

\bibitem[{{Fedorova} et~al.(2015){Fedorova}, {B{\'e}zard}, {Bertaux},
  {Korablev}, and {Wilson}}]{Fedorova2015}
{Fedorova} A, {B{\'e}zard} B, {Bertaux} JL, et~al (2015) {The CO$_{2}$
  continuum absorption in the 1.10- and 1.18-{\ensuremath{\mu}}m windows on
  Venus from Maxwell Montes transits by SPICAV IR onboard Venus express}.
  Planetary and Space Science 113:66--77. \doi{10.1016/j.pss.2014.08.010}

\bibitem[{{Franc{\'e}s-Monerris} et~al.(2022){Franc{\'e}s-Monerris},
  {Carmona-Garc{\'\i}a}, {Trabelsi}, {Saiz-Lopez}, {Lyons}, {Francisco}, and
  {Roca-Sanju{\'a}n}}]{Frances-Monerris2022}
{Franc{\'e}s-Monerris} A, {Carmona-Garc{\'\i}a} J, {Trabelsi} T, et~al (2022)
  {Photochemical and thermochemical pathways to S$_{2}$ and polysulfur
  formation in the atmosphere of Venus}. Nature Communications 13:4425.
  \doi{10.1038/s41467-022-32170-x}

\bibitem[{{Frandsen} et~al.(2016){Frandsen}, {Wennberg}, and
  {Kjaergaard}}]{Frandsen2016}
{Frandsen} BN, {Wennberg} PO, {Kjaergaard} HG (2016) {Identification of OSSO as
  a near-UV absorber in the Venusian atmosphere}. Geophysical Research Letters
  43(21):11,146--11,155. \doi{10.1002/2016GL070916}

\bibitem[{{Frandsen} et~al.(2020){Frandsen}, {Farahani}, {Vogt}, {Lane}, and
  {Kjaergaard}}]{Frandsen2020}
{Frandsen} BN, {Farahani} S, {Vogt} E, et~al (2020) {Spectroscopy of OSSO and
  Other Sulfur Compounds Thought to be Present in the Venus Atmosphere}.
  Journal of Physical Chemistry A 124(35):7047--7059.
  \doi{10.1021/acs.jpca.0c04388}

\bibitem[{{Frisch} and {Mueller}(2013)}]{Frisch+Mueller2013}
{Frisch} PC, {Mueller} HR (2013) {Time-Variability in the Interstellar Boundary
  Conditions of the Heliosphere: Effect of the Solar Journey on the Galactic
  Cosmic Ray Flux at Earth}. Space Science Reviews 176(1-4):21--34.
  \doi{10.1007/s11214-011-9776-x},
  {\href{https://arxiv.org/abs/1010.4507}{{arXiv:1010.4507}}} {[astro-ph.GA]}

\bibitem[{{Gao} et~al.(2014){Gao}, {Zhang}, {Crisp}, {Bardeen}, and
  {Yung}}]{Gao2014}
{Gao} P, {Zhang} X, {Crisp} D, et~al (2014) {Bimodal distribution of sulfuric
  acid aerosols in the upper haze of Venus}. Icarus 231:83--98.
  \doi{10.1016/j.icarus.2013.10.013},
  {\href{https://arxiv.org/abs/1312.3750}{{arXiv:1312.3750}}} {[astro-ph.EP]}

\bibitem[{{Garvin} et~al.(2022){Garvin}, {Getty}, {Arney}, {Johnson}, {Kohler},
  {Schwer}, {Sekerak}, {Bartels}, {Saylor}, {Elliott}, {Goodloe}, {Garrison},
  {Cottini}, {Izenberg}, {Lorenz}, {Malespin}, {Ravine}, {Webster}, {Atkinson},
  {Aslam}, {Atreya}, {Bos}, {Brinckerhoff}, {Campbell}, {Crisp}, {Filiberto},
  {Forget}, {Gilmore}, {Gorius}, {Grinspoon}, {Hofmann}, {Kane}, {Kiefer},
  {Lebonnois}, {Mahaffy}, {Pavlov}, {Trainer}, {Zahnle}, and
  {Zolotov}}]{Garvin2022}
{Garvin} JB, {Getty} SA, {Arney} GN, et~al (2022) {Revealing the Mysteries of
  Venus: The DAVINCI Mission}. The Planetary Science Journal 3(5):117.
  \doi{10.3847/PSJ/ac63c2},
  {\href{https://arxiv.org/abs/2206.07211}{{arXiv:2206.07211}}} {[astro-ph.EP]}

\bibitem[{{Gierasch}(1987)}]{Gierasch1987}
{Gierasch} PJ (1987) {Waves in the atmosphere of Venus}. Nature
  328(6130):510--512. \doi{10.1038/328510a0}

\bibitem[{{Gillmann} et~al.(2009){Gillmann}, {Chassefi{\`e}re}, and
  {Lognonn{\'e}}}]{Gillmann2009}
{Gillmann} C, {Chassefi{\`e}re} E, {Lognonn{\'e}} P (2009) {A consistent
  picture of early hydrodynamic escape of Venus atmosphere explaining present
  Ne and Ar isotopic ratios and low oxygen atmospheric content}. Earth and
  Planetary Science Letters 286(3-4):503--513. \doi{10.1016/j.epsl.2009.07.016}

\bibitem[{{Gillmann} et~al.(2022){Gillmann}, {Way}, {Avice}, {Breuer},
  {Golabek}, {H{\"o}ning}, {Krissansen-Totton}, {Lammer}, {O'Rourke},
  {Persson}, {Plesa}, {Salvador}, {Scherf}, and {Zolotov}}]{Gillmann2022}
{Gillmann} C, {Way} MJ, {Avice} G, et~al (2022) {The Long-Term Evolution of the
  Atmosphere of Venus: Processes and Feedback Mechanisms}. Space Science
  Reviews 218(7):56. \doi{10.1007/s11214-022-00924-0},
  {\href{https://arxiv.org/abs/2204.08540}{{arXiv:2204.08540}}} {[astro-ph.EP]}

\bibitem[{{Gnedykh} et~al.(1987){Gnedykh}, {Zasova}, {Moroz}, {Moshkin}, and
  {Ekonomov}}]{Gnedykh1987}
{Gnedykh} VI, {Zasova} LV, {Moroz} VI, et~al (1987) {The vertical structure of
  the Venus cloud layer at the landing sites of Vega 1 and Vega 2.}
  Kosmicheskie Issledovaniia 25:707--714

\bibitem[{{Goldblatt} et~al.(2013){Goldblatt}, {Robinson}, {Zahnle}, and
  {Crisp}}]{Goldblatt2013}
{Goldblatt} C, {Robinson} TD, {Zahnle} KJ, et~al (2013) {Low simulated
  radiation limit for runaway greenhouse climates}. Nature Geoscience
  6(8):661--667. \doi{10.1038/ngeo1892}

\bibitem[{{Gordon} et~al.(2017){Gordon}, {Kirkby}, {Baltensperger}, {Bianchi},
  {Breitenlechner}, {Curtius}, {Dias}, {Dommen}, {Donahue}, {Dunne},
  {Duplissy}, {Ehrhart}, {Flagan}, {Frege}, {Fuchs}, {Hansel}, {Hoyle},
  {Kulmala}, {K{\"u}rten}, {Lehtipalo}, {Makhmutov}, {Molteni}, {Rissanen},
  {Stozkhov}, {Tr{\"o}stl}, {Tsagkogeorgas}, {Wagner}, {Williamson}, {Wimmer},
  {Winkler}, {Yan}, and {Carslaw}}]{Gordon2017}
{Gordon} H, {Kirkby} J, {Baltensperger} U, et~al (2017) {Causes and importance
  of new particle formation in the present-day and preindustrial atmospheres}.
  Journal of Geophysical Research (Atmospheres) 122(16):8739--8760.
  \doi{10.1002/2017JD026844}

\bibitem[{{Greaves} et~al.(2021){Greaves}, {Richards}, {Bains}, {Rimmer},
  {Sagawa}, {Clements}, {Seager}, {Petkowski}, {Sousa-Silva}, {Ranjan},
  {Drabek-Maunder}, {Fraser}, {Cartwright}, {Mueller-Wodarg}, {Zhan},
  {Friberg}, {Coulson}, {Lee}, and {Hoge}}]{Greaves2021}
{Greaves} JS, {Richards} AMS, {Bains} W, et~al (2021) {Phosphine gas in the
  cloud decks of Venus}. Nature Astronomy 5:655--664.
  \doi{10.1038/s41550-020-1174-4}

\bibitem[{{Grinspoon}(1997)}]{Grinspoon1997}
{Grinspoon} DH (1997) {Venus revealed : a new look below the clouds of our
  mysterious twin planet}

\bibitem[{{Hallsworth} et~al.(2021){Hallsworth}, {Koop}, {Dallas}, {Zorzano},
  {Burkhardt}, {Golyshina}, {Mart{\'\i}n-Torres}, {Dymond}, {Ball}, and
  {McKay}}]{Hallsworth2021}
{Hallsworth} JE, {Koop} T, {Dallas} TD, et~al (2021) {Water activity in Venus's
  uninhabitable clouds and other planetary atmospheres}. Nature Astronomy
  5:665--675. \doi{10.1038/s41550-021-01391-3}

\bibitem[{{Hamano} et~al.(2013){Hamano}, {Abe}, and {Genda}}]{Hamano2013}
{Hamano} K, {Abe} Y, {Genda} H (2013) {Emergence of two types of terrestrial
  planet on solidification of magma ocean}. Nature 497(7451):607--610.
  \doi{10.1038/nature12163}

\bibitem[{{Hansen} and {Hovenier}(1974)}]{Hansen+Hovenier1974}
{Hansen} JE, {Hovenier} JW (1974) {Interpretation of the polarization of
  Venus.} Journal of the Atmospheric Sciences 31:1137--1160.
  \doi{10.1175/1520-0469(1974)031<1137:IOTPOV>2.0.CO;2}

\bibitem[{{Hapke} and {Nelson}(1975)}]{Hapke+Nelson1975}
{Hapke} B, {Nelson} R (1975) {Evidence for an elemental sulfur component of the
  clouds from Venus spectrophotometry.} Journal of the Atmospheric Sciences
  32:1212--1218. \doi{10.1175/1520-0469(1975)032<1212:EFAESC>2.0.CO;2}

\bibitem[{{Hashimoto} and {Abe}(2005)}]{Hashimoto+Abe2005}
{Hashimoto} GL, {Abe} Y (2005) {Climate control on Venus: Comparison of the
  carbonate and pyrite models}. Planetary and Space Science 53(8):839--848.
  \doi{10.1016/j.pss.2005.01.005}

\bibitem[{{Haus} et~al.(2013){Haus}, {Kappel}, and {Arnold}}]{Haus2013}
{Haus} R, {Kappel} D, {Arnold} G (2013) {Self-consistent retrieval of
  temperature profiles and cloud structure in the northern hemisphere of Venus
  using VIRTIS/VEX and PMV/VENERA-15 radiation measurements}. Planetary and
  Space Science 89:77--101. \doi{10.1016/j.pss.2013.09.020}

\bibitem[{{Haus} et~al.(2014){Haus}, {Kappel}, and {Arnold}}]{Haus2014}
{Haus} R, {Kappel} D, {Arnold} G (2014) {Atmospheric thermal structure and
  cloud features in the southern hemisphere of Venus as retrieved from
  VIRTIS/VEX radiation measurements}. Icarus 232:232--248.
  \doi{10.1016/j.icarus.2014.01.020}

\bibitem[{{Helbert} et~al.(2023){Helbert}, {Alemanno}, {Maturilli}, {van Den
  Neucker}, {Adeli}, and {Dyar}}]{Helbert2023}
{Helbert} J, {Alemanno} G, {Maturilli} A, et~al (2023) {The new Venus spectral
  facility at the DLR Planetary Spectroscopy Laboratory to support the ESA
  EnVision and NASA DACINCI and VERITAS missions}. In: {Strojnik} M, {Helbert}
  J (eds) Infrared Remote Sensing and Instrumentation XXXI, p 1268605,
  \doi{10.1117/12.2676635}

\bibitem[{{Horinouchi} et~al.(2020){Horinouchi}, {Hayashi}, {Watanabe},
  {Yamada}, {Yamazaki}, {Kouyama}, {Taguchi}, {Fukuhara}, {Takagi}, {Ogohara},
  {Murakami}, {Peralta}, {Limaye}, {Imamura}, {Nakamura}, {Sato}, and
  {Satoh}}]{Horinouchi2020}
{Horinouchi} T, {Hayashi} YY, {Watanabe} S, et~al (2020) {How waves and
  turbulence maintain the super-rotation of Venus{\textquoteright} atmosphere}.
  Science 368(6489):405--409. \doi{10.1126/science.aaz4439}

\bibitem[{{H{\"u}bers} et~al.(2023){H{\"u}bers}, {Richter}, {Graf},
  {G{\"u}sten}, {Klein}, {Stutzki}, and {Wiesemeyer}}]{Hubers2023}
{H{\"u}bers} HW, {Richter} H, {Graf} UU, et~al (2023) {Direct detection of
  atomic oxygen on the dayside and nightside of Venus}. Nature Communications
  14:6812. \doi{10.1038/s41467-023-42389-x}

\bibitem[{{Ignatiev} et~al.(2009){Ignatiev}, {Titov}, {Piccioni}, {Drossart},
  {Markiewicz}, {Cottini}, {Roatsch}, {Almeida}, and {Manoel}}]{Ignatiev2009}
{Ignatiev} NI, {Titov} DV, {Piccioni} G, et~al (2009) {Altimetry of the Venus
  cloud tops from the Venus Express observations}. Journal of Geophysical
  Research (Planets) 114(E5):E00B43. \doi{10.1029/2008JE003320}

\bibitem[{{Imamura} and {Hashimoto}(2001)}]{Imamura+Hashimoto2001}
{Imamura} T, {Hashimoto} GL (2001) {Microphysics of Venusian Clouds in Rising
  Tropical Air.} Journal of the Atmospheric Sciences 58(23):3597--3612.
  \doi{10.1175/1520-0469(2001)058<3597:MOVCIR>2.0.CO;2}

\bibitem[{{Imamura} et~al.(2017){Imamura}, {Ando}, {Tellmann}, {P{\"a}tzold},
  {H{\"a}usler}, {Yamazaki}, {Sato}, {Noguchi}, {Futaana}, {Oschlisniok},
  {Limaye}, {Choudhary}, {Murata}, {Takeuchi}, {Hirose}, {Ichikawa}, {Toda},
  {Tomiki}, {Abe}, {Yamamoto}, {Noda}, {Iwata}, {Murakami}, {Satoh},
  {Fukuhara}, {Ogohara}, {Sugiyama}, {Kashimura}, {Ohtsuki}, {Takagi},
  {Yamamoto}, {Hirata}, {Hashimoto}, {Yamada}, {Suzuki}, {Ishii},
  {Hayashiyama}, {Lee}, and {Nakamura}}]{Imamura2017}
{Imamura} T, {Ando} H, {Tellmann} S, et~al (2017) {Initial performance of the
  radio occultation experiment in the Venus orbiter mission Akatsuki}. Earth,
  Planets and Space 69(1):137. \doi{10.1186/s40623-017-0722-3}

\bibitem[{{Ingersoll}(1969)}]{Ingersoll1969}
{Ingersoll} AP (1969) {The Runaway Greenhouse: A History of Water on Venus.}
  Journal of the Atmospheric Sciences 26(6):1191--1198.
  \doi{10.1175/1520-0469(1969)026<1191:TRGAHO>2.0.CO;2}

\bibitem[{{James} et~al.(1997){James}, {Toon}, and {Schubert}}]{James1997}
{James} EP, {Toon} OB, {Schubert} G (1997) {A Numerical Microphysical Model of
  the Condensational Venus Cloud}. Icarus 129(1):147--171.
  \doi{10.1006/icar.1997.5763}

\bibitem[{{Jenkins} et~al.(1994){Jenkins}, {Steffes}, {Hinson}, {Twicken}, and
  {Tyler}}]{Jenkins1994}
{Jenkins} JM, {Steffes} PG, {Hinson} DP, et~al (1994) {Radio Occultation
  Studies of the Venus Atmosphere with the Magellan Spacecraft. 2. Results from
  the October 1991 Experiments}. Icarus 110(1):79--94.
  \doi{10.1006/icar.1994.1108}

\bibitem[{Jiang et~al.(2024)Jiang, Rimmer, Lozano, Tosca, Kufner, Sasselov, and
  Thompson}]{Jiang2024}
Jiang CZ, Rimmer PB, Lozano GG, et~al (2024) Iron-sulfur chemistry can explain
  the ultraviolet absorber in the clouds of venus. Science Advances
  10(1):eadg8826. \doi{10.1126/sciadv.adg8826},
  \urlprefix\url{https://www.science.org/doi/abs/10.1126/sciadv.adg8826},
  {\href{https://arxiv.org/abs/https://www.science.org/doi/pdf/10.1126/sciadv.adg8826}{{https://www.science.org/doi/pdf/10.1126/sciadv.adg8826}}}

\bibitem[{{Jordan} et~al.(2022){Jordan}, {Shorttle}, and {Rimmer}}]{Jordan2022}
{Jordan} S, {Shorttle} O, {Rimmer} PB (2022) {Proposed energy-metabolisms
  cannot explain the atmospheric chemistry of Venus}. Nature Communications
  13:3274. \doi{10.1038/s41467-022-30804-8},
  {\href{https://arxiv.org/abs/2206.06414}{{arXiv:2206.06414}}} {[astro-ph.EP]}

\bibitem[{{Karyu} et~al.(2023){Karyu}, {Kuroda}, {Itoh}, {Nitta}, {Ikeda},
  {Yamamoto}, {Sugimoto}, {Terada}, {Kasaba}, {Takahashi}, and
  {Hartogh}}]{Karyu2023}
{Karyu} H, {Kuroda} T, {Itoh} K, et~al (2023) {Vertical-Wind-Induced Cloud
  Opacity Variation in Low Latitudes Simulated by a Venus GCM}. Journal of
  Geophysical Research (Planets) 128(2):e2022JE007595.
  \doi{10.1029/2022JE007595}

\bibitem[{{Karyu} et~al.(2024){Karyu}, {Kuroda}, {Imamura}, {Terada},
  {Vandaele}, {Mahieux}, and {Viscardy}}]{Karyu2024}
{Karyu} H, {Kuroda} T, {Imamura} T, et~al (2024) {One-dimensional Microphysics
  Model of Venusian Clouds from 40 to 100 km: Impact of the Middle-atmosphere
  Eddy Transport and SOIR Temperature Profile on the Cloud Structure}. The
  Planetary Science Journal 5(3):57. \doi{10.3847/PSJ/ad25f3}

\bibitem[{{Kasting}(1988)}]{Kasting1988}
{Kasting} JF (1988) {Runaway and moist greenhouse atmospheres and the evolution
  of Earth and Venus}. Icarus 74(3):472--494.
  \doi{10.1016/0019-1035(88)90116-9}

\bibitem[{{Kawabata} et~al.(1980){Kawabata}, {Coffeen}, {Hansen}, {Lane},
  {Sato}, and {Travis}}]{Kawabata1980}
{Kawabata} K, {Coffeen} DL, {Hansen} JE, et~al (1980) {Cloud and haze
  properties from Pioneer Venus polarimetry}. Journal of Geophysical Research
  85:8129--8140. \doi{10.1029/JA085iA13p08129}

\bibitem[{{Khatuntsev} et~al.(2013){Khatuntsev}, {Patsaeva}, {Titov},
  {Ignatiev}, {Turin}, {Limaye}, {Markiewicz}, {Almeida}, {Roatsch}, and
  {Moissl}}]{Khatuntsev2013}
{Khatuntsev} IV, {Patsaeva} MV, {Titov} DV, et~al (2013) {Cloud level winds
  from the Venus Express Monitoring Camera imaging}. Icarus 226(1):140--158.
  \doi{10.1016/j.icarus.2013.05.018}

\bibitem[{{Knollenberg} and {Hunten}(1980)}]{Knollenberg+Hunten1980}
{Knollenberg} RG, {Hunten} DM (1980) {The microphysics of the clouds of Venus -
  Results of the Pioneer Venus particle size spectrometer experiment}. Journal
  of Geophysical Research 85:8039--8058. \doi{10.1029/JA085iA13p08039}

\bibitem[{{Kopparla} et~al.(2020){Kopparla}, {Seshadri}, {Imamura}, and
  {Lee}}]{Kopparla2020}
{Kopparla} P, {Seshadri} A, {Imamura} T, et~al (2020) {A Recharge Oscillator
  Model for Interannual Variability in Venus' Clouds}. Journal of Geophysical
  Research (Planets) 125(11):e06568. \doi{10.1029/2020JE006568},
  {\href{https://arxiv.org/abs/2010.16122}{{arXiv:2010.16122}}} {[astro-ph.EP]}

\bibitem[{{Krasnopolsky}(1985)}]{Krasnopolsky1985}
{Krasnopolsky} VA (1985) {Chemical composition of venus clouds}. Planetary and
  Space Science 33(1):109--117. \doi{10.1016/0032-0633(85)90147-3}

\bibitem[{{Krasnopolsky}(2006)}]{Krasnopolsky2006a}
{Krasnopolsky} VA (2006) {Chemical composition of Venus atmosphere and clouds:
  Some unsolved problems}. Planetary and Space Science 54(13-14):1352--1359.
  \doi{10.1016/j.pss.2006.04.019}

\bibitem[{{Krasnopolsky}(2007)}]{Krasnopolsky2007}
{Krasnopolsky} VA (2007) {Chemical kinetic model for the lower atmosphere of
  Venus}. Icarus 191(1):25--37. \doi{10.1016/j.icarus.2007.04.028}

\bibitem[{{Krasnopolsky}(2012)}]{Krasnopolsky2012}
{Krasnopolsky} VA (2012) {A photochemical model for the Venus atmosphere at
  47-112 km}. Icarus 218(1):230--246. \doi{10.1016/j.icarus.2011.11.012}

\bibitem[{{Krasnopolsky}(2013{\natexlab{a}})}]{Krasnopolsky2013a}
{Krasnopolsky} VA (2013{\natexlab{a}}) {Nighttime photochemical model and night
  airglow on Venus}. Planetary and Space Science 85:78--88.
  \doi{10.1016/j.pss.2013.05.022}

\bibitem[{{Krasnopolsky}(2013{\natexlab{b}})}]{Krasnopolsky2013b}
{Krasnopolsky} VA (2013{\natexlab{b}}) {S$_{3}$ and S$_{4}$ abundances and
  improved chemical kinetic model for the lower atmosphere of Venus}. Icarus
  225(1):570--580. \doi{10.1016/j.icarus.2013.04.026}

\bibitem[{{Krasnopolsky}(2015)}]{Krasnopolsky2015}
{Krasnopolsky} VA (2015) {Vertical profiles of H$_{2}$O, H$_{2}$SO$_{4}$, and
  sulfuric acid concentration at 45-75 km on Venus}. Icarus 252:327--333.
  \doi{10.1016/j.icarus.2015.01.024}

\bibitem[{{Krasnopolsky}(2016)}]{Krasnopolsky2016}
{Krasnopolsky} VA (2016) {Sulfur aerosol in the clouds of Venus}. Icarus
  274:33--36. \doi{10.1016/j.icarus.2016.03.010}

\bibitem[{{Krasnopolsky}(2017)}]{Krasnopolsky2017}
{Krasnopolsky} VA (2017) {On the iron chloride aerosol in the clouds of Venus}.
  Icarus 286:134--137. \doi{10.1016/j.icarus.2016.10.003}

\bibitem[{{Krasnopolsky} and {Pollack}(1994)}]{Krasnopolsky+Pollack1994}
{Krasnopolsky} VA, {Pollack} JB (1994) {H$_{2}$O-H$_{2}$SO$_{4}$ System in
  Venus' Clouds and OCS, CO, and H$_{2}$SO$_{4}$ Profiles in Venus'
  Troposphere}. In: Lunar and Planetary Science Conference, Lunar and Planetary
  Science Conference, p 747

\bibitem[{{Laken} et~al.(2012){Laken}, {Pall{\'e}}, {{\v{C}}alogovi{\'c}}, and
  {Dunne}}]{Laken2012}
{Laken} BA, {Pall{\'e}} E, {{\v{C}}alogovi{\'c}} J, et~al (2012) {A cosmic
  ray-climate link and cloud observations}. Journal of Space Weather and Space
  Climate 2:A18. \doi{10.1051/swsc/2012018},
  {\href{https://arxiv.org/abs/1211.5245}{{arXiv:1211.5245}}} {[physics.ao-ph]}

\bibitem[{{Lebonnois} et~al.(2016){Lebonnois}, {Sugimoto}, and
  {Gilli}}]{Lebonnois2016}
{Lebonnois} S, {Sugimoto} N, {Gilli} G (2016) {Wave analysis in the atmosphere
  of Venus below 100-km altitude, simulated by the LMD Venus GCM}. Icarus
  278:38--51. \doi{10.1016/j.icarus.2016.06.004}

\bibitem[{{Lee}(2017)}]{Lee2017}
{Lee} KW (2017) {Daylight Observations of Venus with Naked Eye in the
  Goryeosa}. Journal of Astronomy and Space Sciences 34(1):67--73.
  \doi{10.5140/JASS.2017.34.1.67}

\bibitem[{{Lee} et~al.(2012){Lee}, {Titov}, {Tellmann}, {Piccialli},
  {Ignatiev}, {P{\"a}tzold}, {H{\"a}usler}, {Piccioni}, and
  {Drossart}}]{Lee2012}
{Lee} YJ, {Titov} DV, {Tellmann} S, et~al (2012) {Vertical structure of the
  Venus cloud top from the VeRa and VIRTIS observations onboard Venus Express}.
  Icarus 217(2):599--609. \doi{10.1016/j.icarus.2011.07.001}

\bibitem[{{Lee} et~al.(2015{\natexlab{a}}){Lee}, {Imamura}, {Schr{\"o}der}, and
  {Marcq}}]{Lee2015b}
{Lee} YJ, {Imamura} T, {Schr{\"o}der} SE, et~al (2015{\natexlab{a}}) {Long-term
  variations of the UV contrast on Venus observed by the Venus Monitoring
  Camera on board Venus Express}. Icarus 253:1--15.
  \doi{10.1016/j.icarus.2015.02.015}

\bibitem[{{Lee} et~al.(2015{\natexlab{b}}){Lee}, {Titov}, {Ignatiev},
  {Tellmann}, {P{\"a}tzold}, and {Piccioni}}]{Lee2015a}
{Lee} YJ, {Titov} DV, {Ignatiev} NI, et~al (2015{\natexlab{b}}) {The radiative
  forcing variability caused by the changes of the upper cloud vertical
  structure in the Venus mesosphere}. Planetary and Space Science 113:298--308.
  \doi{10.1016/j.pss.2014.12.006}

\bibitem[{{Lee} et~al.(2016){Lee}, {Sagawa}, {Haus}, {Stefani}, {Imamura},
  {Titov}, and {Piccioni}}]{Lee2016}
{Lee} YJ, {Sagawa} H, {Haus} R, et~al (2016) {Sensitivity of net thermal flux
  to the abundance of trace gases in the lower atmosphere of Venus}. Journal of
  Geophysical Research (Planets) 121(9):1737--1752. \doi{10.1002/2016JE005087}

\bibitem[{{Lee} et~al.(2019){Lee}, {Jessup}, {Perez-Hoyos}, {Titov},
  {Lebonnois}, {Peralta}, {Horinouchi}, {Imamura}, {Limaye}, {Marcq}, {Takagi},
  {Yamazaki}, {Yamada}, {Watanabe}, {Murakami}, {Ogohara}, {McClintock},
  {Holsclaw}, and {Roman}}]{Lee2019}
{Lee} YJ, {Jessup} KL, {Perez-Hoyos} S, et~al (2019) {Long-term Variations of
  Venus{\textquoteright}s 365 nm Albedo Observed by Venus Express, Akatsuki,
  MESSENGER, and the Hubble Space Telescope}. The Astronomical Journal
  158(3):126. \doi{10.3847/1538-3881/ab3120},
  {\href{https://arxiv.org/abs/1907.09683}{{arXiv:1907.09683}}} {[astro-ph.EP]}

\bibitem[{{Lee} et~al.(2021){Lee}, {Garc{\'\i}a Mu{\~n}oz}, {Yamazaki},
  {Yamada}, {Watanabe}, and {Encrenaz}}]{Lee2021}
{Lee} YJ, {Garc{\'\i}a Mu{\~n}oz} A, {Yamazaki} A, et~al (2021) {Investigation
  of UV Absorbers on Venus Using the 283 and 365 nm Phase Curves Obtained From
  Akatsuki}. Geophysical Research Letters 48(7):e90577.
  \doi{10.1029/2020GL090577},
  {\href{https://arxiv.org/abs/2103.09021}{{arXiv:2103.09021}}} {[astro-ph.EP]}

\bibitem[{{Lef{\`e}vre} et~al.(2008){Lef{\`e}vre}, {Bertaux}, {Clancy},
  {Encrenaz}, {Fast}, {Forget}, {Lebonnois}, {Montmessin}, and
  {Perrier}}]{Lefevre2008}
{Lef{\`e}vre} F, {Bertaux} JL, {Clancy} RT, et~al (2008) {Heterogeneous
  chemistry in the atmosphere of Mars}. Nature 454(7207):971--975.
  \doi{10.1038/nature07116}

\bibitem[{{Lef{\`e}vre} et~al.(2022){Lef{\`e}vre}, {Marcq}, and
  {Lef{\`e}vre}}]{Lefevre2022}
{Lef{\`e}vre} M, {Marcq} E, {Lef{\`e}vre} F (2022) {The impact of turbulent
  vertical mixing in the Venus clouds on chemical tracers}. Icarus 386:115148.
  \doi{10.1016/j.icarus.2022.115148},
  {\href{https://arxiv.org/abs/2210.09240}{{arXiv:2210.09240}}} {[astro-ph.EP]}

\bibitem[{{Lef{\`e}vre} et~al.(2024){Lef{\`e}vre}, {Lef{\`e}vre}, {Marcq},
  {M{\"a}{\"a}tt{\"a}nen}, {Stolzenbach}, and {Streel}}]{Lefevre2024}
{Lef{\`e}vre} M, {Lef{\`e}vre} F, {Marcq} E, et~al (2024) {Impact of the
  Turbulent Vertical Mixing on Chemical and Cloud Species in the Venus Cloud
  Layer}. Geophysical Research Letters 51(12):e2024GL108771.
  \doi{10.1029/2024GL108771}

\bibitem[{{Limaye} et~al.(2018){Limaye}, {Grassi}, {Mahieux}, {Migliorini},
  {Tellmann}, and {Titov}}]{Limaye2018}
{Limaye} SS, {Grassi} D, {Mahieux} A, et~al (2018) {Venus Atmospheric Thermal
  Structure and Radiative Balance}. Space Science Reviews 214(5):102.
  \doi{10.1007/s11214-018-0525-2}

\bibitem[{{Limaye} et~al.(2021){Limaye}, {Mogul}, {Baines}, {Bullock},
  {Cockell}, {Cutts}, {Gentry}, {Grinspoon}, {Head}, {Jessup}, {Kompanichenko},
  {Lee}, {Mathies}, {Milojevic}, {Pertzborn}, {Rothschild}, {Sasaki},
  {Schulze-Makuch}, {Smith}, and {Way}}]{Limaye2021}
{Limaye} SS, {Mogul} R, {Baines} KH, et~al (2021) {Venus, an Astrobiology
  Target}. Astrobiology 21(10):1163--1185. \doi{10.1089/ast.2020.2268}

\bibitem[{{Lincowski} et~al.(2021){Lincowski}, {Meadows}, {Crisp}, {Akins},
  {Schwieterman}, {Arney}, {Wong}, {Steffes}, {Parenteau}, and
  {Domagal-Goldman}}]{Lincowski2021}
{Lincowski} AP, {Meadows} VS, {Crisp} D, et~al (2021) {Claimed Detection of
  PH$_{3}$ in the Clouds of Venus Is Consistent with Mesospheric SO$_{2}$}. The
  Astrophysical Journal Letters 908(2):L44. \doi{10.3847/2041-8213/abde47},
  {\href{https://arxiv.org/abs/2101.09837}{{arXiv:2101.09837}}} {[astro-ph.EP]}

\bibitem[{Lo et~al.(2003)Lo, Lee, Wu, Pan, Su, Cheng, Roffler, Chiang, Lee, Wu,
  and Tao}]{Lo2003}
Lo CH, Lee SC, Wu PY, et~al (2003) Antitumor and antimetastatic activity of
  il-231. The Journal of Immunology 171(2):600--607.
  \doi{10.4049/jimmunol.171.2.600},
  \urlprefix\url{https://doi.org/10.4049/jimmunol.171.2.600},
  {\href{https://arxiv.org/abs/https://journals.aai.org/jimmunol/article-pdf/171/2/600/1171201/600.pdf}{{https://journals.aai.org/jimmunol/article-pdf/171/2/600/1171201/600.pdf}}}

\bibitem[{{Luginin} et~al.(2016){Luginin}, {Fedorova}, {Belyaev}, {Montmessin},
  {Wilquet}, {Korablev}, {Bertaux}, and {Vandaele}}]{Luginin2016}
{Luginin} M, {Fedorova} A, {Belyaev} D, et~al (2016) {Aerosol properties in the
  upper haze of Venus from SPICAV IR data}. Icarus 277:154--170.
  \doi{10.1016/j.icarus.2016.05.008}

\bibitem[{{Luginin} et~al.(2024){Luginin}, {Fedorova}, {Belyaev}, {Montmessin},
  {Korablev}, and {Bertaux}}]{Luginin2024}
{Luginin} M, {Fedorova} A, {Belyaev} D, et~al (2024) {Bimodal aerosol
  distribution in Venus' upper haze from joint SPICAV-UV and -IR observations
  on Venus Express}. Icarus 409:115866. \doi{10.1016/j.icarus.2023.115866}

\bibitem[{{M{\"a}{\"a}tt{\"a}nen} et~al.(2023){M{\"a}{\"a}tt{\"a}nen},
  {Guilbon}, {Burgalat}, and {Montmessin}}]{Maattanen2023}
{M{\"a}{\"a}tt{\"a}nen} A, {Guilbon} S, {Burgalat} J, et~al (2023) {Development
  of a new cloud model for Venus (MAD-VenLA) using the Modal Aerosol Dynamics
  approach}. Advances in Space Research 71(1):1116--1136.
  \doi{10.1016/j.asr.2022.09.063}

\bibitem[{{Mahieux} et~al.(2024){Mahieux}, {Viscardy}, {Yelle}, {Karyu},
  {Chamberlain}, {Robert}, {Piccialli}, {Trompet}, {Erwin}, {Ubukata},
  {Nakagawa}, {Koyama}, {Maggiolo}, {Pereira}, {Cessateur}, {Willame}, and
  {Vandaele}}]{Mahieux2024}
{Mahieux} A, {Viscardy} S, {Yelle} RV, et~al (2024) {Unexpected increase of the
  deuterium to hydrogen ratio in the Venus mesosphere}. Proceedings of the
  National Academy of Science 121(34):e2401638121.
  \doi{10.1073/pnas.2401638121}

\bibitem[{{Marcq} et~al.(2013){Marcq}, {Bertaux}, {Montmessin}, and
  {Belyaev}}]{Marcq2013}
{Marcq} E, {Bertaux} JL, {Montmessin} F, et~al (2013) {Variations of sulphur
  dioxide at the cloud top of Venus's dynamic atmosphere}. Nature Geoscience
  6(1):25--28. \doi{10.1038/ngeo1650}

\bibitem[{{Marcq} et~al.(2018){Marcq}, {Mills}, {Parkinson}, and
  {Vandaele}}]{Marcq2018}
{Marcq} E, {Mills} FP, {Parkinson} CD, et~al (2018) {Composition and Chemistry
  of the Neutral Atmosphere of Venus}. Space Science Reviews 214(1):10.
  \doi{10.1007/s11214-017-0438-5}

\bibitem[{{Marcq} et~al.(2019){Marcq}, {Baggio}, {Lef{\`e}vre}, {Stolzenbach},
  {Montmessin}, {Belyaev}, {Korablev}, and {Bertaux}}]{Marcq2019}
{Marcq} E, {Baggio} L, {Lef{\`e}vre} F, et~al (2019) {Discovery of cloud top
  ozone on Venus}. Icarus 319:491--498. \doi{10.1016/j.icarus.2018.10.006}

\bibitem[{{Marcq} et~al.(2020){Marcq}, {Lea Jessup}, {Baggio}, {Encrenaz},
  {Lee}, {Montmessin}, {Belyaev}, {Korablev}, and {Bertaux}}]{Marcq2020}
{Marcq} E, {Lea Jessup} K, {Baggio} L, et~al (2020) {Climatology of SO$_{2}$
  and UV absorber at Venus' cloud top from SPICAV-UV nadir dataset}. Icarus
  335:113368. \doi{10.1016/j.icarus.2019.07.002}

\bibitem[{{Marcq} et~al.(2023){Marcq}, {B{\'e}zard}, {Reess}, {Henry},
  {{\'E}rard}, {Robert}, {Montmessin}, {Lef{\`e}vre}, {Lef{\`e}vre},
  {Stolzenbach}, {Bertaux}, {Piccioni}, and {Drossart}}]{Marcq2023}
{Marcq} E, {B{\'e}zard} B, {Reess} JM, et~al (2023) {Minor species in Venus'
  night side troposphere as observed by VIRTIS-H/Venus Express}. Icarus
  405:115714. \doi{10.1016/j.icarus.2023.115714}

\bibitem[{{Markiewicz} et~al.(2018){Markiewicz}, {Petrova}, and
  {Shalygina}}]{Markiewicz2018}
{Markiewicz} WJ, {Petrova} EV, {Shalygina} OS (2018) {Aerosol properties in the
  upper clouds of Venus from glory observations by the Venus Monitoring Camera
  (Venus Express mission)}. Icarus 299:272--293.
  \doi{10.1016/j.icarus.2017.08.011}

\bibitem[{{Marsh} and {Svensmark}(2000)}]{Marsh+Svensmark2000}
{Marsh} N, {Svensmark} H (2000) {Cosmic Rays, Clouds, and Climate}. Space
  Science Reviews 94:215--230. \doi{10.1023/A:1026723423896}

\bibitem[{{McElroy} et~al.(1982){McElroy}, {Prather}, and
  {Rodriguez}}]{McElroy1982}
{McElroy} MB, {Prather} MJ, {Rodriguez} JM (1982) {Escape of Hydrogen from
  Venus}. Science 215(4540):1614--1615. \doi{10.1126/science.215.4540.1614}

\bibitem[{{McGouldrick}(2017)}]{McGouldrick2017}
{McGouldrick} K (2017) {Effects of variation in coagulation and photochemistry
  parameters on the particle size distributions in the Venus clouds}. Earth,
  Planets and Space 69(1):161. \doi{10.1186/s40623-017-0744-x}

\bibitem[{{McGouldrick} and {Barth}(2023)}]{McGouldrick+Barth2023}
{McGouldrick} K, {Barth} EL (2023) {The Influence of Cloud Condensation Nucleus
  Coagulation on the Venus Cloud Structure}. The Planetary Science Journal
  4(3):50. \doi{10.3847/PSJ/acbdf8}

\bibitem[{{McGouldrick} and {Toon}(2007)}]{McGouldrick+Toon2007}
{McGouldrick} K, {Toon} OB (2007) {An investigation of possible causes of the
  holes in the condensational Venus cloud using a microphysical cloud model
  with a radiative-dynamical feedback}. Icarus 191(1):1--24.
  \doi{10.1016/j.icarus.2007.04.007}

\bibitem[{{McGouldrick} et~al.(2021){McGouldrick}, {Peralta}, {Barstow}, and
  {Tsang}}]{McGouldrick2021}
{McGouldrick} K, {Peralta} J, {Barstow} JK, et~al (2021) {Using VIRTIS on Venus
  Express to Constrain the Properties of the Giant Dark Cloud Observed in
  Images of Venus by IR2 on Akatsuki}. The Planetary Science Journal 2(4):153.
  \doi{10.3847/PSJ/ac0e39}

\bibitem[{{Molaverdikhani} et~al.(2012){Molaverdikhani}, {McGouldrick}, and
  {Esposito}}]{Molaverdikhani2012}
{Molaverdikhani} K, {McGouldrick} K, {Esposito} LW (2012) {The abundance and
  vertical distribution of the unknown ultraviolet absorber in the venusian
  atmosphere from analysis of Venus Monitoring Camera images}. Icarus
  217(2):648--660. \doi{10.1016/j.icarus.2011.08.008}

\bibitem[{{Moroz}(1981)}]{Moroz1981}
{Moroz} VI (1981) {The Atmosphere of Venus}. Space Science Reviews
  29(1):3--127. \doi{10.1007/BF00177144}

\bibitem[{{Mukhin} et~al.(1982){Mukhin}, {Gelman}, {Lamonov}, {Melnikov},
  {Nenarokov}, {Okhotnikov}, {Rotin}, and {Khokhlov}}]{Mukhin1982}
{Mukhin} LM, {Gelman} BG, {Lamonov} NI, et~al (1982) {VENERA-13 and VENERA-14
  Gas Chromatography Analysis of the Venus Atmosphere Composition}. Soviet
  Astronomy Letters 8:216--218

\bibitem[{{Nakamura} et~al.(2016){Nakamura}, {Imamura}, {Ishii}, {Abe},
  {Kawakatsu}, {Hirose}, {Satoh}, {Suzuki}, {Ueno}, {Yamazaki}, {Iwagami},
  {Watanabe}, {Taguchi}, {Fukuhara}, {Takahashi}, {Yamada}, {Imai}, {Ohtsuki},
  {Uemizu}, {Hashimoto}, {Takagi}, {Matsuda}, {Ogohara}, {Sato}, {Kasaba},
  {Kouyama}, {Hirata}, {Nakamura}, {Yamamoto}, {Horinouchi}, {Yamamoto},
  {Hayashi}, {Kashimura}, {Sugiyama}, {Sakanoi}, {Ando}, {Murakami}, {Sato},
  {Takagi}, {Nakajima}, {Peralta}, {Lee}, {Nakatsuka}, {Ichikawa}, {Inoue},
  {Toda}, {Toyota}, {Tachikawa}, {Narita}, {Hayashiyama}, {Hasegawa}, and
  {Kamata}}]{Nakamura2016}
{Nakamura} M, {Imamura} T, {Ishii} N, et~al (2016) {AKATSUKI returns to Venus}.
  Earth, Planets and Space 68(1):75. \doi{10.1186/s40623-016-0457-6}

\bibitem[{{Nordheim} et~al.(2015){Nordheim}, {Dartnell}, {Desorgher}, {Coates},
  and {Jones}}]{Nordheim2015}
{Nordheim} TA, {Dartnell} LR, {Desorgher} L, et~al (2015) {Ionization of the
  venusian atmosphere from solar and galactic cosmic rays}. Icarus 245:80--86.
  \doi{10.1016/j.icarus.2014.09.032}

\bibitem[{{Oschlisniok} et~al.(2012){Oschlisniok}, {H{\"a}usler},
  {P{\"a}tzold}, {Tyler}, {Bird}, {Tellmann}, {Remus}, and
  {Andert}}]{Oschlisniok2012}
{Oschlisniok} J, {H{\"a}usler} B, {P{\"a}tzold} M, et~al (2012) {Microwave
  absorptivity by sulfuric acid in the Venus atmosphere: First results from the
  Venus Express Radio Science experiment VeRa}. Icarus 221(2):940--948.
  \doi{10.1016/j.icarus.2012.09.029}

\bibitem[{{Oschlisniok} et~al.(2021){Oschlisniok}, {H{\"a}usler},
  {P{\"a}tzold}, {Tellmann}, {Bird}, {Peter}, and {Andert}}]{Oschlisniok2021}
{Oschlisniok} J, {H{\"a}usler} B, {P{\"a}tzold} M, et~al (2021) {Sulfuric acid
  vapor and sulfur dioxide in the atmosphere of Venus as observed by the Venus
  Express radio science experiment VeRa}. Icarus 362:114405.
  \doi{10.1016/j.icarus.2021.114405}

\bibitem[{{Oyama} et~al.(1980){Oyama}, {Carle}, {Woeller}, {Pollack},
  {Reynolds}, and {Craig}}]{Oyama1980}
{Oyama} VI, {Carle} GC, {Woeller} F, et~al (1980) {Pioneer Venus gas
  chromatography of the lower atmosphere of Venus}. Journal of Geophysical
  Research 85:7891--7902. \doi{10.1029/JA085iA13p07891}

\bibitem[{{Palmer} and {Williams}(1975)}]{Palmer+Williams1975}
{Palmer} KF, {Williams} D (1975) {Optical constants of sulfuric acid;
  application to the clouds of Venus?} Applied Optics 14(1):208--219.
  \doi{10.1364/AO.14.000208}

\bibitem[{{Parkinson} et~al.(2015){Parkinson}, {Gao}, {Schulte}, {Bougher},
  {Yung}, {Bardeen}, {Wilquet}, {Vandaele}, {Mahieux}, {Tellmann}, and
  {P{\"a}tzold}}]{Parkinson2015a}
{Parkinson} CD, {Gao} P, {Schulte} R, et~al (2015) {Distribution of sulphuric
  acid aerosols in the clouds and upper haze of Venus using Venus Express VAST
  and VeRa temperature profiles}. Planetary and Space Science 113:205--218.
  \doi{10.1016/j.pss.2015.01.023}

\bibitem[{{Patsaeva} et~al.(2019){Patsaeva}, {Khatuntsev}, {Zasova},
  {Hauchecorne}, {Titov}, and {Bertaux}}]{Patsaeva2019}
{Patsaeva} MV, {Khatuntsev} IV, {Zasova} LV, et~al (2019) {Solar-Related
  Variations of the Cloud Top Circulation Above Aphrodite Terra From VMC/Venus
  Express Wind Fields}. Journal of Geophysical Research (Planets)
  124(7):1864--1879. \doi{10.1029/2018JE005620}

\bibitem[{{Patsaeva} et~al.(2024){Patsaeva}, {Khatuntsev}, {Titov}, {Ignatiev},
  {Zasova}, {Gorinov}, and {Turin}}]{Patsaeva2024}
{Patsaeva} MV, {Khatuntsev} IV, {Titov} DV, et~al (2024) {Wind Speed Variations
  at the Venus Cloud Top above Aphrodite Terra According to Long-term UV
  Observations by VMC/VENUS Express and UVI/AKATSUKI}. Solar System Research
  58(2):148--162. \doi{10.1134/S0038094623700053}

\bibitem[{{P{\'e}rez-Hoyos} et~al.(2018){P{\'e}rez-Hoyos},
  {S{\'a}nchez-Lavega}, {Garc{\'\i}a-Mu{\~n}oz}, {Irwin}, {Peralta},
  {Holsclaw}, {McClintock}, and {Sanz-Requena}}]{Perez-Hoyos2018}
{P{\'e}rez-Hoyos} S, {S{\'a}nchez-Lavega} A, {Garc{\'\i}a-Mu{\~n}oz} A, et~al
  (2018) {Venus Upper Clouds and the UV Absorber From MESSENGER/MASCS
  Observations}. Journal of Geophysical Research (Planets) 123(1):145--162.
  \doi{10.1002/2017JE005406},
  {\href{https://arxiv.org/abs/1801.03820}{{arXiv:1801.03820}}} {[astro-ph.EP]}

\bibitem[{{Petkowski} et~al.(2024){Petkowski}, {Seager}, {Grinspoon}, {Bains},
  {Ranjan}, {Rimmer}, {Buchanan}, {Agrawal}, {Mogul}, and
  {Carr}}]{Petkowski2024}
{Petkowski} JJ, {Seager} S, {Grinspoon} DH, et~al (2024) {Astrobiological
  Potential of Venus Atmosphere Chemical Anomalies and Other Unexplained Cloud
  Properties}. Astrobiology 24(4):343--370. \doi{10.1089/ast.2022.0060},
  {\href{https://arxiv.org/abs/2401.04708}{{arXiv:2401.04708}}} {[astro-ph.EP]}

\bibitem[{{Petrova}(2018)}]{Petrova2018}
{Petrova} EV (2018) {Glory on Venus and selection among the unknown UV
  absorbers}. Icarus 306:163--170. \doi{10.1016/j.icarus.2018.02.016}

\bibitem[{{Petrova} et~al.(2015){Petrova}, {Shalygina}, and
  {Markiewicz}}]{Petrova2015}
{Petrova} EV, {Shalygina} OS, {Markiewicz} WJ (2015) {UV contrasts and
  microphysical properties of the upper clouds of Venus from the UV and NIR
  VMC/VEx images}. Icarus 260:190--204. \doi{10.1016/j.icarus.2015.07.015}

\bibitem[{{Pierrehumbert}(2010)}]{Pierrehumbert2010}
{Pierrehumbert} RT (2010) {Principles of Planetary Climate}

\bibitem[{{Pollack} et~al.(1980){Pollack}, {Toon}, {Whitten}, {Boese},
  {Ragent}, {Tomasko}, {Eposito}, {Travis}, and {Wiedman}}]{Pollack1980}
{Pollack} JB, {Toon} OB, {Whitten} RC, et~al (1980) {Distribution and source of
  the UV absorption in Venus' atmosphere}. Journal of Geophysical Research
  85:8141--8150. \doi{10.1029/JA085iA13p08141}

\bibitem[{{Rimbot} et~al.(2024){Rimbot}, {Witasse}, {Pinto}, {Knutsen},
  {S{\'a}nchez-Cano}, {Wood}, {Tremolizzo}, and {Exner}}]{Rimbot2024}
{Rimbot} T, {Witasse} O, {Pinto} M, et~al (2024) {Galactic cosmic rays at 0.7
  A.U. with Venus Express housekeeping data}. Planetary and Space Science
  242:105867. \doi{10.1016/j.pss.2024.105867}

\bibitem[{{Rimmer} et~al.(2021){Rimmer}, {Jordan}, {Constantinou}, {Woitke},
  {Shorttle}, {Hobbs}, and {Paschodimas}}]{Rimmer2021}
{Rimmer} PB, {Jordan} S, {Constantinou} T, et~al (2021) {Hydroxide Salts in the
  Clouds of Venus: Their Effect on the Sulfur Cycle and Cloud Droplet pH}. The
  Planetary Science Journal 2(4):133. \doi{10.3847/PSJ/ac0156},
  {\href{https://arxiv.org/abs/2101.08582}{{arXiv:2101.08582}}} {[astro-ph.EP]}

\bibitem[{{Rossi} et~al.(2015){Rossi}, {Marcq}, {Montmessin}, {Fedorova},
  {Stam}, {Bertaux}, and {Korablev}}]{Rossi2015}
{Rossi} L, {Marcq} E, {Montmessin} F, et~al (2015) {Preliminary study of Venus
  cloud layers with polarimetric data from SPICAV/VEx}. Planetary and Space
  Science 113:159--168. \doi{10.1016/j.pss.2014.11.011}

\bibitem[{{Rossow} and {Williams}(1979)}]{Rossow+Williams1979}
{Rossow} WB, {Williams} GP (1979) {Large-scale motion in the Venus
  stratosphere.} Journal of the Atmospheric Sciences 36:377--389.
  \doi{10.1175/1520-0469(1979)036<0377:LSMITV>2.0.CO;2}

\bibitem[{{Salvador} et~al.(2017){Salvador}, {Massol}, {Davaille}, {Marcq},
  {Sarda}, and {Chassefi{\`e}re}}]{Salvador2017}
{Salvador} A, {Massol} H, {Davaille} A, et~al (2017) {The relative influence of
  H$_{2}$O and CO$_{2}$ on the primitive surface conditions and evolution of
  rocky planets}. Journal of Geophysical Research (Planets) 122(7):1458--1486.
  \doi{10.1002/2017JE005286}

\bibitem[{{Salvador} et~al.(2023){Salvador}, {Avice}, {Breuer}, {Gillmann},
  {Lammer}, {Marcq}, {Raymond}, {Sakuraba}, {Scherf}, and {Way}}]{Salvador2023}
{Salvador} A, {Avice} G, {Breuer} D, et~al (2023) {Magma Ocean, Water, and the
  Early Atmosphere of Venus}. Space Science Reviews 219(7):51.
  \doi{10.1007/s11214-023-00995-7}

\bibitem[{{S{\'a}nchez-Lavega} et~al.(2017){S{\'a}nchez-Lavega}, {Lebonnois},
  {Imamura}, {Read}, and {Luz}}]{Sanchez-Lavega2017}
{S{\'a}nchez-Lavega} A, {Lebonnois} S, {Imamura} T, et~al (2017) {The
  Atmospheric Dynamics of Venus}. Space Science Reviews 212(3-4):1541--1616.
  \doi{10.1007/s11214-017-0389-x}

\bibitem[{{Seager} et~al.(2021){Seager}, {Petkowski}, {Gao}, {Bains}, {Bryan},
  {Ranjan}, and {Greaves}}]{Seager2021}
{Seager} S, {Petkowski} JJ, {Gao} P, et~al (2021) {The Venusian Lower
  Atmosphere Haze as a Depot for Desiccated Microbial Life: A Proposed Life
  Cycle for Persistence of the Venusian Aerial Biosphere}. Astrobiology
  21(10):1206--1223. \doi{10.1089/ast.2020.2244},
  {\href{https://arxiv.org/abs/2009.06474}{{arXiv:2009.06474}}} {[astro-ph.EP]}

\bibitem[{{Seager} et~al.(2023){Seager}, {Petkowski}, {Seager}, {Grimes},
  {Zinsli}, {Vollmer-Snarr}, {Abd El-Rahman}, {Wishart}, {Lee}, {Gautam},
  {Herrington}, {Bains}, and {Darrow}}]{Seager2023}
{Seager} S, {Petkowski} JJ, {Seager} MD, et~al (2023) {Stability of nucleic
  acid bases in concentrated sulfuric acid: Implications for the habitability
  of Venus' clouds}. Proceedings of the National Academy of Science
  120(25):e2220007120. \doi{10.1073/pnas.2220007120},
  {\href{https://arxiv.org/abs/2306.17182}{{arXiv:2306.17182}}}
  {[physics.chem-ph]}

\bibitem[{{Seiff} et~al.(1985){Seiff}, {Schofield}, {Kliore}, {Taylor},
  {Limaye}, {Revercomb}, {Sromovsky}, {Kerzhanovich}, {Moroz}, and
  {Marov}}]{Seiff1985}
{Seiff} A, {Schofield} JT, {Kliore} AJ, et~al (1985) {Models of the structure
  of the atmosphere of Venus from the surface to 100 kilometers altitude}.
  Advances in Space Research 5(11):3--58. \doi{10.1016/0273-1177(85)90197-8}

\bibitem[{{Seinfeld} and {Pandis}(2016)}]{Seinfeld+Pandis2016}
{Seinfeld} JH, {Pandis} SN (2016) Atmospheric chemistry and physics: from air
  pollution to climate change. John Wiley \& Sons, Hoboken, NJ, US

\bibitem[{{Shalygina} et~al.(2015){Shalygina}, {Petrova}, {Markiewicz},
  {Ignatiev}, and {Shalygin}}]{Shalygina2015}
{Shalygina} OS, {Petrova} EV, {Markiewicz} WJ, et~al (2015) {Optical properties
  of the Venus upper clouds from the data obtained by Venus Monitoring Camera
  on-board the Venus Express}. Planetary and Space Science 113:135--158.
  \doi{10.1016/j.pss.2014.11.012}

\bibitem[{{Shao} et~al.(2020){Shao}, {Zhang}, {Bierson}, and
  {Encrenaz}}]{Shao2020}
{Shao} WD, {Zhang} X, {Bierson} CJ, et~al (2020) {Revisiting the Sulfur-Water
  Chemical System in the Middle Atmosphere of Venus}. Journal of Geophysical
  Research (Planets) 125(8):e06195. \doi{10.1029/2019JE006195},
  {\href{https://arxiv.org/abs/2006.09522}{{arXiv:2006.09522}}} {[astro-ph.EP]}

\bibitem[{{Shao} et~al.(2024){Shao}, {Mendon{\c{c}}a}, and {Dai}}]{Shao2024}
{Shao} WD, {Mendon{\c{c}}a} JM, {Dai} L (2024) {Three-Dimensional Venus Cloud
  Structure Simulated by a General Circulation Model}. Journal of Geophysical
  Research (Planets) 129(7):e2023JE008088. \doi{10.1029/2023JE008088},
  {\href{https://arxiv.org/abs/2407.15966}{{arXiv:2407.15966}}} {[astro-ph.EP]}

\bibitem[{{Spacek}(2021)}]{Spacek2021}
{Spacek} J (2021) {Organic Carbon Cycle in the Atmosphere of Venus}. arXiv
  e-prints arXiv:2108.02286. \doi{10.48550/arXiv.2108.02286},
  {\href{https://arxiv.org/abs/2108.02286}{{arXiv:2108.02286}}} {[astro-ph.EP]}

\bibitem[{{Spacek} and {Benner}(2021)}]{Spacek+Benner2021}
{Spacek} J, {Benner} SA (2021) {The Organic Carbon Cycle in the Atmosphere of
  Venus and Evolving Red Oil}. In: Venera-D: Venus Cloud Habitability System
  Workshop, p 4052

\bibitem[{{Spacek} et~al.(2024){Spacek}, {Rimmer}, {Owens}, {Cady}, {Sharma},
  and {Benner}}]{Spacek2024}
{Spacek} J, {Rimmer} P, {Owens} GE, et~al (2024) {Production and Reactions of
  Organic Molecules in Clouds of Venus}. ACS Earth and Space Chemistry
  8(1):89--98. \doi{10.1021/acsearthspacechem.3c00261}

\bibitem[{{Stephens} et~al.(2015){Stephens}, {O'Brien}, {Webster}, {Pilewski},
  {Kato}, and {Li}}]{Stephens2015}
{Stephens} GL, {O'Brien} D, {Webster} PJ, et~al (2015) {The albedo of Earth}.
  Reviews of Geophysics 53(1):141--163. \doi{10.1002/2014RG000449}

\bibitem[{{Stolzenbach} et~al.(2023){Stolzenbach}, {Lef{\`e}vre}, {Lebonnois},
  and {M{\"a}{\"a}tt{\"a}nen}}]{Stolzenbach2023}
{Stolzenbach} A, {Lef{\`e}vre} F, {Lebonnois} S, et~al (2023)
  {Three-dimensional modeling of Venus photochemistry and clouds}. Icarus
  395:115447. \doi{10.1016/j.icarus.2023.115447}

\bibitem[{{Svedhem} et~al.(2007){Svedhem}, {Titov}, {McCoy}, {Lebreton},
  {Barabash}, {Bertaux}, {Drossart}, {Formisano}, {H{\"a}usler}, {Korablev},
  {Markiewicz}, {Nevejans}, {P{\"a}tzold}, {Piccioni}, {Zhang}, {Taylor},
  {Lellouch}, {Koschny}, {Witasse}, {Eggel}, {Warhaut}, {Accomazzo},
  {Rodriguez-Canabal}, {Fabrega}, {Schirmann}, {Clochet}, and
  {Coradini}}]{Svedhem2007}
{Svedhem} H, {Titov} DV, {McCoy} D, et~al (2007) {Venus Express{\textemdash}The
  first European mission to Venus}. Planetary and Space Science
  55(12):1636--1652. \doi{10.1016/j.pss.2007.01.013}

\bibitem[{{Svensmark} et~al.(2009){Svensmark}, {Bondo}, and
  {Svensmark}}]{Svensmark2009}
{Svensmark} H, {Bondo} T, {Svensmark} J (2009) {Cosmic ray decreases affect
  atmospheric aerosols and clouds}. Geophysical Research Letters 36(15):L15101.
  \doi{10.1029/2009GL038429}

\bibitem[{{Thompson}(2021)}]{Thompson2021}
{Thompson} MA (2021) {The statistical reliability of 267-GHz JCMT observations
  of Venus: no significant evidence for phosphine absorption}. Monthly Notices
  of the Royal Astronomical Society: Letters 501(1):L18--L22.
  \doi{10.1093/mnrasl/slaa187},
  {\href{https://arxiv.org/abs/2010.15188}{{arXiv:2010.15188}}} {[astro-ph.EP]}

\bibitem[{{Titov} et~al.(2008){Titov}, {Taylor}, {Svedhem}, {Ignatiev},
  {Markiewicz}, {Piccioni}, and {Drossart}}]{Titov2008}
{Titov} DV, {Taylor} FW, {Svedhem} H, et~al (2008) {Atmospheric structure and
  dynamics as the cause of ultraviolet markings in the clouds of Venus}. Nature
  456(7222):620--623. \doi{10.1038/nature07466}

\bibitem[{{Titov} et~al.(2012){Titov}, {Markiewicz}, {Ignatiev}, {Song},
  {Limaye}, {Sanchez-Lavega}, {Hesemann}, {Almeida}, {Roatsch}, {Matz},
  {Scholten}, {Crisp}, {Esposito}, {Hviid}, {Jaumann}, {Keller}, and
  {Moissl}}]{Titov2012}
{Titov} DV, {Markiewicz} WJ, {Ignatiev} NI, et~al (2012) {Morphology of the
  cloud tops as observed by the Venus Express Monitoring Camera}. Icarus
  217(2):682--701. \doi{10.1016/j.icarus.2011.06.020}

\bibitem[{{Titov} et~al.(2018){Titov}, {Ignatiev}, {McGouldrick}, {Wilquet},
  and {Wilson}}]{Titov2018}
{Titov} DV, {Ignatiev} NI, {McGouldrick} K, et~al (2018) {Clouds and Hazes of
  Venus}. Space Science Reviews 214(8):126. \doi{10.1007/s11214-018-0552-z}

\bibitem[{{Tomasko} et~al.(1979){Tomasko}, {Doose}, and {Smith}}]{Tomasko1979}
{Tomasko} MG, {Doose} LR, {Smith} PH (1979) {Absorption of Sunlight in the
  Atmosphere of Venus}. Science 205(4401):80--82.
  \doi{10.1126/science.205.4401.80}

\bibitem[{{Tomasko} et~al.(1980){Tomasko}, {Smith}, {Suomi}, {Sromovsky},
  {Revercomb}, {Taylor}, {Martonchik}, {Seiff}, {Boese}, {Pollack},
  {Ingersoll}, {Schubert}, and {Covey}}]{Tomasko1980}
{Tomasko} MG, {Smith} PH, {Suomi} VE, et~al (1980) {The thermal balance of
  Venus in light of the Pioneer Venus mission}. Journal of Geophysical Research
  85:8187--8199. \doi{10.1029/JA085iA13p08187}

\bibitem[{{Toon} et~al.(1984){Toon}, {Ragent}, {Colburn}, {Blamont}, and
  {Cot}}]{Toon1984}
{Toon} OB, {Ragent} B, {Colburn} D, et~al (1984) {Large, solid particles in the
  clouds of Venus: Do they exist?} Icarus 57(2):143--160.
  \doi{10.1016/0019-1035(84)90063-0}

\bibitem[{{Toon} et~al.(1988){Toon}, {Turco}, {Westphal}, {Malone}, and
  {Liu}}]{Toon1988}
{Toon} OB, {Turco} RP, {Westphal} D, et~al (1988) {A multidimensional model for
  aerosols - Description of computational analogs}. Journal of the Atmospheric
  Sciences 45:2123--2143. \doi{10.1175/1520-0469(1988)045<2123:AMMFAD>2.0.CO;2}

\bibitem[{{Trauger} and {Lunine}(1983)}]{Trauger+Lunine1983}
{Trauger} JT, {Lunine} JI (1983) {Spectroscopy of molecular oxygen in the
  Atmospheres of Venus and Mars}. Icarus 55(2):272--281.
  \doi{10.1016/0019-1035(83)90082-9}

\bibitem[{{Trompet} et~al.(2021){Trompet}, {Robert}, {Mahieux}, {Schmidt},
  {Erwin}, and {Vandaele}}]{Trompet2021}
{Trompet} L, {Robert} S, {Mahieux} A, et~al (2021) {Phosphine in Venus'
  atmosphere: Detection attempts and upper limits above the cloud top assessed
  from the SOIR/VEx spectra}. Astronomy \& Astrophysics 645:L4.
  \doi{10.1051/0004-6361/202039932}

\bibitem[{{Turbet} et~al.(2021){Turbet}, {Bolmont}, {Chaverot}, {Ehrenreich},
  {Leconte}, and {Marcq}}]{Turbet2021}
{Turbet} M, {Bolmont} E, {Chaverot} G, et~al (2021) {Day-night cloud asymmetry
  prevents early oceans on Venus but not on Earth}. Nature 598(7880):276--280.
  \doi{10.1038/s41586-021-03873-w},
  {\href{https://arxiv.org/abs/2110.08801}{{arXiv:2110.08801}}} {[astro-ph.EP]}

\bibitem[{{Turbet} et~al.(2023){Turbet}, {Fauchez}, {Leconte}, {Bolmont},
  {Chaverot}, {Forget}, {Millour}, {Selsis}, {Charnay}, {Ducrot}, {Gillon},
  {Maurel}, and {Villanueva}}]{Turbet2023}
{Turbet} M, {Fauchez} TJ, {Leconte} J, et~al (2023) {Water condensation zones
  around main sequence stars}. Astronomy \& Astrophysics 679:A126.
  \doi{10.1051/0004-6361/202347539},
  {\href{https://arxiv.org/abs/2308.15110}{{arXiv:2308.15110}}} {[astro-ph.EP]}

\bibitem[{{Vandaele} et~al.(2017){Vandaele}, {Korablev}, {Belyaev},
  {Chamberlain}, {Evdokimova}, {Encrenaz}, {Esposito}, {Jessup}, {Lef{\`e}vre},
  {Limaye}, {Mahieux}, {Marcq}, {Mills}, {Montmessin}, {Parkinson}, {Robert},
  {Roman}, {Sandor}, {Stolzenbach}, {Wilson}, and {Wilquet}}]{Vandaele2017a}
{Vandaele} AC, {Korablev} O, {Belyaev} D, et~al (2017) {Sulfur dioxide in the
  Venus atmosphere: I. Vertical distribution and variability}. Icarus
  295:16--33. \doi{10.1016/j.icarus.2017.05.003}

\bibitem[{{Villanueva} et~al.(2021){Villanueva}, {Cordiner}, {Irwin}, {de
  Pater}, {Butler}, {Gurwell}, {Milam}, {Nixon}, {Luszcz-Cook}, {Wilson},
  {Kofman}, {Liuzzi}, {Faggi}, {Fauchez}, {Lippi}, {Cosentino}, {Thelen},
  {Moullet}, {Hartogh}, {Molter}, {Charnley}, {Arney}, {Mandell}, {Biver},
  {Vandaele}, {de Kleer}, and {Kopparapu}}]{Villanueva2021}
{Villanueva} GL, {Cordiner} M, {Irwin} PGJ, et~al (2021) {No evidence of
  phosphine in the atmosphere of Venus from independent analyses}. Nature
  Astronomy 5:631--635. \doi{10.1038/s41550-021-01422-z}

\bibitem[{{Way} and {Del Genio}(2020)}]{Way+DelGenio2020}
{Way} MJ, {Del Genio} AD (2020) {Venusian Habitable Climate Scenarios: Modeling
  Venus Through Time and Applications to Slowly Rotating Venus-Like
  Exoplanets}. Journal of Geophysical Research (Planets) 125(5):e06276.
  \doi{10.1029/2019JE00627610.1002/essoar.10501118.3}

\bibitem[{{Way} et~al.(2016){Way}, {Del Genio}, {Kiang}, {Sohl}, {Grinspoon},
  {Aleinov}, {Kelley}, and {Clune}}]{Way2016}
{Way} MJ, {Del Genio} AD, {Kiang} NY, et~al (2016) {Was Venus the first
  habitable world of our solar system?} Geophysical Research Letters
  43(16):8376--8383. \doi{10.1002/2016GL069790},
  {\href{https://arxiv.org/abs/1608.00706}{{arXiv:1608.00706}}} {[astro-ph.EP]}

\bibitem[{{Wilquet} et~al.(2009){Wilquet}, {Fedorova}, {Montmessin},
  {Drummond}, {Mahieux}, {Vandaele}, {Villard}, {Korablev}, and
  {Bertaux}}]{Wilquet2009}
{Wilquet} V, {Fedorova} A, {Montmessin} F, et~al (2009) {Preliminary
  characterization of the upper haze by SPICAV/SOIR solar occultation in UV to
  mid-IR onboard Venus Express}. Journal of Geophysical Research (Planets)
  114(12):E00B42. \doi{10.1029/2008JE003186}

\bibitem[{{Yang} et~al.(2013){Yang}, {Cowan}, and {Abbot}}]{Yang2013}
{Yang} J, {Cowan} NB, {Abbot} DS (2013) {Stabilizing Cloud Feedback
  Dramatically Expands the Habitable Zone of Tidally Locked Planets}. The
  Astrophysical Journal Letters 771(2):L45. \doi{10.1088/2041-8205/771/2/L45},
  {\href{https://arxiv.org/abs/1307.0515}{{arXiv:1307.0515}}} {[astro-ph.EP]}

\bibitem[{{Yang} et~al.(2014){Yang}, {Bou{\'e}}, {Fabrycky}, and
  {Abbot}}]{Yang2014}
{Yang} J, {Bou{\'e}} G, {Fabrycky} DC, et~al (2014) {Strong Dependence of the
  Inner Edge of the Habitable Zone on Planetary Rotation Rate}. The
  Astrophysical Journal Letters 787(1):L2. \doi{10.1088/2041-8205/787/1/L2},
  {\href{https://arxiv.org/abs/1404.4992}{{arXiv:1404.4992}}} {[astro-ph.EP]}

\bibitem[{{Yung} and {Demore}(1982)}]{Yung+Demore1982}
{Yung} YL, {Demore} WB (1982) {Photochemistry of the stratosphere of Venus:
  Implications for atmospheric evolution}. Icarus 51(2):199--247.
  \doi{10.1016/0019-1035(82)90080-X}

\bibitem[{{Yung} et~al.(2009){Yung}, {Liang}, {Jiang}, {Shia}, {Lee},
  {B{\'e}zard}, and {Marcq}}]{Yung2009}
{Yung} YL, {Liang} MC, {Jiang} X, et~al (2009) {Evidence for carbonyl sulfide
  (OCS) conversion to CO in the lower atmosphere of Venus}. Journal of
  Geophysical Research (Planets) 114(16):E00B34. \doi{10.1029/2008JE003094}

\bibitem[{{Zasova} et~al.(1981){Zasova}, {Krasnopolskii}, and
  {Moroz}}]{Zasova1981}
{Zasova} LV, {Krasnopolskii} VA, {Moroz} VI (1981) {Vertical distribution of
  SO$_{2}$ in upper cloud layer of Venus and Origin of U.V.-absorption}.
  Advances in Space Research 1(9):13--16. \doi{10.1016/0273-1177(81)90213-1}

\bibitem[{{Zeleznik}(1991)}]{Zeleznik1991}
{Zeleznik} FJ (1991) {Thermodynamic Properties of the Aqueous Sulfuric Acid
  System to 350 K}. Journal of Physical and Chemical Reference Data
  20(6):1157--1200. \doi{10.1063/1.555899}

\bibitem[{{Zhang} and {Showman}(2018)}]{Zhang+Showman2018}
{Zhang} X, {Showman} AP (2018) {Global-mean Vertical Tracer Mixing in Planetary
  Atmospheres. I. Theory and Fast-rotating Planets}. The Astrophysical Journal
  866(1):1. \doi{10.3847/1538-4357/aada85},
  {\href{https://arxiv.org/abs/1803.09149}{{arXiv:1803.09149}}} {[astro-ph.EP]}

\bibitem[{{Zhang} et~al.(2012){Zhang}, {Liang}, {Mills}, {Belyaev}, and
  {Yung}}]{Zhang2012a}
{Zhang} X, {Liang} MC, {Mills} FP, et~al (2012) {Sulfur chemistry in the middle
  atmosphere of Venus}. Icarus 217(2):714--739.
  \doi{10.1016/j.icarus.2011.06.016}

\end{thebibliography}

\end{document}